\documentclass[prb,aps,twocolumn,sshowpacs,floatfix]{revtex4}
\usepackage{graphicx}
\usepackage{amssymb}
\usepackage{amsmath}
\usepackage{xspace} 
\usepackage{color}

\renewcommand{\vec}[1]{\mathbf{#1}}

\begin{document}
\title{Ground-state and spectral signatures of cavity exciton-polariton condensates}
\author{Van-Nham Phan}
\affiliation{Institute of Research and Development, Duy Tan University, K7/25 Quang Trung, Danang, Vietnam}
\author{Klaus W. Becker}
\affiliation{
Institut f{\"u}r Theoretische Physik, Technische Universit{\"a}t Dresden, D-01062 Dresden, Germany}
\author{Holger Fehske}
\affiliation{
Institut f{\"u}r Physik, Ernst-Moritz-Arndt-Universit{\"a}t Greifswald, D-17489 Greifswald, Germany}
%\pacs{71.45.Lr, 71.35.Lk, 63.20.kk, 71.30.+h, 71.28.+d}

\begin{abstract}
We propose a projector-based renormalization framework to study exciton-polariton Bose-Einstein condensation in a microcavity matter-light system. Treating Coulomb interaction and electron-hole/photon coupling effects on an equal footing, we analyze the ground-state properties
of the exciton-polariton model according to the detuning and the excitation density. We demonstrate that the condensate by its nature shows a crossover from an excitonic insulator (of Bose-Einstein, respectively, BCS type) to a polariton and finally photonic condensed state as the excitation density increases at large detuning. If  the detuning is weak, polariton or photonic phases dominate. While in both cases a notable  renormalization of the quasiparticle band structure occurs that strongly  affects the coherent part of the excitonic luminescence, 
the incoherent wave-vector-resolved luminescence spectrum develops a flat bottom only for small detuning.  
\end{abstract}
\date{\today}
\maketitle
\section{Introduction}
For several decades, there has been a considerable research effort to find Bose-Einstein condensation (BEC) in a solid-state system~\cite{GSS95,MS00}. That excitons in semiconductors might condense into a macroscopic phase-coherent ground state was theoretically proposed about 50 years ago~\cite{BBB62,Mo62}.  Experimentally, this has proved challenging, mainly because excitons are normally formed by optical excitations, and a cold degenerate Bose gas of sufficiently high density needs to be prepared on a shorter time scale than the excitons can decay in~\cite{LEKMSS04}.  At high densities, however, very efficient exciton-exciton annihilation processes set  in whose rates scale with the square of the exciton density.  As a result, to date, all attempts to create a dense gas of excitons in a bulk crystal, e.g., in $\rm Cu_2O$,  or in a potential trap did not demonstrate conclusively excitonic BEC (for a recent review, see, e.g.~Ref.~\onlinecite{SSKSKSSNKF12}). 

Different from optically created exciton condensates, the exciton insulator (EI) constitutes a quantum condensed state  in equilibrium~\cite{Mo61,Kno63,HR68a}. In this case, at low temperatures, electronic correlations can cause an anomaly at the semimetal-semiconductor transition that triggers an excitonic instability where  the conventional ground state of the crystal becomes unstable with respect to the spontaneous formation of excitons. Depending on from which side of the semimetal-semiconductor transition the EI is approached, the EI typifies either as a BCS condensate of loosely bound electron-hole pairs or as a Bose-Einstein condensate  of preformed tightly bound excitons~\cite{BF06,ZIBF12}.  Although there are some EI materials under debate~\cite{BSW91,WSTMANTKNT09,MSGMDCMBTBA10}, again we have no positive experimental proof of such an excitonic 
condensate.  

In contrast, polaritons in semiconductor microcavities have been observed to exhibit BEC~\cite{DWSBY02,Kaea06}.  
These experiments have been performed in the low-density regime;  the polaritons are nonetheless not ideal (noninteracting) bosons.
Besides, the polariton system is neither conservative nor in thermal equilibrium with the phonon (heat) bath. Even so, semiconductor exciton
polaritons constitute a promising system to explore the physics of Bose gases, but in a stronger interaction regime~\cite{DHY10}. Thereby, the excitonic (bound electron-hole pair) ``matter'' component and the strongly confined (photon-field) ``light'' component should be preferably treated on an equal footing. Likewise, the cases of low- and high-excitation densities should be described in a consistent scheme.  Thereby the relationship between a polariton BEC, polariton, and  photon lasing has to be clarified~\cite{KO11,BKY14}. Here, a natural way is to analyze the luminescence spectrum of the system~\cite{SVG94,La98,SS10}.

 In this work, we investigate a many-body Hamiltonian describing a coupled electron-hole/photon system in a microcavity.
 In addition to the lattice periodic potential, the electrons and holes experience a Coulomb interaction and a coupling to the light field. In the past, mean-field theories have been used to study the limits of low-excitation densities~\cite{EL01}  and high-excitation densities~\cite{AEKY77}  separately.  An extension to the medium-density regime has been addressed more recently by use of a variational (mean-field) 
 treatment~\cite{KO11}. Here, we employ a projector-based renormalization method (PRM)~\cite{BHS02,PBF10,PFB11} that allows to incorporate fluctuation processes beyond mean field  in the entire excitation density range and treats the Coulomb interaction on an equal footing with the light-matter coupling.  Moreover, depending on the 
 bare band structure (semiconducting or semimetallic)  and the detuning, we can address the formation of (BEC- or BCS-type) 
 excitonic (insulator) phases, polariton and photonic condensates. Assuming that the polariton lifetime is longer than the thermalization time, we will first analyze the ground-state properties of the microcavity polariton system~\cite{SKL06,KO11}.  Since the PRM permits the calculation of spectral properties as well, in a second step,  we will evaluate the excitonic luminescence. The paper is organized as follows. In Sec.~II, we will introduce the exciton-polariton model and present its mean-field  solution to set the stage for the more elaborate PRM treatment outlined in Sec. III. Details of the PRM calculation can be found in the Appendixes. The 
 numerical results are discussed in Sec.~IV. Here, in particular, the behavior of the excitonic/photonic order parameters will be diagramed, just as the particle/photon excitation densities. Moreover, the luminescence spectra will be presented, both wave-vector resolved and integrated.   
 Section~V contains a brief summary and our main conclusions.

%--------------------------------------------
\section{Exciton-polariton model}
\label{II} 
%---------------------------------------------

 In the following, we study a model Hamiltonian for a polariton system  
in a semiconductor microcavity, which is in thermal equilibrium.  Although experiments are usually performed  
away from equilibrium, there are reasons also  to study the stationary state of a closed microcavity polariton system which appears to be well described by its ground state~\cite{KO11}. On the one hand, the quality of microcavity fabrication and  of mirrors  will improve, so that the experimental situation becomes closer to thermal equilibrium.  On the other hand, thermal equilibrium may be considered as the limiting case of a 
non-equilibrium situation. This is the case, when the decay rates for the loss of cavity photons and of fermions, for instance  due to phonons or impurities, into  external bath variables become small~\cite{KESL05,SKL06}. 

  A model which is commonly used to describe such a microcavity polariton system is based on the Hamiltonian\cite{KO11} 
\begin{eqnarray}
\label{1}
\mathcal{H}=  \mathcal{H}_\textrm{el}+ \mathcal H_\textrm{ph} +\mathcal{H}_\textrm{el-ph}
+\mathcal{H}_\textrm{el-el}.
\end{eqnarray}
The first term $\mathcal{H}_\textrm{el}$   considers spinless free conduction electrons and valence holes 
with creation and annihilation operators 
$e^{(\dag)}_{\vec k}$, $h^{(\dag)}_{\vec k}$:
\begin{eqnarray}
\label{2}
&&  \mathcal{H}_\textrm{el} =  \sum_{\vec{k}}\varepsilon_{\vec{k}}^{e} e_{\vec{k}}^{\dagger}e_{\vec{k}}^ {}+\sum_{\vec{k}}\varepsilon_{\vec{k}}^{h}h_{\vec{k}}^{\dagger}h_{\vec{k}}^ {}\,,  \\
\label{3}
&& \varepsilon_{\vec{k}}^{e} = -2t \sum_i^{D}  \cos k_i+\frac{E_g+4tD-\mu}{2} = \varepsilon_{\vec{k}}^{h} \, ,
\end{eqnarray}
where  symmetric tight-binding dispersions $\varepsilon^e_{\vec k}= \varepsilon^h_{\vec k}$
for the respective excitation energies were assumed. In \eqref{3}, $t$ denotes the particle transfer amplitude, $E_g$ gives the minimum distance (gap) between the bare electron and hole bands, and $D$ is the dimension of the hypercubic lattice. Note that a semimetallic setting  occurs when $E_g<0$.

The second term $\mathcal H_{\rm ph}$ is the free photon Hamiltonian with photon 
creation (annihilation) operators $\psi^{\dag}_{\vec q}$ ($\psi_{\vec q}$): 
\begin{eqnarray}
\label{4}
&& \mathcal H_\textrm{ph} =  \sum_{\vec{q}}\omega_{\vec{q}}\psi_{\vec{q}}^{\dagger}\psi_{\vec{q}}^{},
\\
\label{5}
&& \omega_{\vec{q}}=\sqrt{(c{\vec q})^{2}+ \omega_{c}^{2}}-\mu \, .
\end{eqnarray}
Here, $\omega_{\vec q}$ is the photonic excitation energy with  a zero-point cavity frequency $\omega_c$,  and $c$ is the speed of light in the microcavity. 

The last two terms in Hamiltonian \eqref{1} are  a local (attractive) Coulomb interaction between electrons and holes 
and a local interaction between the electron-hole system and photons with coupling constant $g$:
\begin{eqnarray}
\label{6}
&& \mathcal{H}_\textrm{el-el} = -\frac{U}{N} \sum_{\vec k} \rho^e_{\vec k} \rho^h_{-\vec k}\,, \\
\label{7}
&& \mathcal{H}_\textrm{el-ph}   = -\frac{g}{\sqrt{N}}\sum_{\vec{q}\vec{k}}
[e_{\vec{k}+\vec{q}}^{\dagger}h_{-\vec{k}}^{\dagger}\psi_{\vec{q}}^ {}+\textrm{H.c.}] \, ,
\end{eqnarray}
where densities for electrons and holes have been introduced 
$\rho^e_{\vec k}=\sum_{\vec k_1} e^\dag_{\vec k + \vec k_1} e_{\vec k_1}^{}$ and  
$\rho^h_{\vec k}= \sum_{\vec k_1} h^\dag_{\vec k + \vec k_1} h_{\vec k_1}^{}$.  
In principle, additional  electron-electron and hole-hole Coulomb interactions  
might have been taken into account in Eq.~\eqref{6}.  However, they only lead to mere shifts in the one-particle dispersions $\varepsilon^e_{\bf k}$ and $\varepsilon^h_{\bf k}$, since spinless electrons and holes    
as well as a wave-vector independent Coulomb coupling  $U$ are considered in model \eqref{1}. 

Note that in Eqs.~\eqref{3} and \eqref{5}  a chemical potential $\mu$ was included 
to ensure that the total number of excitations 
\begin{equation}
\label{8} 
\mathcal N_\textrm{exc} = \sum_{\bf q}  \psi_{\vec{q}}^{\dagger}\psi_{\vec{q}}^{} + \frac{1}{2} \sum_{\bf k}
(e_{\vec{k}}^{\dagger}e_{\vec{k}}^ {}+ h_{\vec{k}}^{\dagger}h_{\vec{k}}^ {}) 
\end{equation}
is fixed. Clearly, $\mathcal N_\textrm{exc}$  is conserved for Hamiltonian $\mathcal H$. 

\begin{figure}[t]
\includegraphics[width=0.4\textwidth]{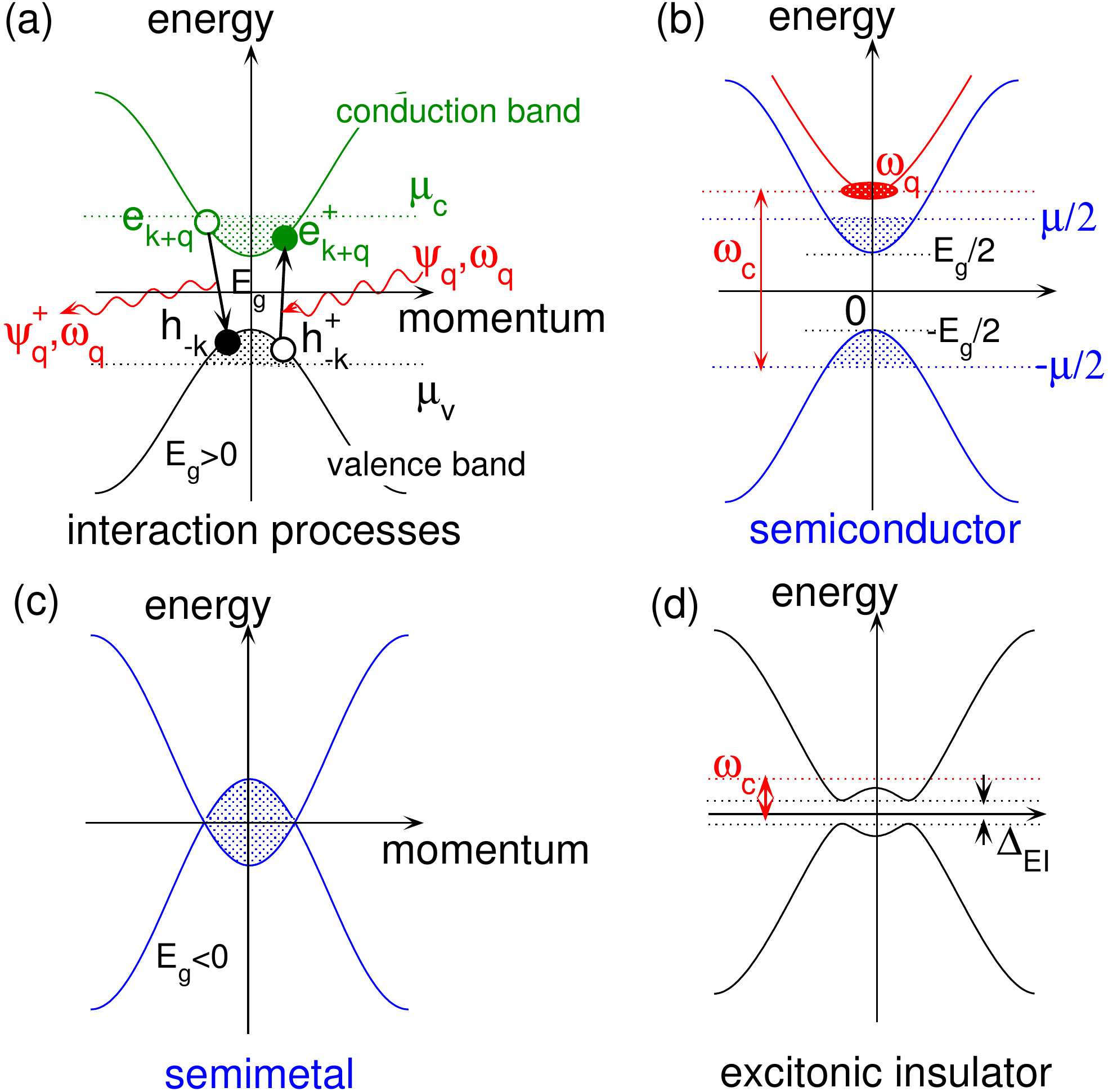}
\caption{(Color online) Microcavity exciton-polariton model~\eqref{1} studied in this work. Panel (a) represents the light-matter interaction processes taken into account. Panel (b)  gives the band structure and relevant energy scales for a semiconducting situation. In panels (c) and (d), a semimetal with $E_g <0$ (overlapping bands) is realized which
might exhibit an excitonic instability that transforms the systems into an excitonic insulator~\cite{PBF10}.}
\label{fig1}
\end{figure}

Apparently, the influence of
 $\mathcal{H}_\textrm{el-ph}$ becomes most important, when the excitation energy of a particle-hole pair
 roughly agrees with a photon excitation. Therefore, for later interpretation of this effect one best 
 introduces the so-called detuning parameter~\cite{KO11}
 \begin{equation}
 \label{9}
 d= \omega_c - E_g  \, .
\end{equation}
Figure~\ref{fig1} illustrates the model under consideration.

%---------------------------------------------------------
%\subsection{Mean-field Hamiltonian}

Let us proceed by separating the mean-field approximation from model \eqref{1}. Introducing the normal ordering for the operator expressions in $\mathcal H_\textrm{el-el}$ and   $\mathcal H_\textrm{el-ph}$,
\begin{eqnarray}
\label{10}
&& :e_{\vec{k}_{1}+\vec k}^{\dagger}e_{\vec{k}_{1}}h_{\vec{k}_{2}-\vec k}^{\dagger}
h^{}_{\vec{k}_{2}}:
=e_{\vec{k}_{1}+\vec k}^{\dagger}e_{\vec{k}_{1}}h_{\vec{k}_{2}-\vec k}^{\dagger}h_{\vec{k}_{2}}
\nonumber\\
&& \quad -\delta_{\vec{k}, 0}(n^e_{{\vec k}_1}:h_{\vec{k}_{2}}^{\dagger}
h_{\vec{k}_{2}}:   
+n^h_{{\vec k}_2 } :e_{\vec{k}_{1} }^{\dagger} e_{\vec{k}_{1} }:) \nonumber \\
&&\quad  
-\delta_{\vec{k}_{1},-\vec{k}_{2}}(d^{}_{{\vec k}_1 +\vec k}
:h^{}_{-\vec{k}_{1}} e{}_{\vec{k}_{1}}:
+d^{}_{{\vec k}_1}
:e_{\vec{k}_{1}+\vec k}^{\dagger} h_{-\vec{k}_{1}-\vec k}^{\dagger}:)\,,\nonumber \\
&&  \\
\label{11}
&&:e_{\vec{q}+\vec{k}}^{\dagger}h_{-\vec{k}}^{\dagger}\psi_{\vec{q}}:
=e_{\vec{q}+\vec{k}}^{\dagger}h_{-\vec{k}}^{\dagger}\psi_{\vec{q}}\nonumber \\
&&\qquad -\delta_{\vec{q},0}(d_{\vec k}:\psi_{0}:+\langle\psi_{0}\rangle
:e_{\vec{k}}^{\dagger}h_{-\vec{k}}^{\dagger}:)\,,
\end{eqnarray}
Hamiltonian \eqref{1} is rewritten as 
\begin{equation}
\label{12}
\mathcal H= \mathcal H_0 + \mathcal H_1
\end{equation}
with
\begin{eqnarray}
\label{13}
\mathcal{H}_0 & = & \sum_{\vec{k}}\hat{\varepsilon}_{\vec{k}}^{e}e_{\vec{k}}^{\dagger}e_{\vec{k}}^{}+\sum_{\vec{k}}
\hat{\varepsilon}_{\vec{k}}^{h}h_{\vec{k}}^{\dagger}h_{\vec{k}}^{}
+\sum_{\vec{q}}\omega_{\vec{q}}\psi_{\vec{q}}^{\dagger}\psi_{\vec{q}}^{}\nonumber \\
&+&\Delta\sum_{\vec{k}}(e_{\vec{k}}^{\dagger}h_{-\vec{k}}^{\dagger}+\textrm{H.c.})
+(\sqrt{N}\Gamma\psi_{0}^{\dagger}+\textrm{H.c.}),   \\
\label{14}
\mathcal H_1 &=& 
-\frac{g}{\sqrt{N}}\sum_{\vec{k}\vec{q}}(:e_{\vec{q}+\vec{k}}^{\dagger}
h_{-\vec{k}}^{\dagger}\psi^{}_{\vec{q}}:+\textrm{H.c.})\nonumber \\
&-&\frac{U}{N}\sum_{\vec{k}_{1}\vec{k}_{2}\vec{k}}
:e_{\vec{k}_{1}+\vec k}^{\dagger}e^{}_{\vec{k}_{1}} h_{\vec{k}_{2}-\vec k}^{\dagger}h^{}_{\vec{k}_2}: \, .
\end{eqnarray}
Additional constants have been neglected. In $\mathcal H_0$ the electronic 
excitation energies have acquired Hartree shifts
\begin{eqnarray}
\label{15}
  && \hat{\varepsilon}_{\vec{k}}^{e}=\varepsilon_{\vec{k}}^{e}
    - \frac{U}{N}\sum_{\vec{q}}n^h_{\vec q},   \\
 &&   \hat{\varepsilon}_{\vec{k}}^{h}=\varepsilon_{\vec{k}}^{h}-\frac{U}{N}\sum_{\vec{q}}n^e_{\vec q},
\end{eqnarray}
with 
\begin{equation}
\label{16}
 n^e_{\vec k}=\langle e^\dagger_{\vec k}e^{}_{\vec k} \rangle, \quad 
n^h_{\vec k}=\langle h^\dagger_{\vec k}h^{}_{\vec k} \rangle \,  .
\end{equation}
The last two contributions in $\mathcal H_0$ are additional fields with prefactors
which will act below as order parameters  for the exciton-polariton condensate:
\begin{eqnarray}
\label{17}
&&  \Delta=-\frac{g}{\sqrt{N}}\langle\psi_{0}\rangle-\frac{U}{N}\sum_{\vec{k}}d_{\vec k}  \,,  \\
 \label{18}
 &&  \Gamma=-\frac{g}{N}\sum_{\vec{k}}d_{\vec k} \, ,  \\
\label{19} 
&& d_{\vec k}=\langle e^\dagger_{\vec k}h^\dagger_{-\vec k} \rangle=\langle h^{}_{-\vec k}e^{}_{\vec k} \rangle=
d_{\vec k}^*\,.
\end{eqnarray}
Note that Hamiltonian $\mathcal H= \mathcal H_0 + \mathcal H_1$,  with $\mathcal H_0$ and
 $\mathcal H_1$ given by Eqs.~\eqref{13} and \eqref{14}, is still exact. 
 The mean-field approximation is obtained by completely neglecting the fluctuation part $\mathcal H_1$, 
 i.e.,~$\mathcal H_{\rm MF} = \mathcal H_0$. However, in the following we are mostly
 interested in the influence of fluctuation contributions  to the physical behavior of an exciton-polariton 
 condensate. Therefore, Hamiltonian $\mathcal H_1$ has to be taken into account.   
 
Expression \eqref{13} for $\mathcal H_0$ can be further  simplified since 
the terms $\propto\psi_0^\dag$ and $\propto\psi_0$ 
can be eliminated by defining new displaced photon operators 
\begin{equation}
\label{20}
\Psi_{{\bf q}}^{\dag}=\psi_{{\bf q}}^{\dag}+\frac{\sqrt{N}\Gamma}{\omega_{\bf q =0}}\delta_{{\bf q,0}}\,.
\end{equation}
Then,
\begin{eqnarray}
\label{21}
\mathcal{H}_0 & = & \sum_{\vec{k}}\hat{\varepsilon}_{\vec{k}}^{e}e_{\vec{k}}^{\dagger}e_{\vec{k}}^{}+\sum_{\vec{k}}\hat{\varepsilon}_{\vec{k}}^{h}h_{\vec{k}}^{\dagger}h_{\vec{k}}^{}
+\sum_{\vec{q}}\omega_{\vec{q}}\Psi_{\vec{q}}^{\dagger}\Psi_{\vec{q}}^{}\nonumber \\
&+&\Delta\sum_{\vec{k}}(e_{\vec{k}}^{\dagger}h_{-\vec{k}}^{\dagger}+\textrm{H.c.})
\end{eqnarray}
and 
\begin{eqnarray}
\label{22}
\mathcal H_1 &=& 
-\frac{g}{\sqrt{N}}\sum_{\vec{k}\vec{q}}[:e_{\vec{q}+\vec{k}}^{\dagger}
h_{-\vec{k}}^{\dagger}\Psi^{}_{\vec{q}}:+\textrm{H.c.}]\nonumber \\
&-&\frac{U}{N}\sum_{\vec{k}_{1}\vec{k}_{2}\vec{k}}
:e_{\vec{k}_{1}+\vec k}^{\dagger}e^{}_{\vec{k}_{1}} h_{\vec{k}_{2}-\vec k}^{\dagger}h^{}_{\vec{k}_2}: \, ,
\end{eqnarray}
where the shift from Eq.~\eqref{20} cancels in the first normal order product term of $\mathcal H_1$. 
Moreover, the electronic part of $\mathcal H_0$ can be diagonalized by means of  a Bogoliubov transformation
(compare Appendix \ref{MF}). 

%----------------------------------------------------------------------------
\section{Influence of fluctuation processes}
\label{III}
%------------------------------------------------------------------------------

In mean-field treatment fluctuation processes from $\mathcal H_1$ are
completely neglected. In the following, we apply the projective renormalization method~\cite{BHS02}  (PRM)
in order to evaluate the order parameters, the electron and photon densities,   and the response functions $A(\vec k,  \omega)$ and $B(\vec q, \omega)$ of the exciton polarization and  the cavity photon mode, respectively,  
for the case that $\mathcal H_1$ is included. The technical details of this calculation are   
shifted to Appendix \ref{A}.  The general concept of the PRM is as follows: 
The presence of the interaction $\mathcal H_1$ usually
prevents a straightforward solution of the Hamiltonian $\mathcal H= \mathcal H_0 + \mathcal H_1$. 
However, by integrating out the interaction $\mathcal H_1$, the Hamiltonian can be transformed 
into a diagonal (or at least quasi-diagonal) form by applying
a sequence of small unitary transformations to $\mathcal H$. 
Denoting for a moment the corresponding generator   
of the whole sequence by $X= -X^\dag$, it is shown  in Appendix~\ref{A} how one arrives 
at an effective Hamiltonian  $\tilde{\mathcal H}= e^X \mathcal He^{-X}$, which has  
the same operator structure as Hamiltonian $\mathcal H_0$  from Eq.~\eqref{21}, 
\begin{eqnarray}
\label{23}
\tilde{\mathcal{H}} & = & \sum_{\vec{k}}\tilde{\varepsilon}_{\vec{k}}^{e}e_{\vec{k}}^{\dagger}e_{\vec{k}}^{}+\sum_{\vec{k}}\tilde{\varepsilon}_{\vec{k}}^{h}h_{\vec{k}}^{\dagger}h_{\vec{k}}^{}
+\sum_{\vec{q}}\tilde\omega_{\vec{q}} \tilde{\Psi}_{\vec{q}}^{\dagger}\tilde{\Psi}_{\vec{q}}^{}\nonumber \\
&+&\sum_{\vec{k}}\tilde \Delta_{\vec k}(e_{\vec{k}}^{\dagger}h_{-\vec{k}}^{\dagger}+\textrm{H.c.})\,.
\end{eqnarray}
Here,  $\tilde{\Psi}^\dag_{\mathbf q}$ is  defined by $\tilde{\Psi}^\dag_{\mathbf q}= \psi^\dag_{\bf q} 
+ ({\sqrt{N} \tilde{\Gamma}}/{\tilde{\omega}_{\bf q=0}}) \delta_{\bf q , 0}$  and $\tilde{\varepsilon}^e_{\mathbf{k}}$, $\tilde{\varepsilon}^h_{\mathbf{k}}$, $\tilde{\omega}_{\bf q}$, 
and $\tilde{\Delta}_{\vec k}$ are parameters which are renormalized in the elimination process. 
They have to be determined self-consistently by taking into account contributions 
to infinite order in the interaction $\mathcal H_1$. The  PRM ensures a well-controlled 
disentanglement of higher-order interaction terms within the elimination procedure. 

We  would like to emphasize that the renormalized quantities $\tilde\Delta_{\vec k}$ just as  $\tilde \Gamma$
in $\Psi_{\vec q}$ play the role of exciton-polariton order parameters for the full system~\eqref{1}. Thereby,
both types of interactions contribute. In particular, both $\mathcal H_\textrm{el-ph}$ and $\mathcal H_\textrm{el-el}$ 
make contributions to $\tilde\Delta_{\vec k}$, where their mutual influence in the formation of a
condensate will be of interest. On the other hand, the shift $\sim \tilde \Gamma$ 
in $\Psi_{\vec q}$ alone leads to a polarization of the photonic subsystem.  
 In case the detuning parameter  $d$ [Eq.~\eqref{9}] is small the tendency for the 
 formation of a photonic condensate is expected to be enhanced. In contrast, for large $d$ the photonic contribution to $\tilde \Delta$  should be small, at least for a not too large excitation density $n_\textrm{exc}=\tfrac{1}{N}\langle \mathcal N_\textrm{exc}\rangle$.          

The PRM also allows to evaluate expectation values $\langle  \mathcal A \rangle$, 
formed with the full Hamiltonian $\mathcal H$. 
Thereby, one uses the property of unitary invariance of operator expressions 
under a trace. Employing the same unitary transformation to $\mathcal A$ as before to
the Hamiltonian, one finds $\langle {\mathcal A} \rangle =
 \langle \tilde{\mathcal A} \rangle_{\tilde{\mathcal H}}$,  where the expectation value on the right-hand side 
 is now formed with $\tilde{\mathcal H}$,  and $\tilde{\mathcal A} = e^X \mathcal A e^{-X}$.  
Just as $\mathcal H_{0}$ before also Hamiltonian $\tilde{\mathcal H}$ can be transformed 
into a diagonal form by a Bogoliubov transformation. Therefore, any expectation value, 
formed with  $\tilde{\mathcal H}$, can be evaluated. 

As a  first example, let us consider  the response function for the excitonic polarization   
$A(\vec k,  \omega)$, which is defined by the following linear response
\begin{eqnarray}
\label{24}
A(\vec k,  \omega) &=&  \frac{1}{2\pi} \int_{-\infty}^\infty 
\langle [{b}_{\bf k}(t), {b}^\dag_{\bf k }]_-\rangle \,  e^{i\omega t} dt  \, ,
\end{eqnarray}
with respect to an external $\vec k$- and $\omega$-dependent field. 
Here, $ {b}^\dag_{\bf k }$ is the excitonic creation operator  
 \begin{equation}
 \label{25}
b_{\vec{k}}^{\dagger}=\frac{1}{\sqrt{N}}\sum_{\vec{q}}e_{\vec{k}+\vec{q}}^{\dagger}h_{-\vec{q}}^{\dagger}
\, .
\end{equation}
Applying the  
unitary invariance of operator expressions under a trace,  $A({\bf k}, \omega)$ is rewritten as   
 \begin{eqnarray}
\label{26}
A({\bf k}, \omega) &=&  \frac{1}{2\pi} \int_{-\infty}^\infty 
\langle [\tilde{b}_{\bf k}(t), \tilde{b}^\dag_{\bf k}]_-\rangle_{\tilde{\mathcal H}} \,  e^{i\omega t} dt  \, ,
\end{eqnarray}
where the expectation value is now formed with  $\tilde{\mathcal H}$ instead of with $\mathcal H$.
Correspondingly, $\tilde{b}^{(\dag)}_{\bf k}$ are the transformed electron operators, 
$\tilde{b}^{(\dag)}_{\bf k}= e^X {b}^{(\dag)}_{\bf k} e^{-X}$, and the time dependence in Eq.~\eqref{26}
is  governed by $\tilde{\mathcal H}$ as well. Explicit expressions for both coherent and 
incoherent contributions to  $A({\bf k}, \omega)$ are derived in Appendix~B.

We note that $A(\vec k, \omega)$ is not a positive-definite
spectral function. However,  $A(\vec k, \omega)$ divided 
by $\omega$ has a positive sign for all $\omega$, i.e.,~$A(\vec k, \omega)/ \omega \ge 0$.  
The quantity $A(\vec k, \omega)$  has the advantage that it fulfills a simple 
sum rule
 \begin{equation}
 \label{27}
\int_{-\infty}^\infty A(\vec k, \omega) d\omega= \frac{1}{N} \sum_{\bf k'} \big[1 -(n^e_{\vec k'} + n^h_{\vec k'}) \big]
\end{equation}
(independent of $\vec k$), which will be used in the following to check the outcome of the numerics. 

 As a second example, we will  evaluate the response function of the cavity photon mode, 
which is sometimes called just luminescence function
\begin{eqnarray}
\label{27a}
B(\vec q,  \omega) &=&  \frac{1}{2\pi} \int_{-\infty}^\infty 
\langle [{\psi}_{\bf q}(t), {\psi}^\dag_{\bf q }]_-\rangle \,  e^{i\omega t} dt  \, .
\end{eqnarray} 
Applying the unitary transformation it can  be written as
  \begin{eqnarray}
\label{27b}
B(\vec q,  \omega) &=&  \frac{1}{2\pi} \int_{-\infty}^\infty 
\langle [\tilde {\psi}_{\bf q}(t), \tilde{\psi}^\dag_{\bf q }]_-\rangle_{\tilde{\mathcal H}} \,  e^{i\omega t} dt  \, ,
\end{eqnarray} 
where $\tilde{\psi}^\dag_{\bf q }$ is the fully transformed photon mode. $B(\vec q, \omega)$ will be evaluated in Appendix \ref{A}
as well. Note that  $B(\vec q, \omega)$ obeys the sum rule $\int_{-\infty}^\infty B(\vec q, \omega) d\omega =1$.

%--------------------------------------------------------------------------------------------------
\section{Numerical results}
\label{IV}
%-----------------------------------------------------------------------------------------------

In the numerical evaluation of the various physical quantities from Sec.~\ref{III} one 
has to solve the set of renormalization equations \eqref{A.12}-\eqref{A.17} self-consistently 
together with the expressions~\eqref{A.42a}-\eqref{A.42c},~\eqref{A.47} for the 
expectation values.  Starting with some chosen initial values for $n^e_{\mathbf{k}}$, $n^h_{\mathbf{k}}$, $n^\psi_{\mathbf{q}}$, and 
$\Delta_{\vec k} = \Gamma =0^+$, the renormalization equations are 
integrated in small steps $\Delta\lambda$ until at $\lambda=0$ the Hamiltonian is completely renormalized. 
Then, the expectation values can be recalculated and
the  renormalization process is restarted again. Convergence is achieved if all quantities are 
determined within some relative error of, for instance, less than $10^{-5}$. To simplify the numerics, we consider a one-dimensional setting hereafter, and limit the number of lattice sites to $N=160$. 
Nevertheless, the results presented in the framework of  the PRM approximation should also give a qualitative account of what happens in a higher-dimensional 
microcavity polariton system.

%----------------------------------------------------
\subsection{Ground-state properties}

Assuming a quasi-equilibrium situation, the ground-state of the system can be determined for a fixed excitation density $n_\textrm{exc}$ at zero temperature in dependence on the model parameters, i.e., according to the detuning $d$,  the electron-hole Coulomb attraction $U$, and the light-matter coupling strength $g$.  
Here and in what follows all energies are given in units of the particle transfer amplitude $t$  and the wave vectors in units of the lattice constant $a$. For the explicit evaluation one best introduces a dimensionless 
speed of  light $\bar c$ using $\hbar \omega_{\vec q}/t = [{\bar c}^2 ({\vec q}/\pi)^2 + (\hbar \omega_c/t)^2]^{(1/2)} -
(\hbar \mu/t)$ where $\bar c=(\hbar \pi/ a t)  c $. Taking $\bar c=80$ and typical values for $a \simeq  5 \textrm{\AA}$ 
and $t \simeq 2\textrm{eV}$ one is led  to a value of $c \simeq  0.4\,  c_0$ for the speed of light of the microcavity, which  is about half  the speed of light $c_0$ in vacuum. However, as we have noticed, most of the physical properties only slightly depend on the actual value of $c$.

%% ------------------------------------Fig 2 --------------------------
\begin{figure}[t]
\includegraphics[width=0.48\textwidth]{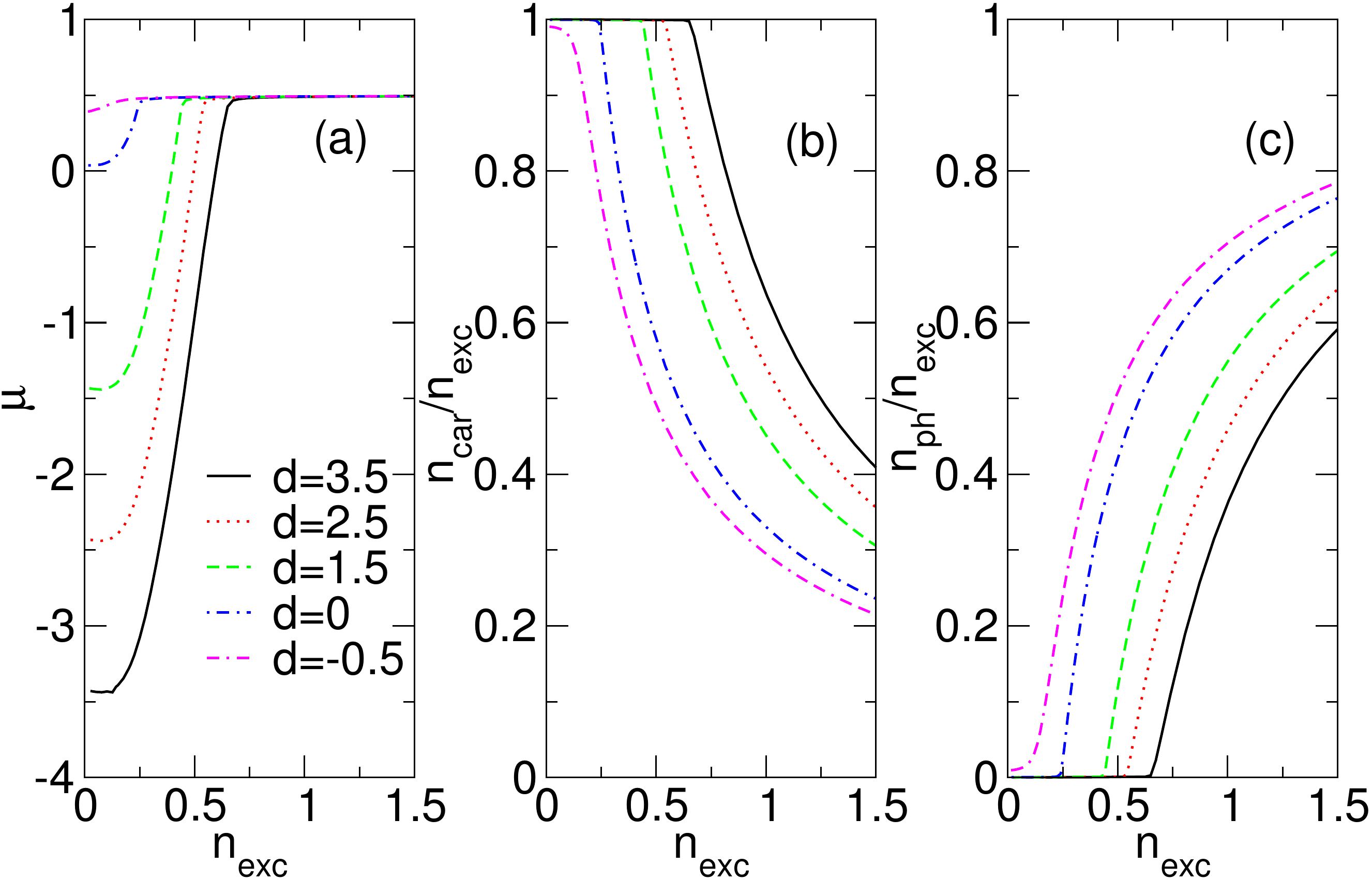}
\caption{(Color online) Chemical potential $\mu$ (a), density of excitons $n_{X}$ (b), and density of photons $n_\textrm{ph}$ (c), as a function of the total excitation density for various  detunings $d$. Model parameters are:
$U=2$, $g=0.2$, and $\omega_c=0.5$.}
\label{fig2}
\end{figure}

%% ---------------------------------- Fig. 3 -----------
\begin{figure}[t]
\includegraphics[width=0.4\textwidth]{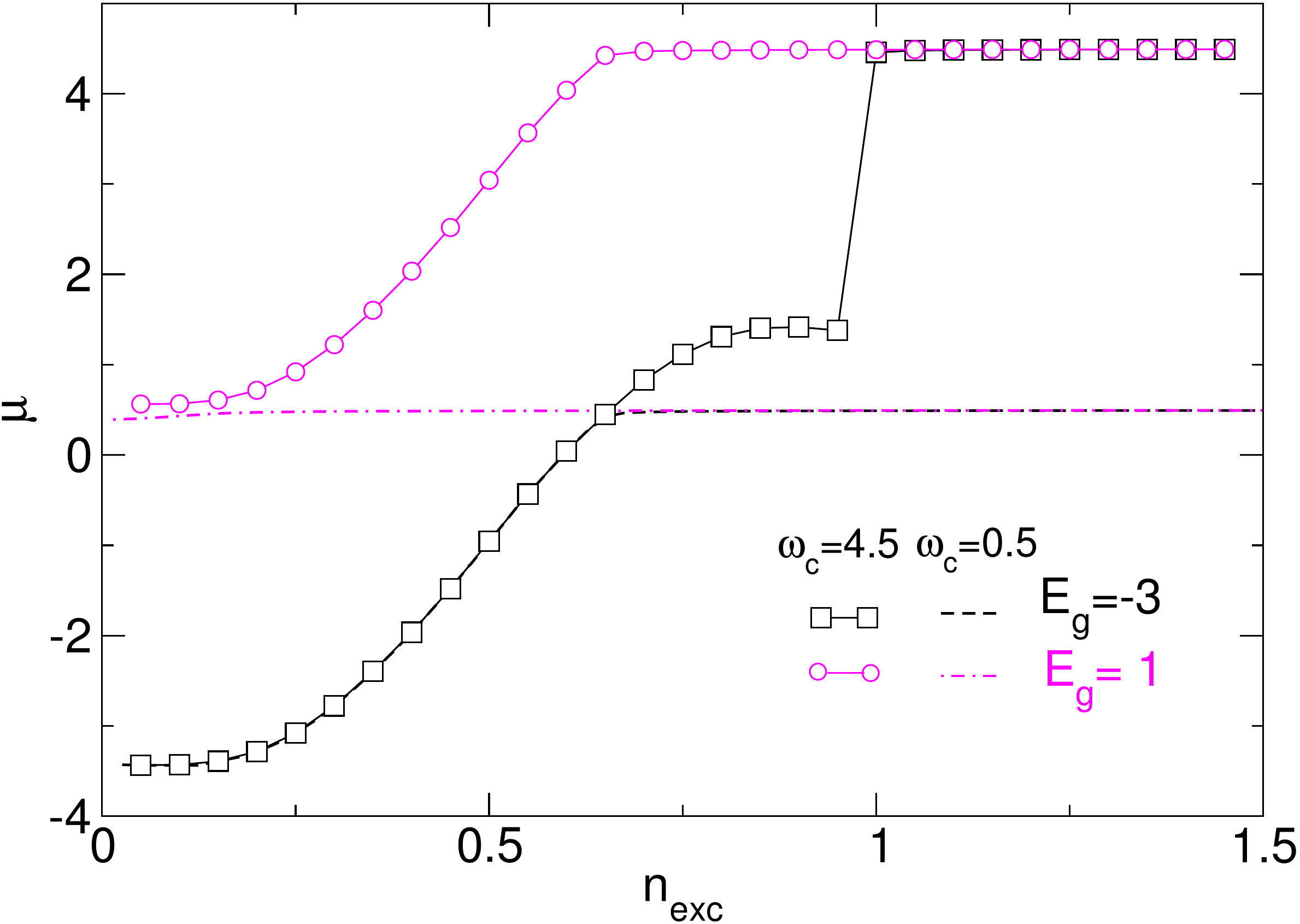}
\caption{(Color online) Chemical potential $\mu$ as a function of the excitation density $n_\textrm{exc}$ at fixed values of $E_g=1$ and $E_g=-3$.
Dashed lines mark the results for $\omega_c=0.5$ [cf. Fig.~\ref{fig2}(a)]; solid lines with symbols give the new data for 
$\omega_c=4.5$, where  $d=3.5$ and  $d=7.5$ result for $E_g=1$ and $E_g=-3$, respectively.  The interaction parameters are $U=2$ and $g=0.2$.}
\label{fig2-ii}
\end{figure}

Figure~\ref{fig2} shows how the chemical potential, the partial densities of carriers and photons
vary as the total number (density) of  excitations changes at $\omega_c=0.5$), for detunings 
ranging from $d=3.5$ ($E_g=-3$) to $d=-0.5$  ($E_g=1$). Recall negative (positive) values of $E_g$ lead 
to a semimetallic (semiconducting) bare band structure. As a matter of course, the chemical potential 
increases as the number of excitations increases [see Fig~\ref{fig2}~(a)]. 
The weak variation at small $n_\textrm{exc}$ is an effect of the van Hove singularity of the one-dimensional (1D) density of states, while the almost constant $\mu$ at large $n_\textrm{exc}$  can be traced back  to conduction electron phase-space filling: 
If $\mu$ reaches $\omega_c$, any further excitation (that minimizes the ground-state energy) will be photonic. The partial excitation densities of carriers and photons shown in Figs.~\ref{fig2}~(b) and (c), respectively, corroborate this scenario. We see that for large detuning the excitations in the low-density regime are basically electron-hole excitations. Thereby, the electrons and holes form an electron-hole plasma at weak-to-moderate values of $U$, or might bound into excitons in the strong-coupling regime. Increasing $n_\textrm{exc}$, above a certain threshold value a sharp onset of photon excitations takes place, signaling laser-like behavior~\cite{KO11}. The electron-hole plasma, respectively, excitonic domain appearing at low density shrinks as the detuning becomes smaller  and finally a very gradual (but still opposing) variation of $n_\textrm{car}$ and $n_\textrm{ph}$ is  observed as $n_\textrm{exc}$ increases. Obviously, now the quasiparticle excitations are a mixture of excitons and photons, i.e., they can be viewed as polaritons. 

This scenario is corroborated by Fig.~\ref{fig2-ii}, which compares  the variation of $\mu$  with $n_\textrm{exc}$ for small ($\omega_c=0.5$) and large ($\omega_c=4.5$) values of the cavity frequency when the gap parameter $E_g$  is kept fixed. For $E_g=-3$, yielding a large detuning in both 
low-$\omega_c$ and high-$\omega_c$ cases, the (continuous) $\mu(n_\textrm{exc})$ dependence is almost the same until $\mu$ intersects the photon energy.  As becomes clear from Fig.~\ref{fig2}(a) 
 for $\omega_c=0.5$ no photon excitations  are involved in the small  $n_\textrm{exc}$ regime below this intersection, which is also true for $\omega_c=4.5$.  Due to the same $E_g$ and thus the same dispersion 
$\varepsilon^e_{\vec k} =\varepsilon^h_{\vec k}$  for both cases the curves $\mu$ as a function  $n_\textrm{exc}$ should be the same as long as $\mu$ is smaller than $\omega_c=0.5$.   
 If the cavity frequency is (much) larger than the width of the bare band structure, we observe a jump at $n_\textrm{exc}=1$. Here all available electrons and holes are bound into excitons, i.e., any further excitation is purely photonic by their nature.

%% --------------------------------- Fig. 4 ------------------------------
\begin{figure}[t]
\includegraphics[width=0.44\textwidth]{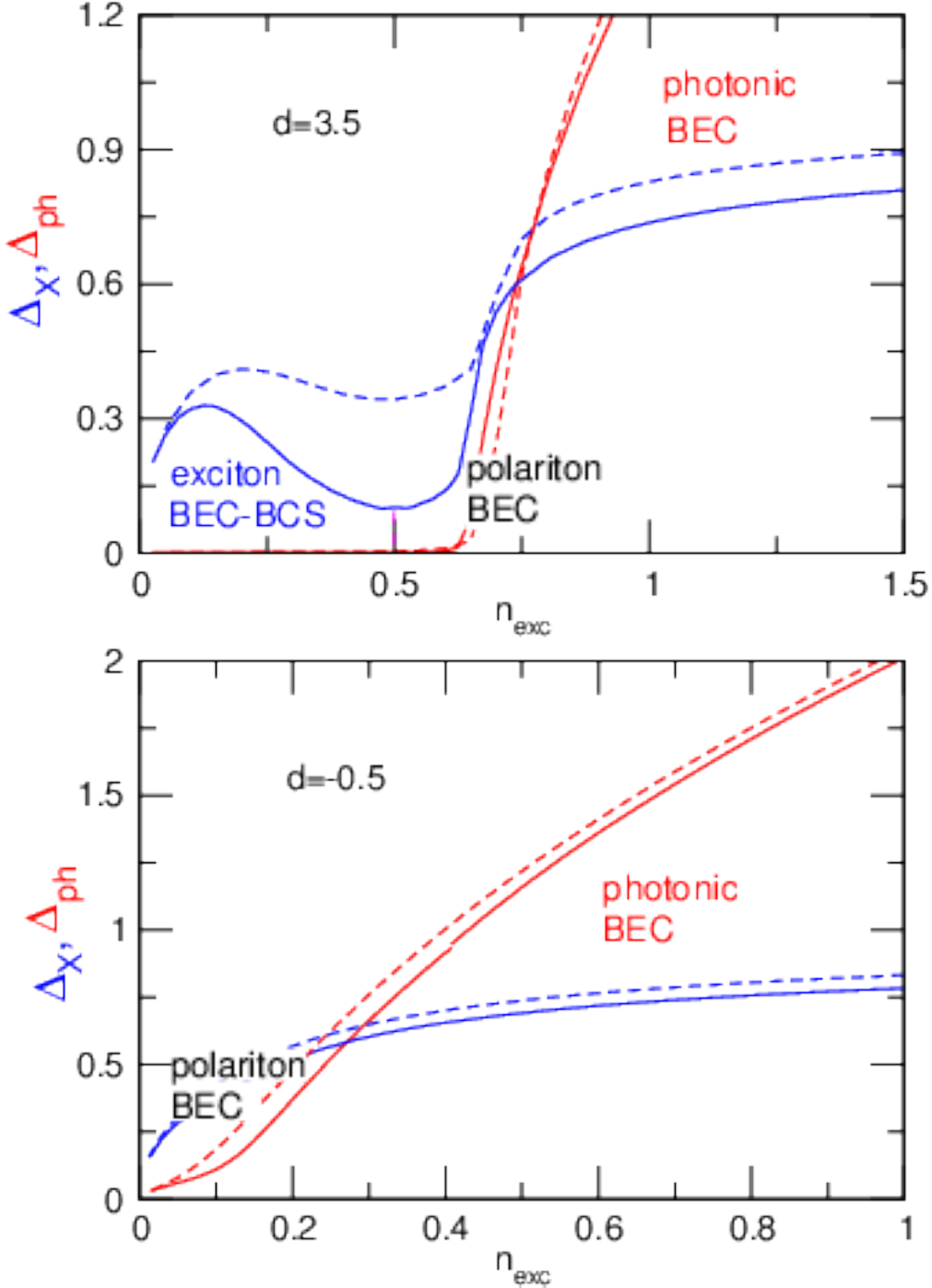}
\caption{(Color online) Excitonic ($\Delta_X$) and photonic ($\Delta_\textrm{ph}$) order parameters  as a function of the excitation density $n_\textrm{exc}$ at large (upper panel) and small (lower panel) detuning $d$. Different phases refer to the predominant nature of the condensate. Parameters are $U=2$, $g=0.2$, and $\omega_c=0.5$. The dashed lines give the corresponding results in the 
mean-field approximation (see Appendix \ref{MF}), which naturally overestimates the tendency towards the formation of condensed phases.}
\label{fig3}
\end{figure}
%

%% ------------------------------- Fig. 5 -------------------------
\begin{figure}[t]
\includegraphics[width=0.23\textwidth]{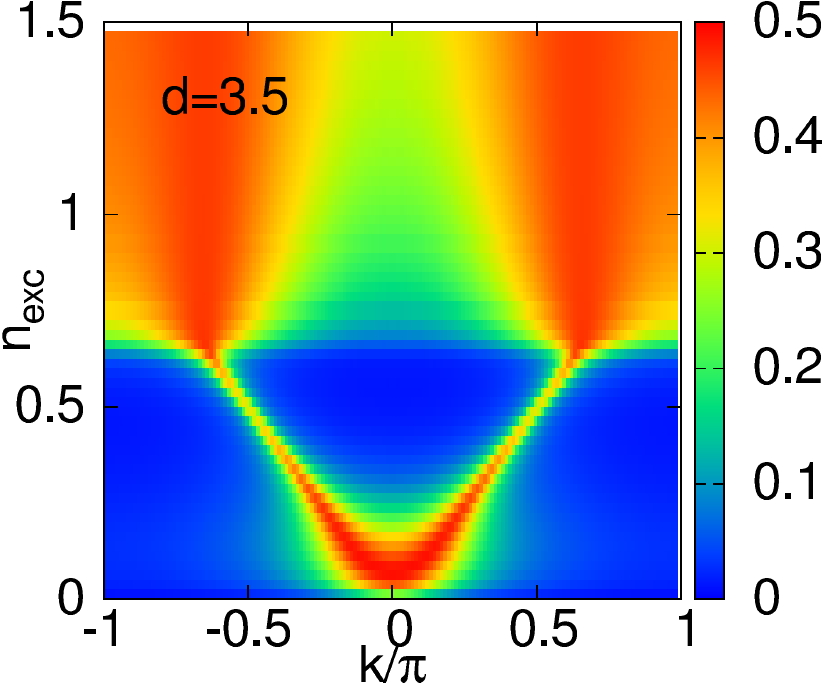}
\includegraphics[width=0.23\textwidth]{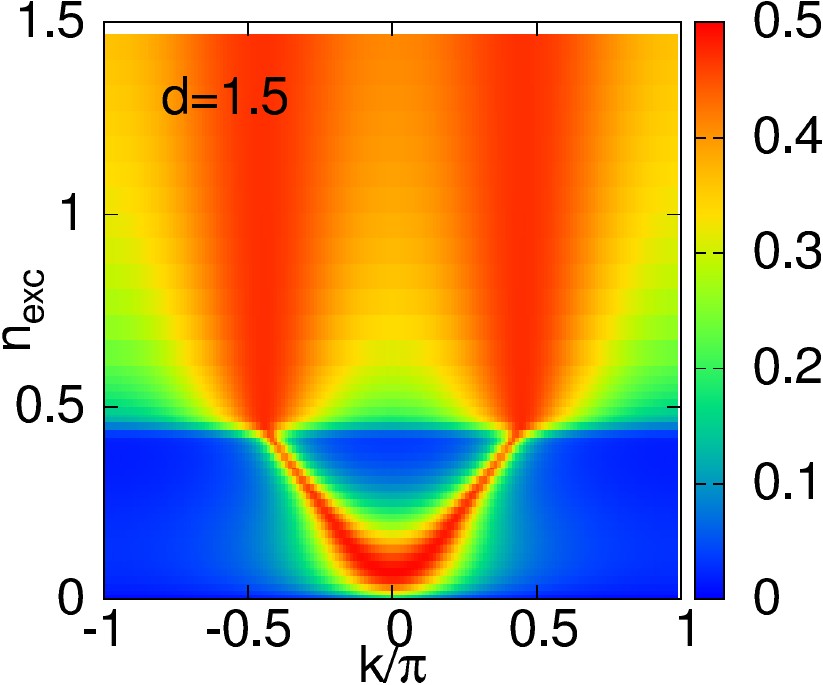}\\
\includegraphics[width=0.23\textwidth]{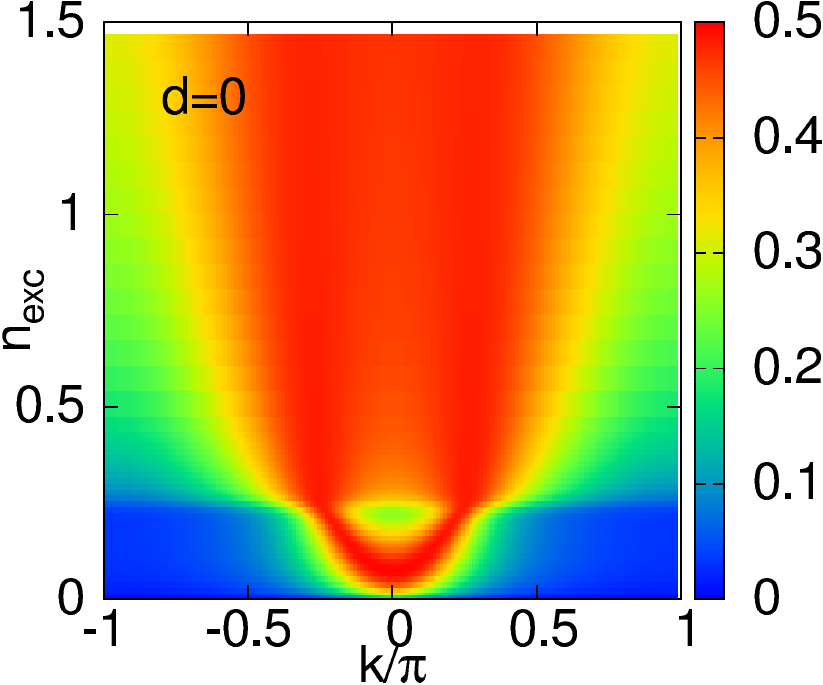}
\includegraphics[width=0.23\textwidth]{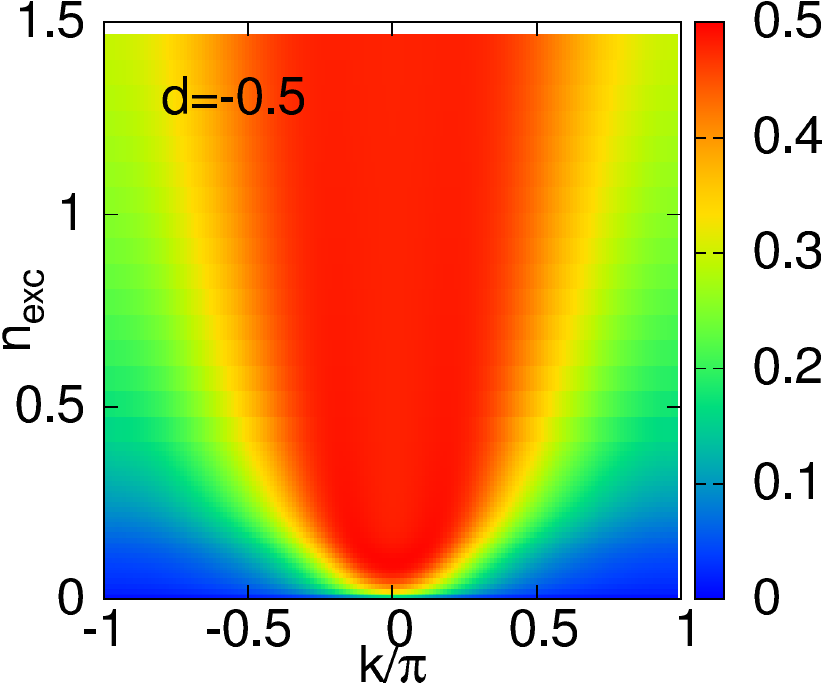}
\caption{(Color online)  Intensity plot of the electron-hole-pairing amplitude $d_{\bf k}$ in the momentum-density-plane at  different detuning $d$, for $U=2$, $g=0.2$, and $\omega_c=0.5$.}
\label{fig4}
\end{figure}

In order to analyze how Coulomb and light-matter interactions operate together establishing a quantum condensed state, we have separately determined the two (excitonic and photonic) contributions to the order parameter $\Delta$ on the right-hand side of Eq.~\eqref{17}: $\Delta_X=-\tfrac{U}{N}\sum_{\bf k} d_{\bf k}$ and $\Delta_\textrm{ph}=-\tfrac{g}{\sqrt{N}}\langle\psi_0\rangle$. The results are shown in Fig.~\ref{fig3}. For large detuning (upper panel), an excitonic condensate is formed at low densities (note that the photonic order parameter vanishes). For the $U$ value $U=2$ considered here it typifies a BEC of preformed  electron-hole pairs. As the excitation density increases phase-space (Pauli blocking) effects become more and more  important (see below) and the condensate  becomes BCS-type; but still the light-component is negligible.  Increasing the density further photonic effects came into play. As a result the condensate turns from excitonic to polaritonic. At even higher excitation densities the excitonic component saturates, whereas the photonic order parameter continues its increase. This classifies a photonic condensate.  For smaller detuning but fixed $\omega_c$, 
both excitonic and photonic order parameters are intimately connected in the whole low-to-intermediate 
excitation density regime, indicating a polariton BEC,  which  again gives way to photonic BEC at very large $n_\textrm{exc}$.  Of course, by their nature, all these transitions are crossovers. 

%% ------------------------------ Fig. 6 --------------------
\begin{figure}[t]
\includegraphics[width=0.23\textwidth]{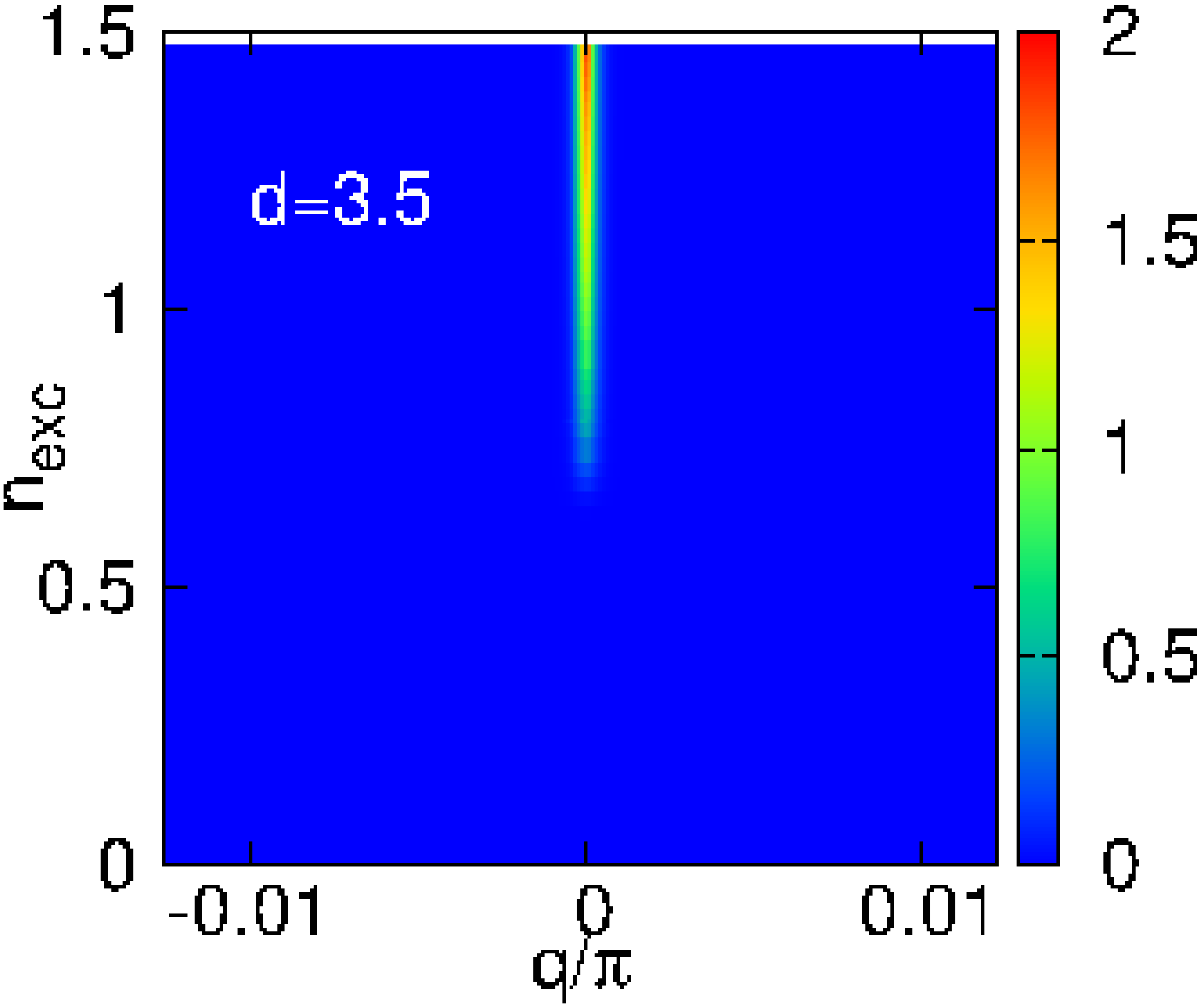}
\includegraphics[width=0.23\textwidth]{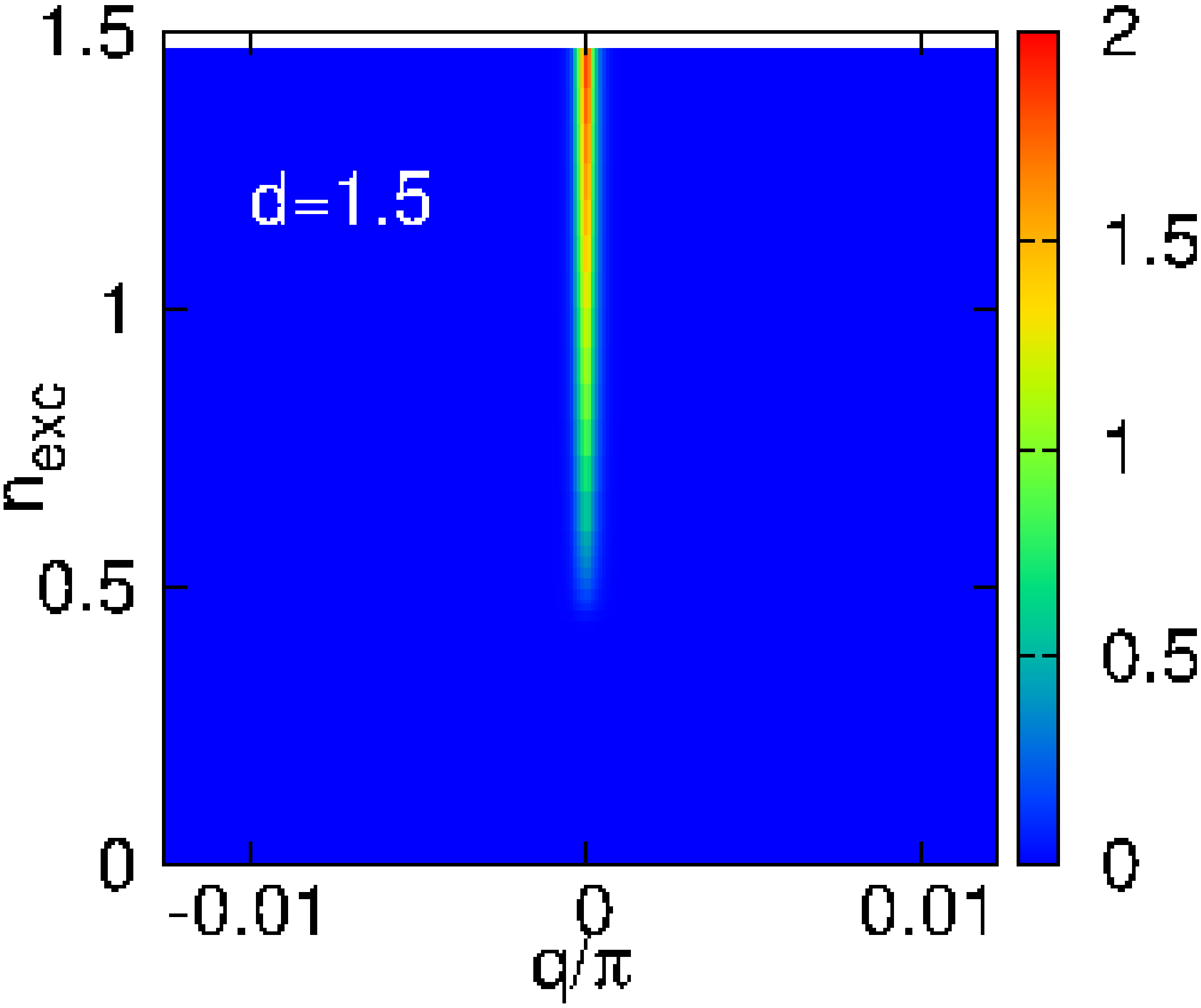}\\
\includegraphics[width=0.23\textwidth]{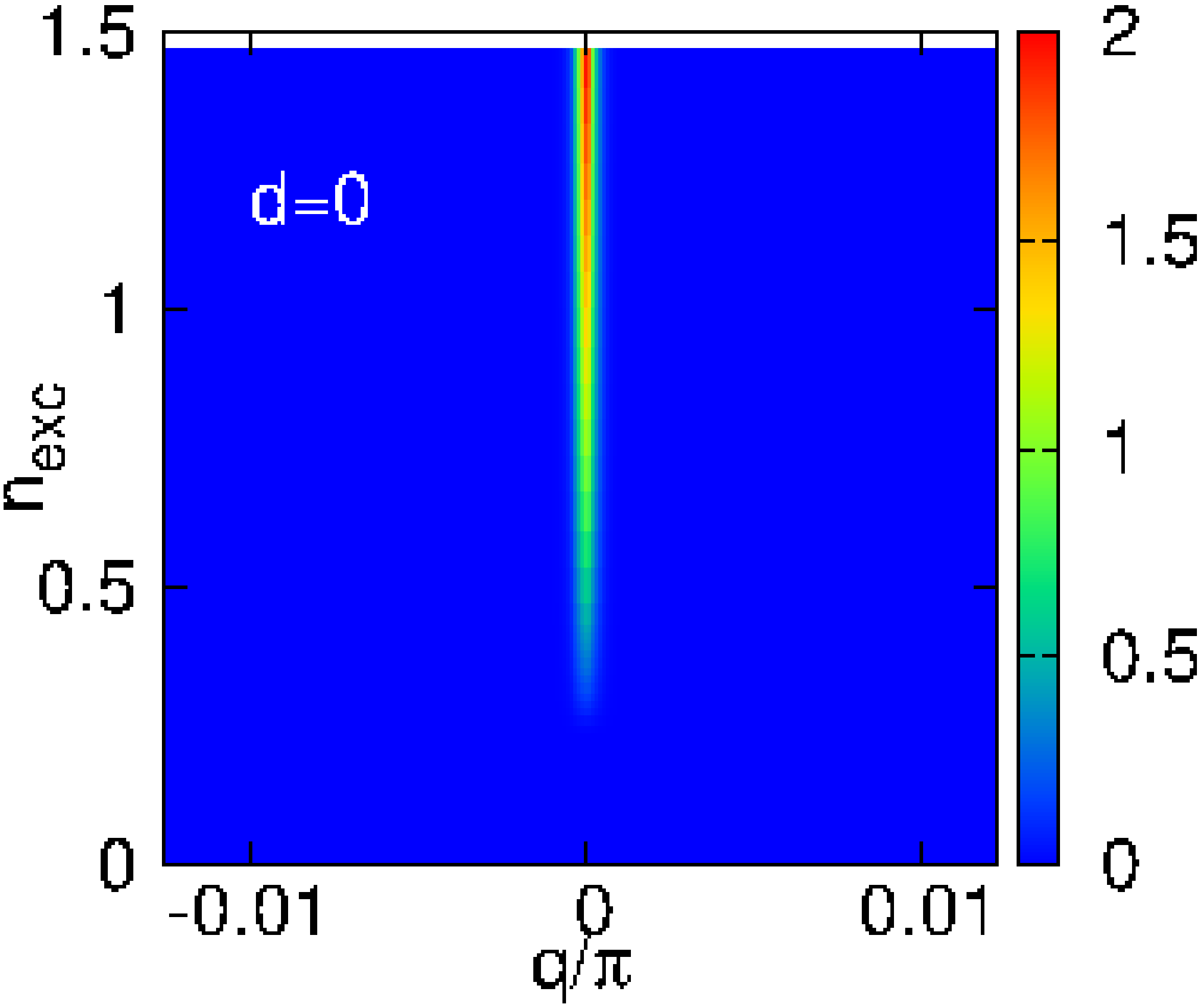}
\includegraphics[width=0.23\textwidth]{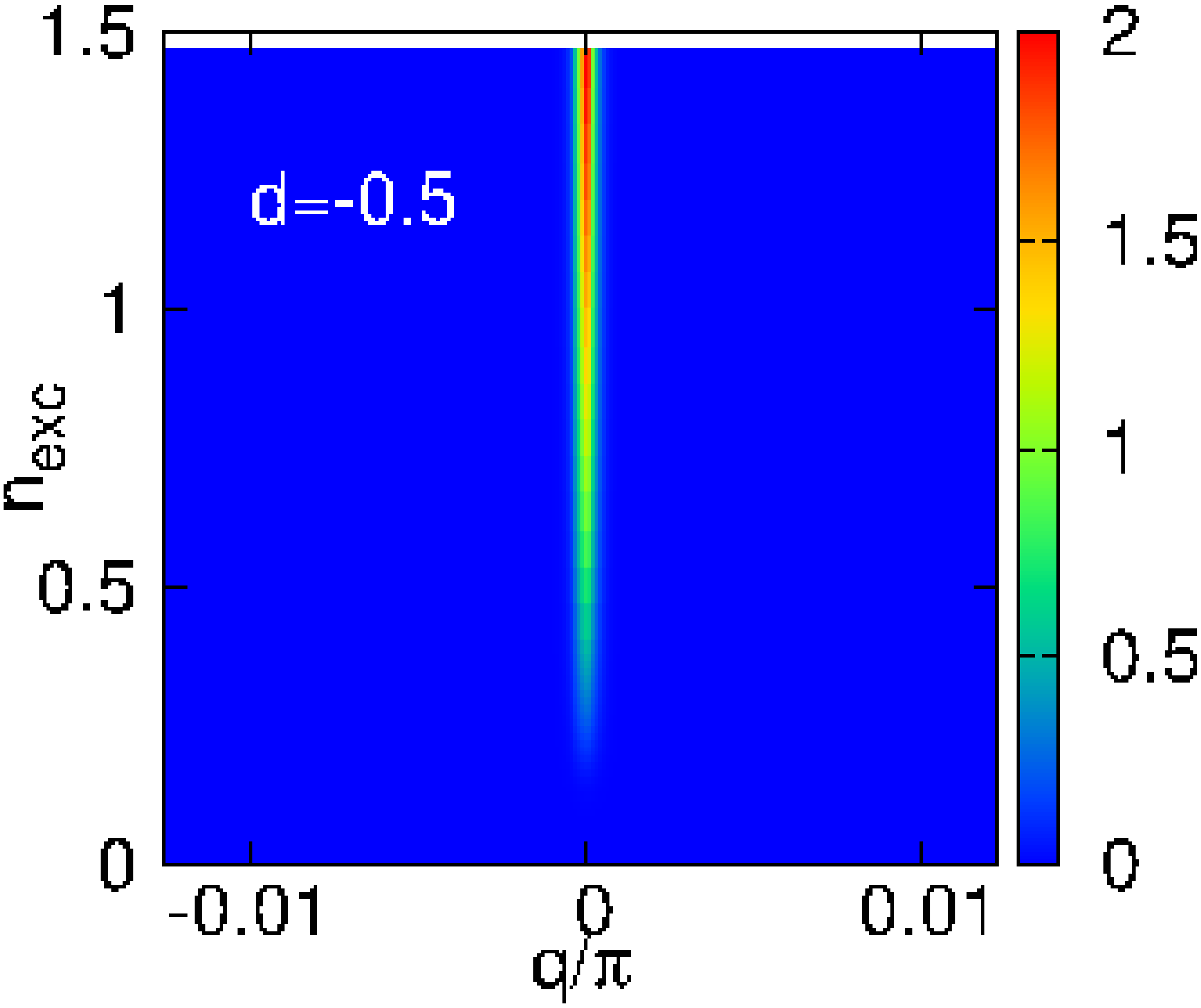}
\caption{(Color online) Intensity plot of the photon density $\langle \psi_{\bf q}^\dagger \psi^{}_{\bf q}\rangle$  in the momentum-density-plane at different detuning $d$, for  $U=2$, $g=0.2$, $\omega_c=0.5$, and $T=0.01$.}
\label{fig5}
\end{figure}

%% --------------------------- Fig. 7 ----------------------
\begin{figure}[t]
\includegraphics[width=0.4\textwidth]{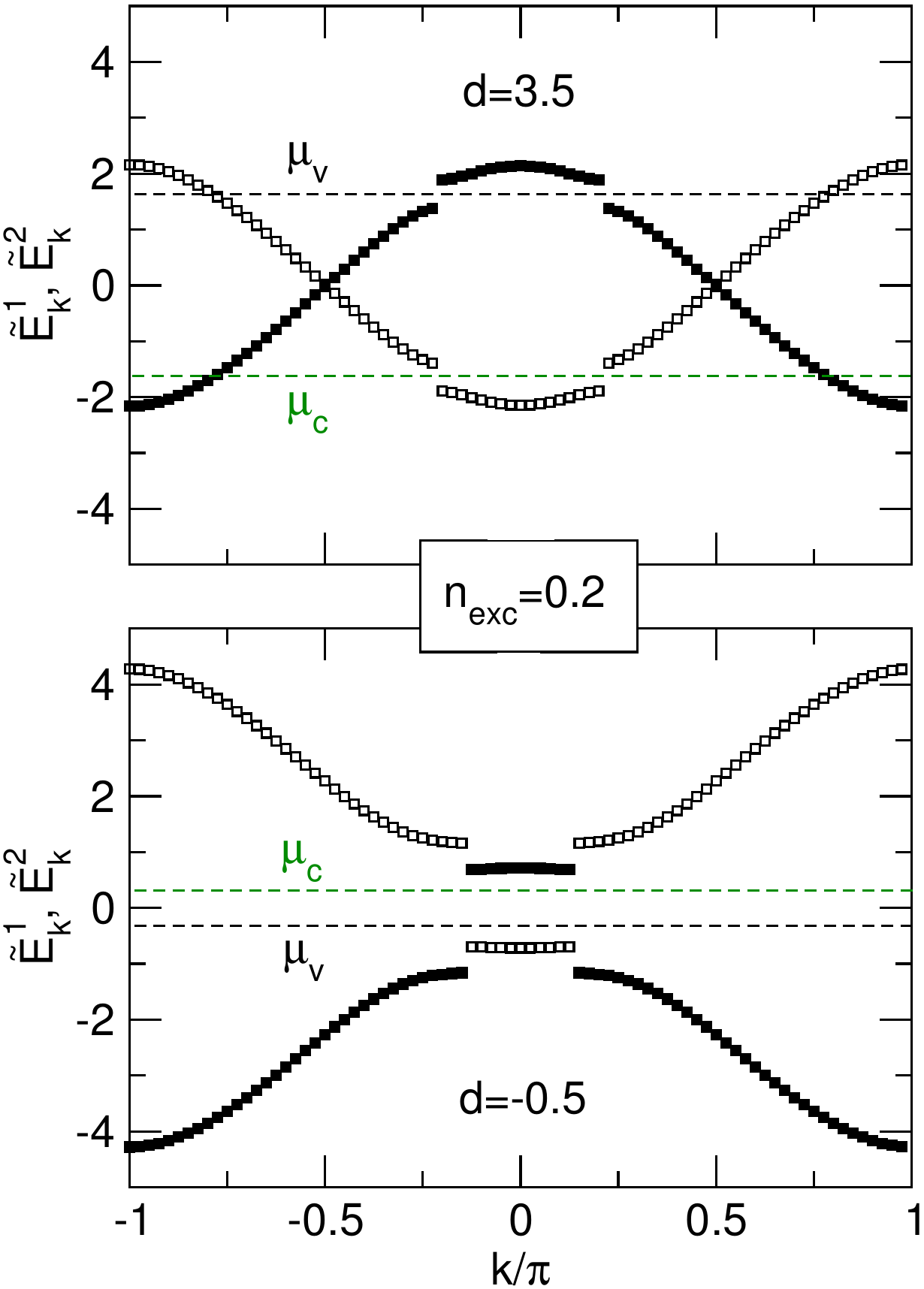}
\caption{(Color online) Quasiparticle energies~\eqref{A.27}  for large (upper panel) and small (lower panel) detuning at $n_\textrm{exc}=0.2$, where, $U=2$, $g=0.2$, and $\omega_c=0.5$. Filled (open) symbols mark the valence (conduction) band with chemical potentials $\mu_v= -\mu/2$ ($\mu_c=\mu/2 $).}
\label{fig6a}
\end{figure}
%

%% ------------------------- Fig. 8 ---------------------------
\begin{figure}[t]
\includegraphics[width=0.4\textwidth]{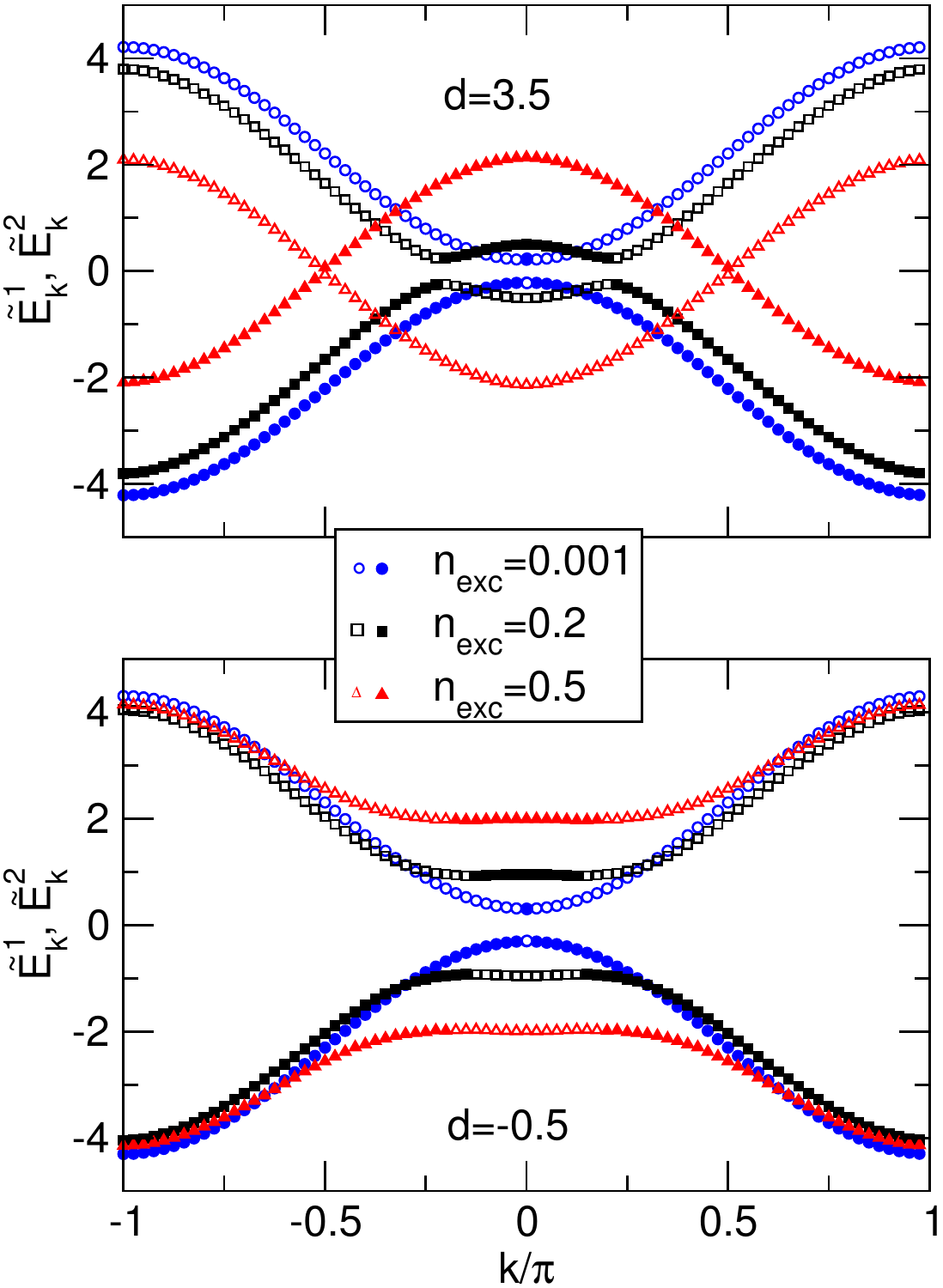}
\caption{(Color online) Renormalized band structure~\eqref{A.27} for large (upper panel) and small (lower panel) detuning at various $n_\textrm{exc}$. 
Note that now the energies  of the valence bands (filled symbols) and conduction bands (open symbols) are measured from $\mu_v$ and $\mu_c$, respectively.  Again,  $U=2$, $g=0.2$, and $\omega_c=0.5$. }
\label{fig6b}
\end{figure}

Figures~\ref{fig4} and~\ref{fig5} give the wavevector-resolved intensity of the electron-hole pair order-parameter function $d_{\bf k}$ [Eq.~\eqref{A.41}] and the photon density $\langle \psi_{\bf q}^\dagger \psi^{}_{\bf q}\rangle$ [Eq. \eqref{A.44}], respectively. For large detuning $d=3.5$, Fig.~\ref{fig4}  indicates how the maximum 
of the pairing amplitude $d_{\bf k}$ is continuously shifted from  $k=0$ at $n_\textrm{exc}\ll1$ to larger values of $k$ as  $n_\textrm{exc}$ is raised, which reveals finite density (Pauli blocking) effects. Above a `critical' density $n_\textrm{exc}\simeq 0.66$ (cf. also Fig.~\ref{3}), where $\mu\simeq\omega_c$, the photon field
comes into play (cf. Fig.~\ref{fig5} left upper panel). Simultaneously, the renormalization of the band structure due to the Coulomb interaction (see following) leads to a high intensity of $d_{\bf k}$  at large momenta ($|{\bf k}|>\pi/2$). 
 For small detuning  $d=-0.5$, the light-matter coupling affects the behavior of $d_{\bf k}$ from the very beginning ($n_\textrm{exc}\to 0$), yielding a strong polariton signature around $k=0$ which broadens at higher excitation densities. Clearly the intensity of the photon field is always peaked around $q=0$ and comes up at larger excitation density  the larger the detuning is (see Fig.~\ref{6}).

Starting from the bare band structure~\eqref{3}, it will be interesting to look how the quasiparticle bands~\eqref{A.27} evolve, which 
are renormalized on account of Coulomb and light-matter interaction effects. Figure~\ref{fig6a} gives $\tilde{E}_{\bf k}^{1,2}$ for $n_\textrm{exc}=0.2$. 
For large detuning ($d=3.5$, $E_g=-3$), the bare bands inter-penetrate [cf. Fig.~1(c)].  Here, basically all excitations are excitons (formed by the electrons and holes in the central part of the Brillouin zone).  For small detuning ($d=-0.5$, $E_g=1$), the (bare) semiconductor band structure [cf. Fig.~1(b)] is preserved.  Again, excitonic bound states occur but  not as many as for $d=3.5$;
instead, more photonic states contribute to $n_\textrm{exc}$.

Figure~\ref{fig6b} shows the renormalized ``band structure''  for different excitation densities; here valence and conduction bands were shifted 
by $-\mu_v$, respectively, $-\mu_c$. Of course, at  $n_\textrm{exc}=0.001$ the dispersions are barely changed from those of the bare bands. 
However, in order to realize such a very small excitation densities at $d=3.5$, $E_g=-3$, i.e., for strongly overlapping bare bands, 
a large negative value of $\mu$ arises [cf. Fig.~\ref{fig2}~(a)].  Increasing $n_\textrm{exc}$, the location of the gap is shifted from $k=0$, (as was the case for $n_\textrm{exc}=0.001$), to a finite $k$-value. We find a band structure as for a BCS-type exciton insulator state~\cite{PBF10} [cf. Fig.~\ref{fig1}~(d)].  For $n_\textrm{exc}=0.5$, a complete back folding of the bands (doubling of the Brillouin zone) takes place. For this effect, the attractive Coulomb interaction between electrons and holes is responsible. The situation significantly changes at small detuning. Here, always a semiconductor band structure is observed, although 
the particle-photon coupling leads to a flattening of the top of the valence band, respectively, bottom of the conduction band. As a result, the bandwidth of both bands shrinks and the gap  broadens. This  clearly can be  attributed to the hybridization between electronic and photonic degrees of freedom in the course of polariton formation.

\begin{figure}[t]
\includegraphics[width=0.47\textwidth]{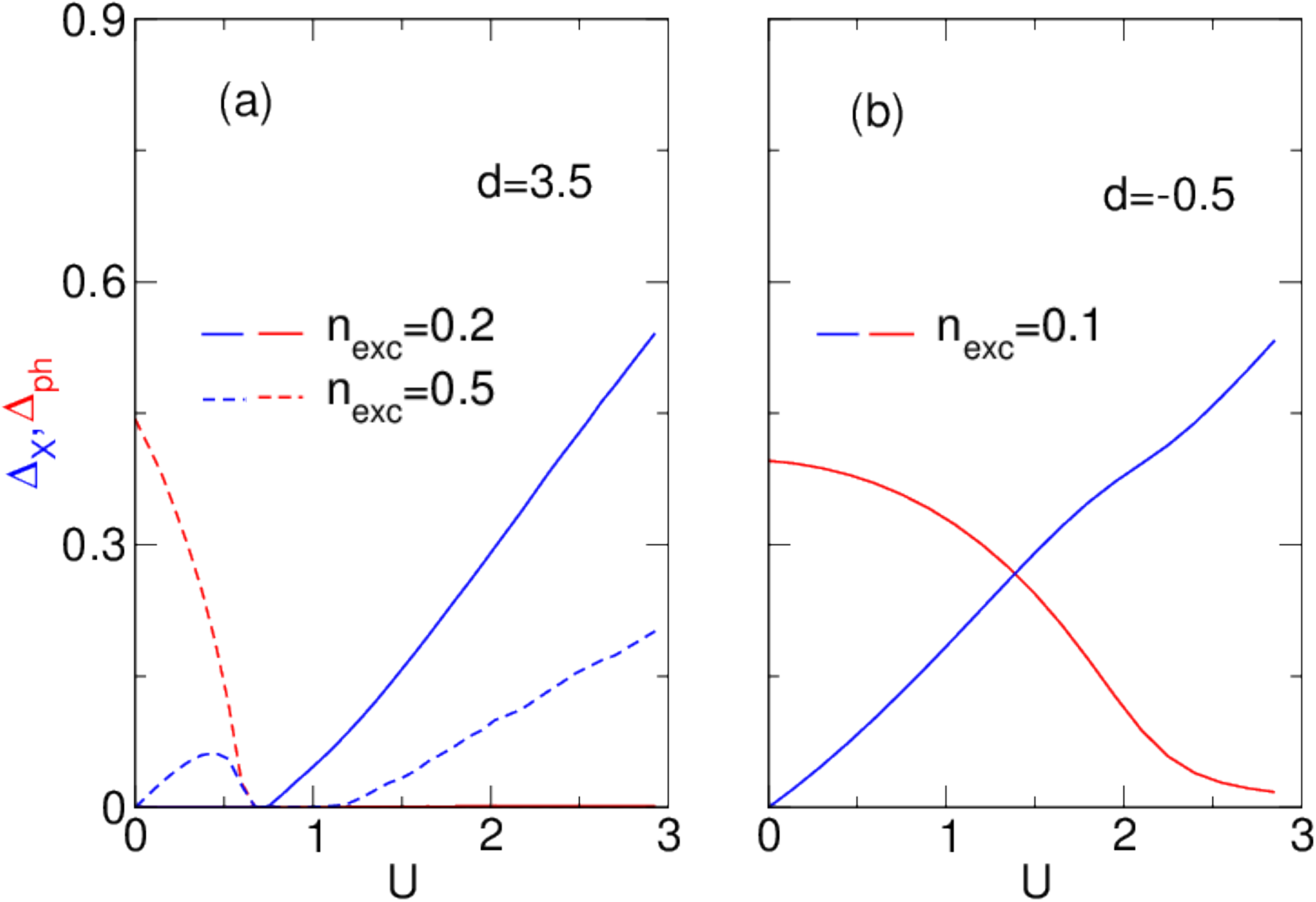}
\caption{(Color online) Excitonic (blue lines) and photonic (red lines) order parameters as a function of $U$ for the case of large detuning, $d=3.5$ (left), and small detuning $d=-0.5$ (right), where $g=0.2$ and $\omega_c=0.5$.}
\label{fig7}
\end{figure}

\begin{figure}[t]
\includegraphics[width=0.47\textwidth]{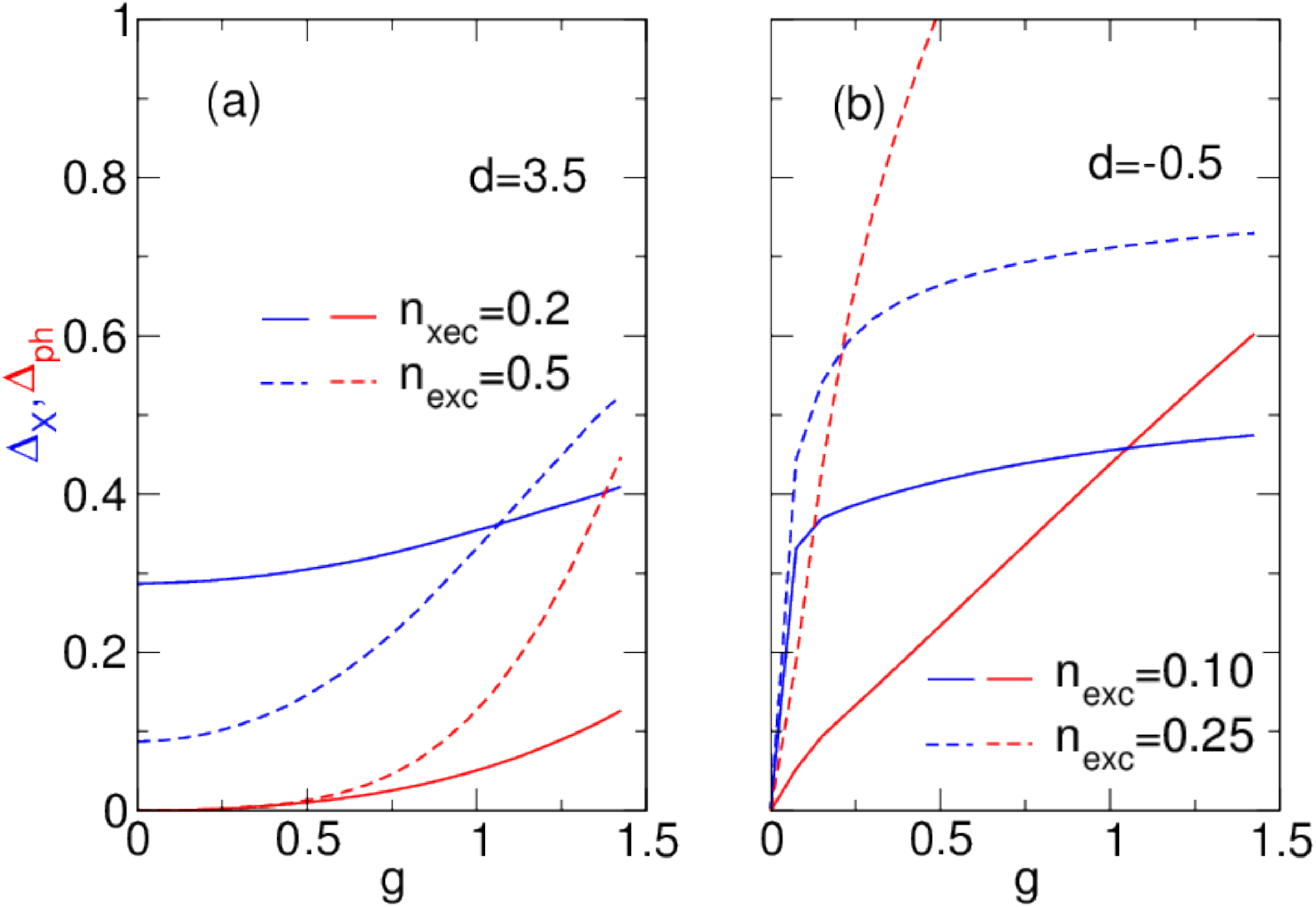}
\caption{(Color online) Excitonic (blue lines) and photonic  (red lines) order parameters as functions of $g$ 
for the cases of large detuning, $d=3.5$ (left), and small detuning $d=-0.5$ (right), where $U=2$ and $\omega_c=0.5$ (note that $U=2$ roughly equates the critical value for exciton formation at $g=0$).}
\label{fig8}
\end{figure}

Let us now discuss the ground-state properties in dependence on the Coulomb and light-matter interaction strengths. Figure~\ref{fig7}  gives the variation of $\Delta_X$ and $\Delta_\textrm{ph}$ with $U$. For large detuning and small excitation density, electron-hole pairing starts above a certain Coulomb interaction threshold with states involved that are close to the Fermi momenta. We find almost no photonic contribution in this case. Hence the coherent state classifies as an excitonic condensate. At larger excitation density polaritons are formed for small values of $U$  (note that for $U=0$ the condensate is completely triggered by the photons). Increasing $U$, the ground state becomes dominated by Coulomb correlations again, and we obtain an ordered state of tightly bound
excitons (reminiscent of the excitonic insulator phase). At small detuning, the polariton BEC features finite excitonic and photonic order parameters, where the former (latter) is  enhanced (suppressed) as $U$ rises at fixed $g$, indicating a crossover from an excitonic to a photonic dominated ground-state wave function.   The $g$ dependence of the order parameters displayed in Fig.~\ref{fig8} 
demonstrates that both pairings, $\Delta_X$ and $\Delta_\textrm{ph}$, are always strengthened by increasing the light-matter coupling for both large and small detunings. In contrast, for decreasing $g\rightarrow 0$ only  $\Delta_\textrm{ph}$ vanishes, whereas   $\Delta_X$ stays finite for large detuning but approaches zero for small detuning because we are in the polariton regime and $U=2 \lesssim U_c$. Moreover, a slow saturation of $\Delta_{X}$ at large values of $g$ is observed. 

%% ---------------------- Fig. 9 ------------------------------

%% ------------------------  Fig. 10 ---------------------------

%-----------------------------------------------------------------------------------------------------------------
\subsection{Spectral properties}
%-----------------------------------------------------------------------------------------------------------------

\begin{figure}[t]
\includegraphics[width=0.23\textwidth]{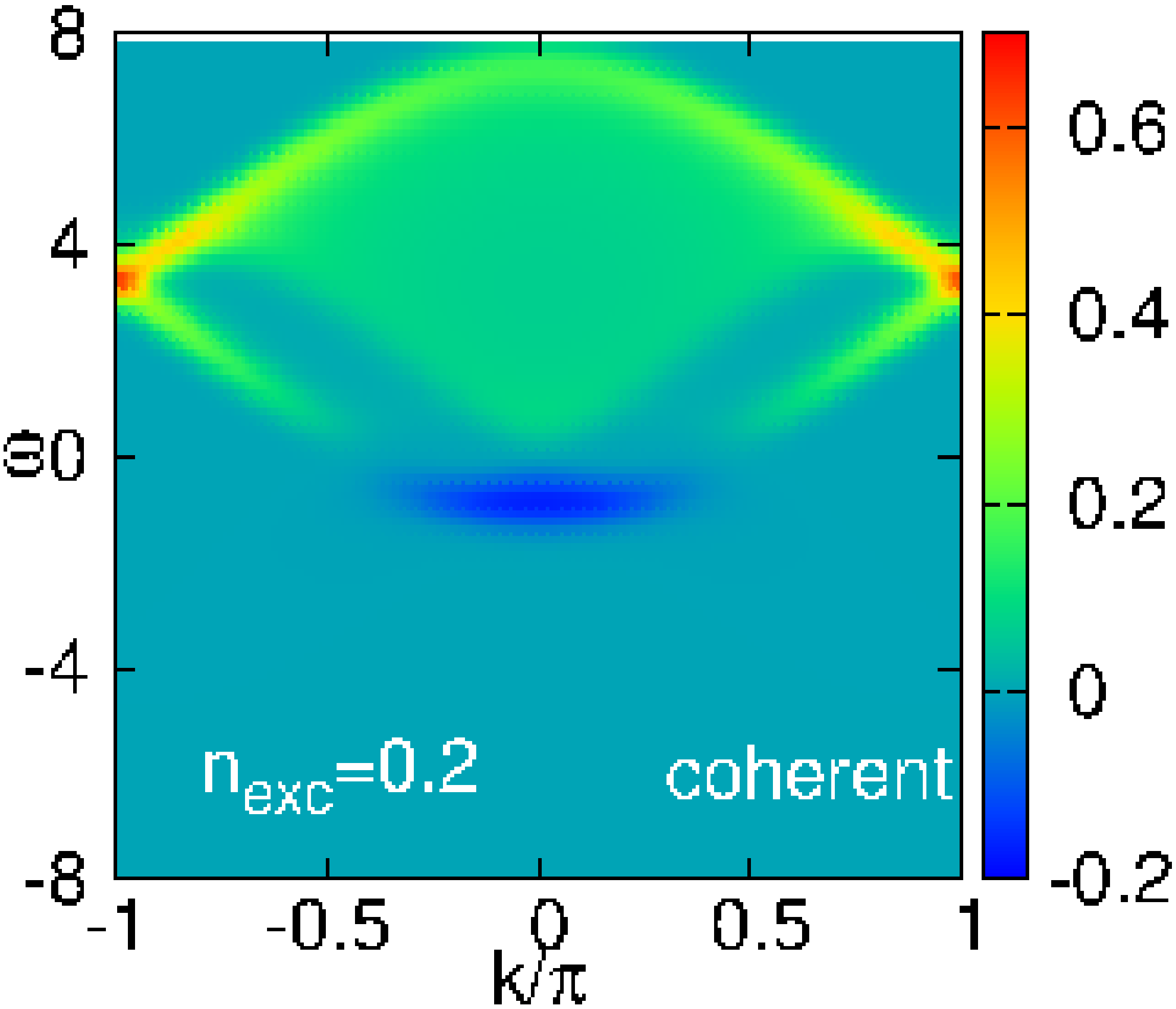}
\includegraphics[width=0.23\textwidth]{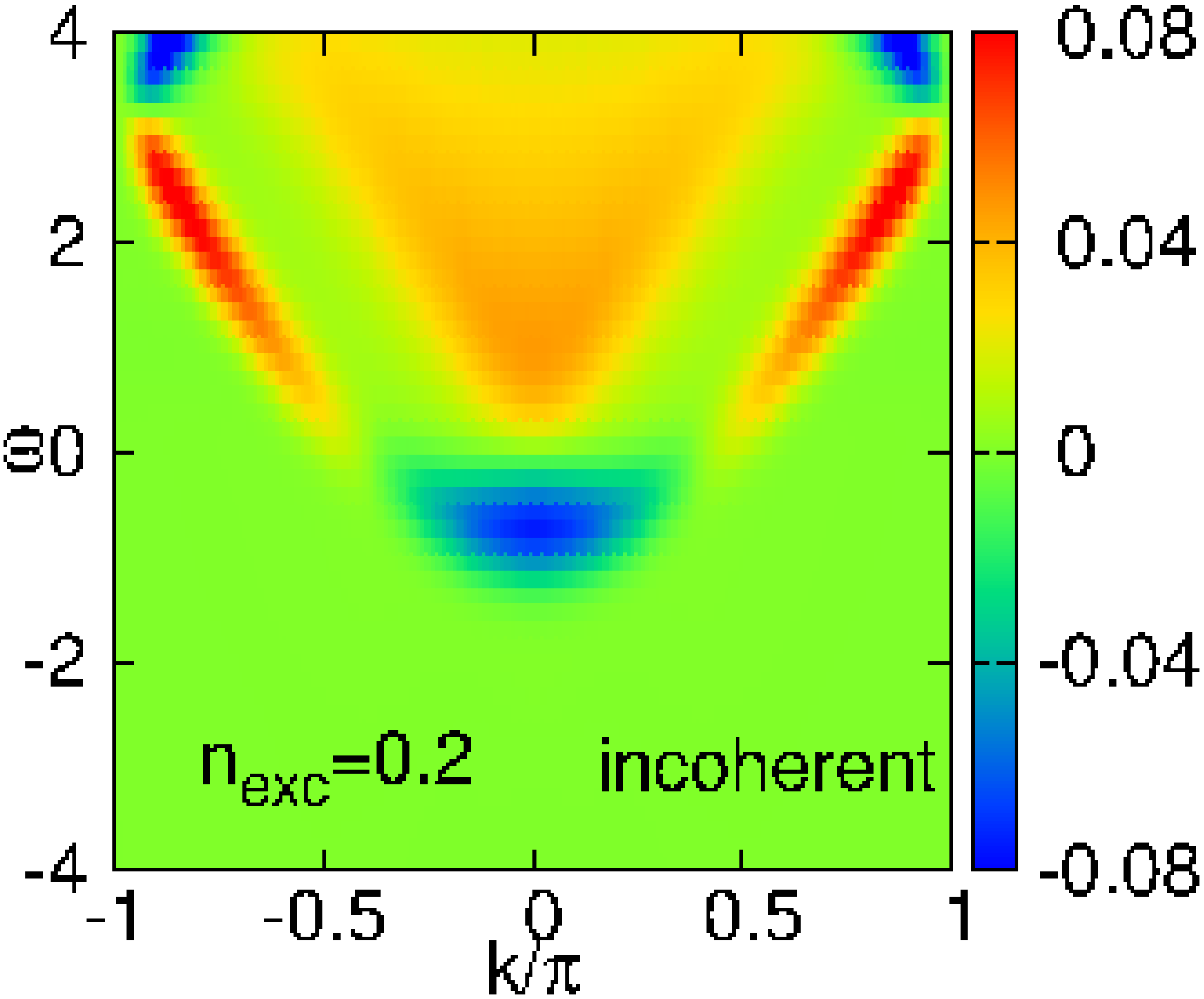}\\
\includegraphics[width=0.23\textwidth]{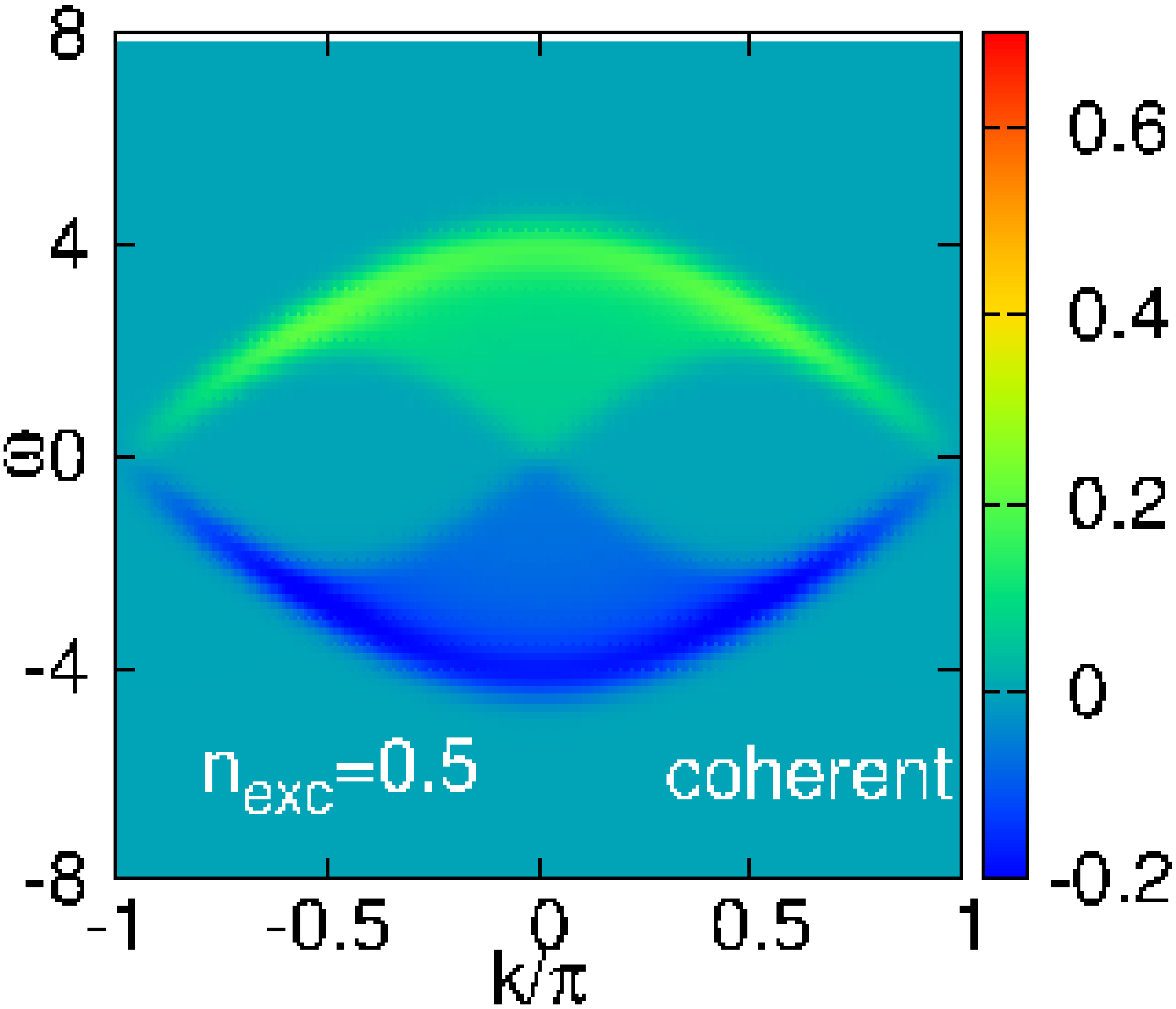}
\includegraphics[width=0.23\textwidth]{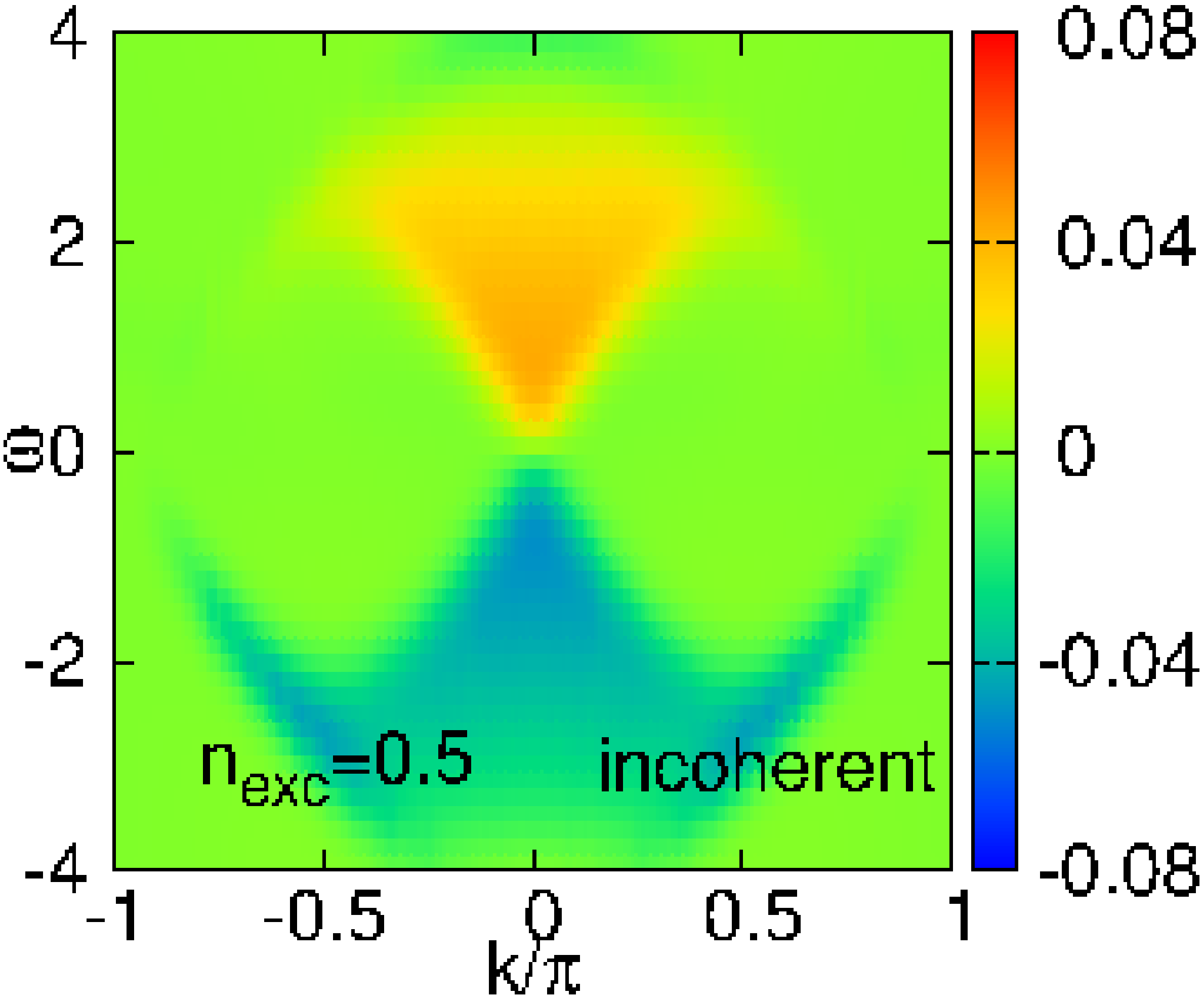}
\caption{(Color online) Intensity plot of the excitonic  polarization $A({\bf k}, \omega)$ for $n_\textrm{exc}=0.2$ (upper panels) and $n_\textrm{exc}=0.5$ (lower panels) at large detuning $d=3.5$. Again  $U=2$, $g=0.2$, and $\omega_c=0.5$. Here the left panels refer to the coherent part $A^\textrm{coh}({\bf k}, \omega)$, the right panels give the incoherent contribution $A^\textrm{inc}({\bf k}, \omega)$.}
\label{fig9}
\end{figure}

\begin{figure}[t]
\includegraphics[width=0.23\textwidth]{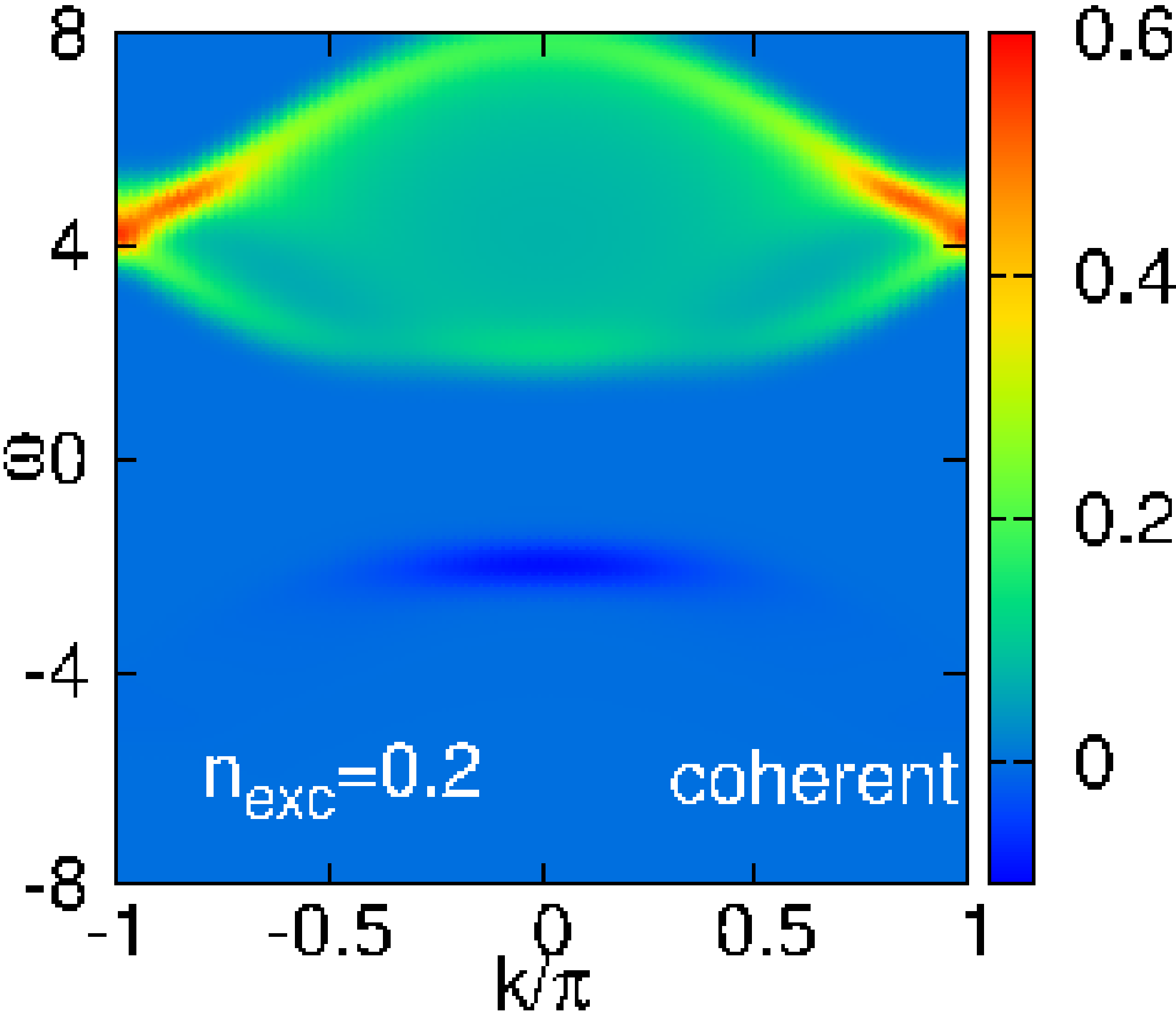}
\includegraphics[width=0.23\textwidth]{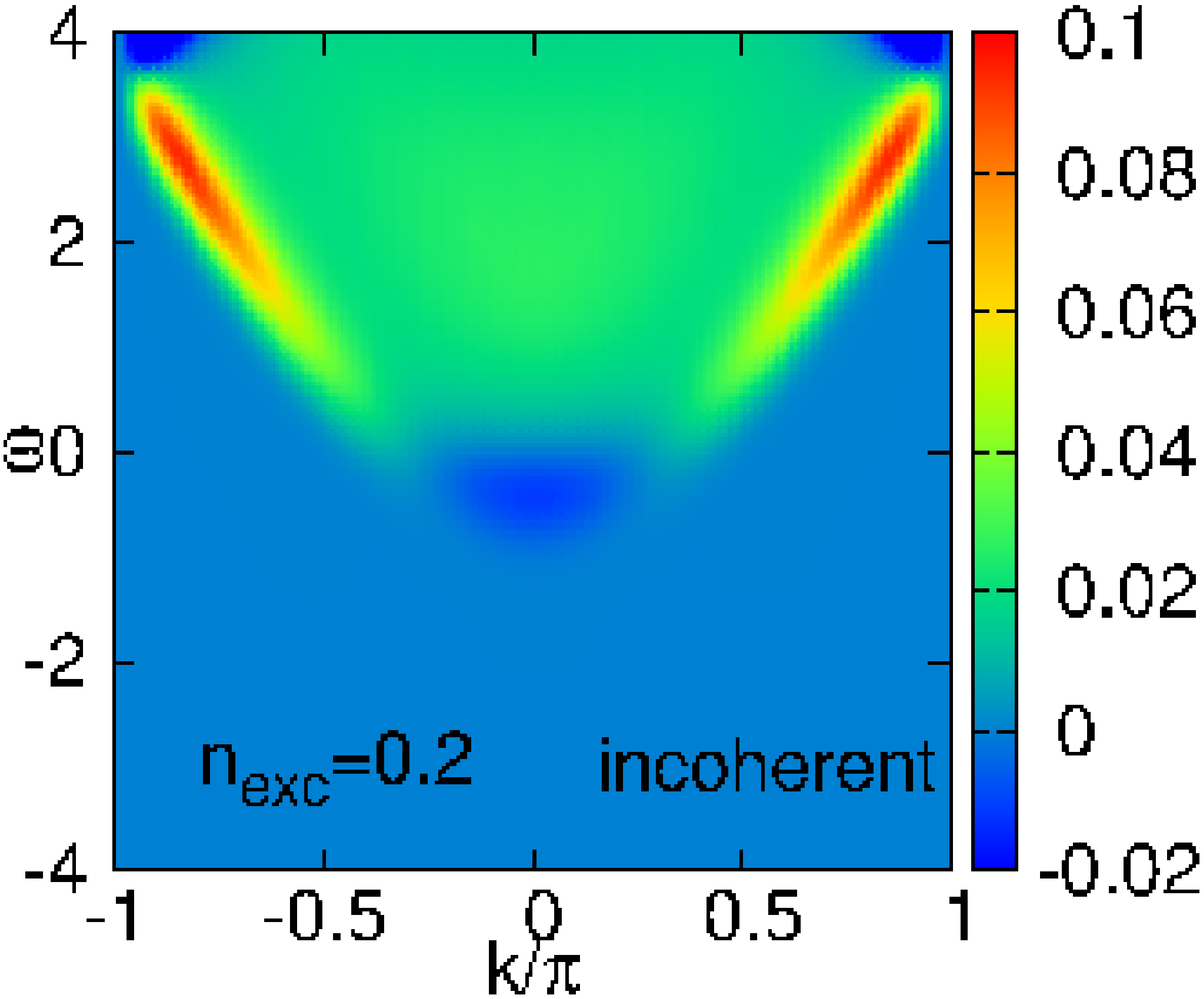}\\
\includegraphics[width=0.23\textwidth]{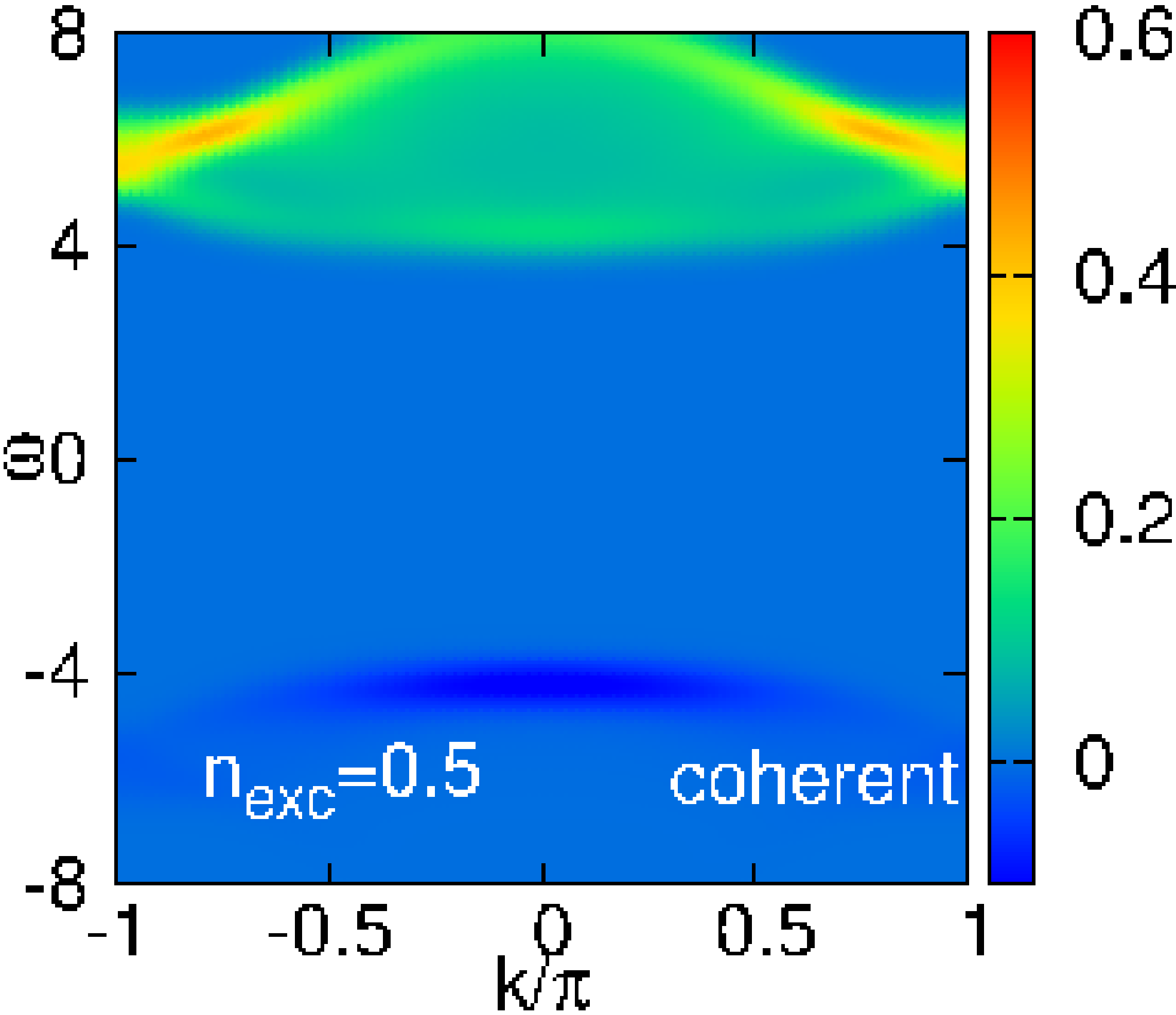}
\includegraphics[width=0.23\textwidth]{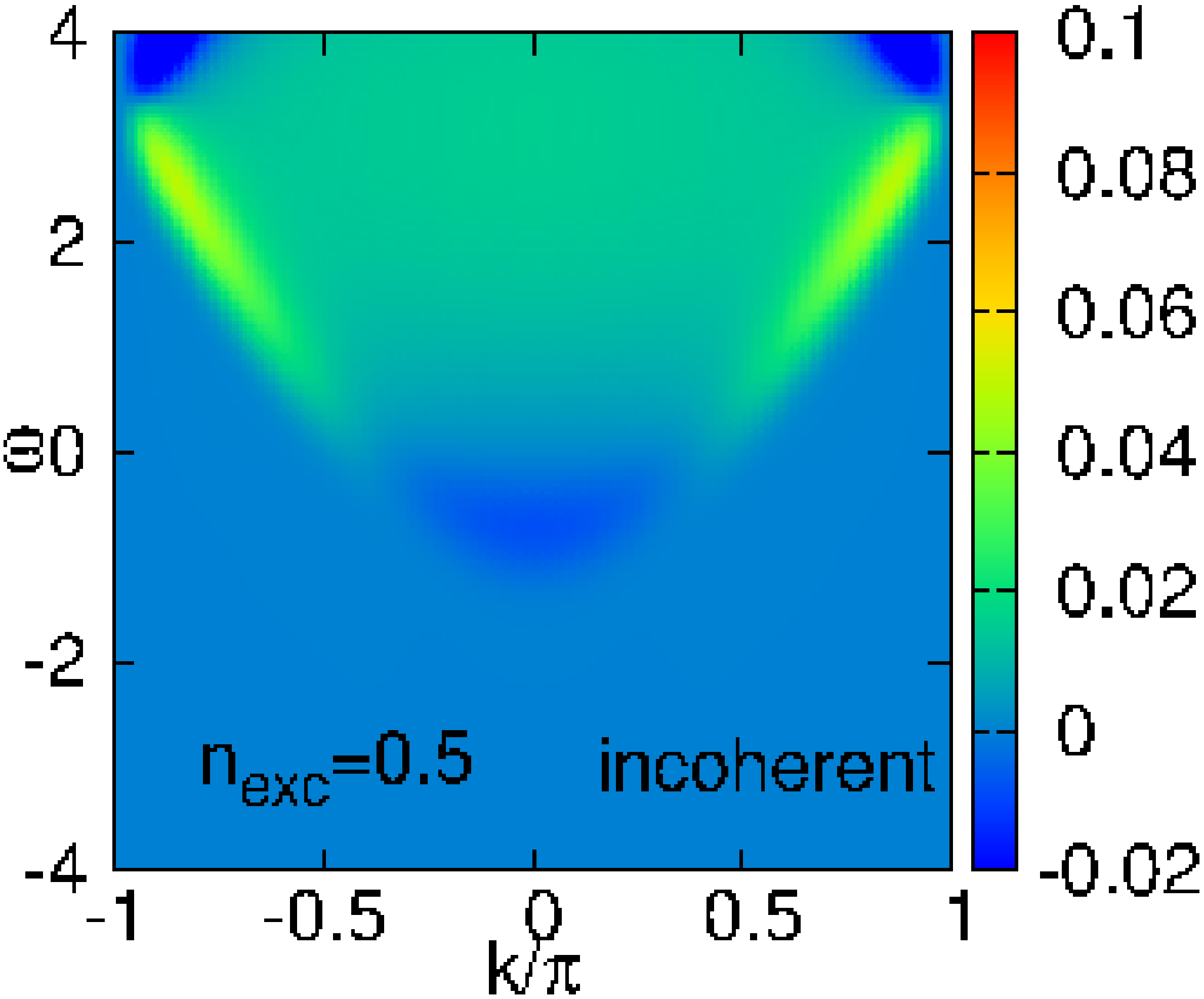}
\caption{(Color online) Intensity plot of the  excitonic polarization $A({\bf k}, \omega)$ for $n_\textrm{exc}=0.2$ (upper panels) and $n_\textrm{exc}=0.5$ (lower panels) at small detuning $d=-0.5$. Other parameters and notations as in Fig.~\ref{fig9}.}
\label{fig10}
\end{figure}

The luminescence of the microcavity exciton-polariton system is  first characterized by the intensity plots 
of $A(k,\omega)$; see Figs.~\ref{fig9} and~\ref{fig10} at $\omega_c=0.5$, for the cases of large and small detuning, respectively. Here $\omega$ and ${\bf k}$ denote the energy and momentum transfer. The left panels display the (dominant) coherent contributions~\eqref{B.6}, resulting from electron-hole pair annihilation and creation processes inside and in between the fully renormalized quasiparticle bands $\tilde{E}^{1,2}_{\bf k}$ [cf. Eq.~\eqref{A.27} and Figs.~\ref{fig6a} and \ref{fig6b}] without any additional photons involved.  The less intense incoherent parts~\eqref{B.7} include higher-order exciton and photon contributions. Special attention deserves the significant flattening of the  excitonic response at small momentum transfer for small detuning, which is caused by a strong light-matter interaction and indicates the formation of an exciton-polariton condensate~\cite{SSKSKSSNKF12}.  

%% ------------------------  FIG.11 --------------------------------
%

%% ------------------------- FIG. 12 -----------------------------

\begin{figure}[t]
\includegraphics[width=0.23\textwidth]{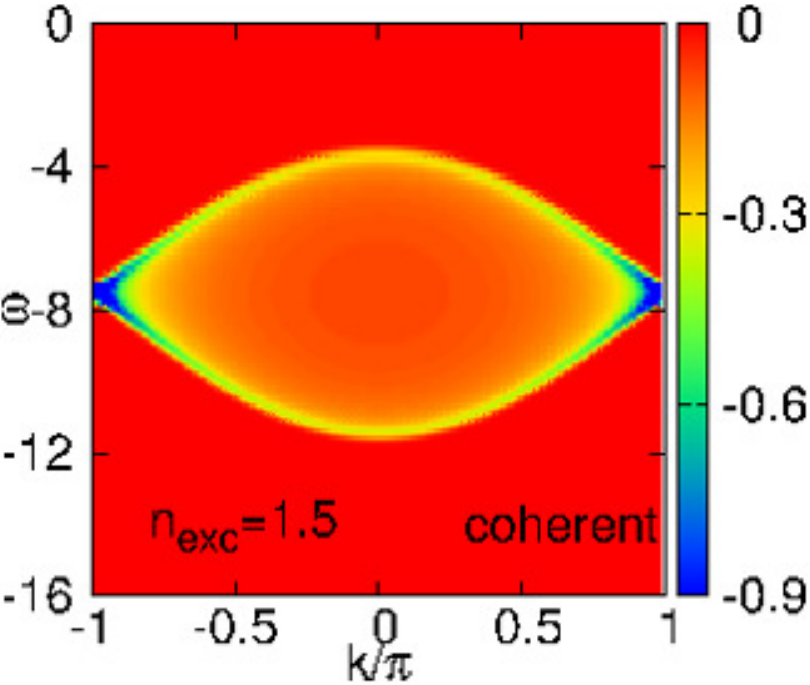}
\includegraphics[width=0.23\textwidth]{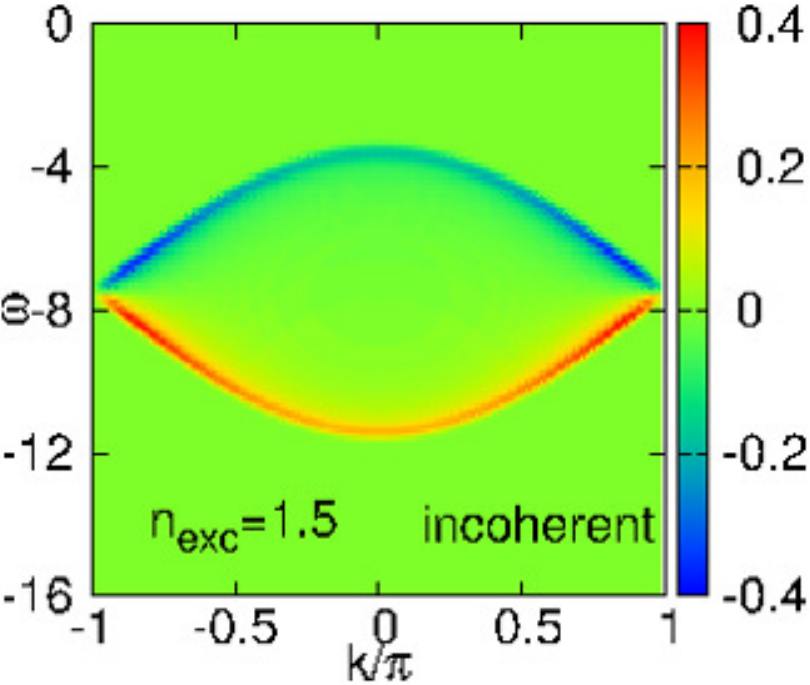}
\caption{(Color online) Intensity plot of the excitonic  polarization $A({\bf k}, \omega)$ for $n_\textrm{exc}=1.5$, where $U=2$, and $g=0.2$.
 Now $\omega_c=4.5$ and $E_g=-3$, resulting in a detuning $d=7.5$.}
\label{fig12}
\end{figure}

If the cavity frequency and the detuning are very large, a coherent signal for the excitonic polarization is obtained for negative $\omega$ only. Figure~\ref{fig12} displays $A({\bf k}, \omega)$ for $\omega_c=4.5$ and $d=7.5$ at large excitation density $n_\textrm{exc}=1.5$. We see that all available electrons and holes are paired into excitons, and the photonic excitations [not directly probed by  $A({\bf k}, \omega)$] are energetically separated 
(cf. Fig.~\ref{fig2-ii}). \\

%% -----------------------  FIG.13 -----------------------------

\begin{figure}[t]
\includegraphics[width=0.4\textwidth]{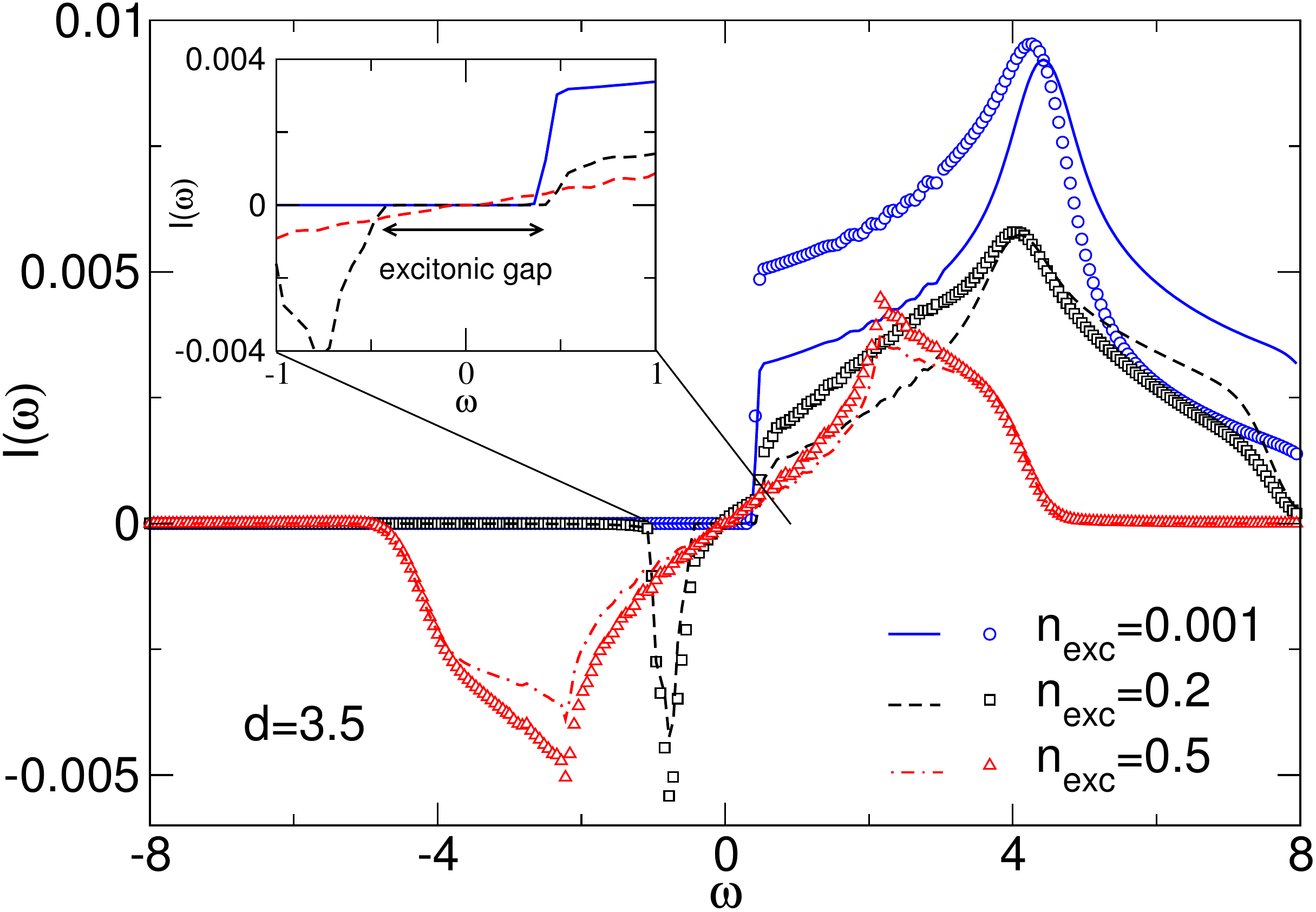}
\includegraphics[width=0.4\textwidth]{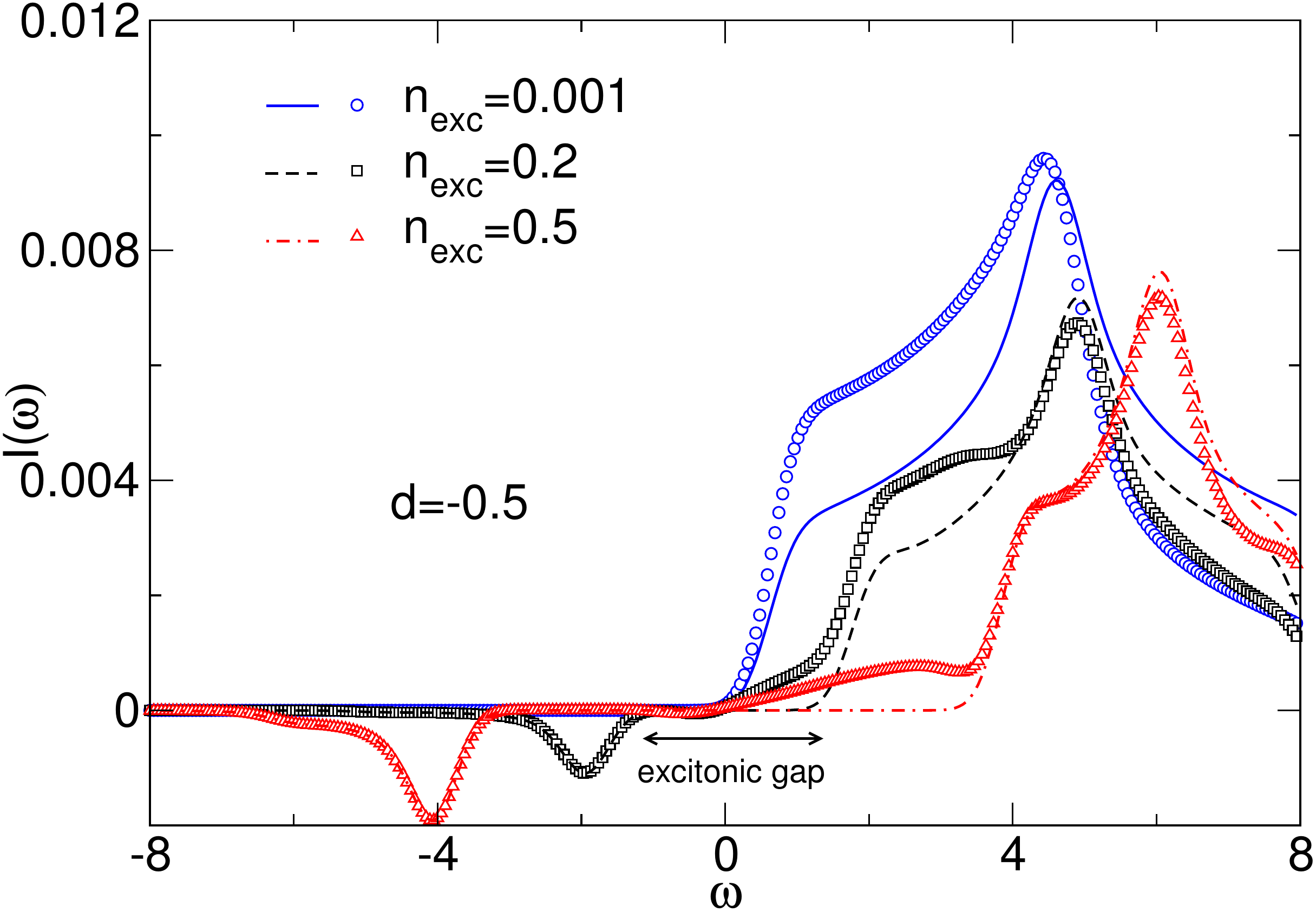}
\caption{(Color online) Total excitonic intensity $I(\omega$) for large detuning ($d=3.5$, upper panel) and 
small detuning ($d=-0.5$, lower panel) at various $n_\textrm{exc}$, where  $U=2$,  $g=0.2$, and $\omega_c=0.5$. 
}
\label{fig11}
\end{figure}

\begin{figure}[t]
\includegraphics[width=0.4\textwidth]{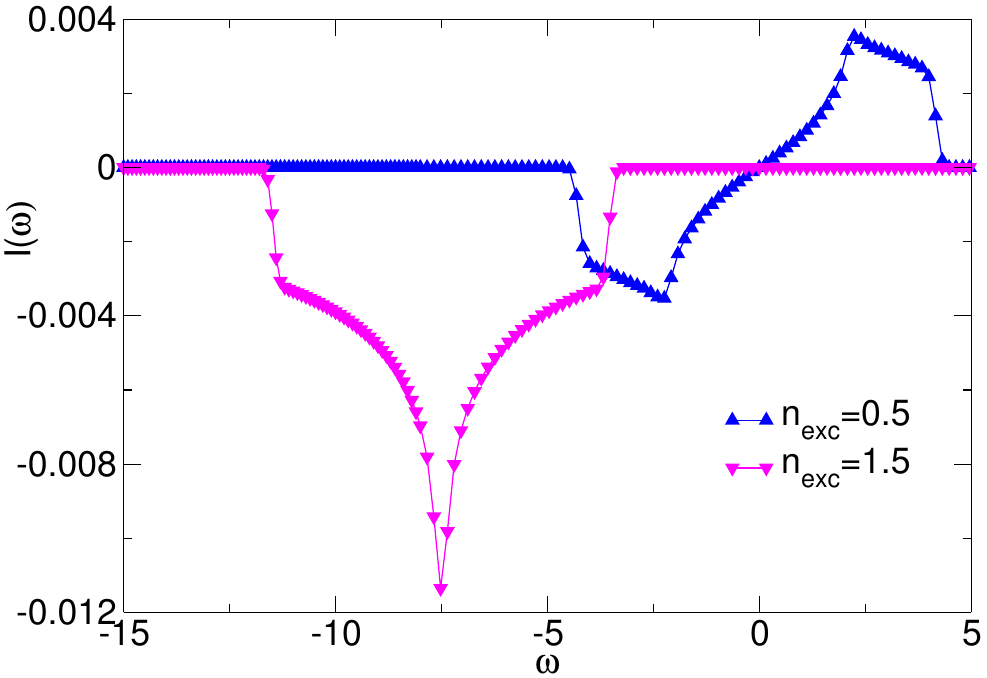}
\caption{(Color online) Total excitonic intensity $I(\omega$) for $n_\textrm{exc}=0.5$ and 1.5  at $\omega_c=4.5$. 
Model parameters are  $U=2$, $g=0.2$, and $E_g=-3$ ($d=7.5$).}
\label{fig13}
\end{figure}

 The total intensity of the excitonic polarization is given by  
\begin{equation}
\label{i_omega}
I(\omega)=\frac{1}{N}\sum_{\bf k} |S({\bf k})|^2 A({\bf k},\omega)\,,
\end{equation}
where the prefactor $ |S({\bf k})|^2$ is proportional to the exciton-photon interaction strength. For 
convenience will be  set $|S({\bf k})|^2 = g^2$.
The quantity $I(\omega)$ is shown in Figs.~\ref{fig11} and~\ref{fig13} for $\omega_c= 0.5$ and $4.5$, respectively for different excitation densities.  Starting in Fig.~\ref{fig11} with small $n_\textrm{exc}$, we observe a distinctly asymmetric line shape (with respect to $\omega\to -\omega$). 
The gap around $\omega=0$ is an evidence for the formation of an exciton-polariton condensate, particularly for small detuning (see Fig.~\ref{fig11}).  
For $\omega_c=4.5$ (Fig.~\ref{fig13}), excitonic and photonic excitations are well separated and the  excitonic polarization intensity acquires a symmetric line shape.  
Note that $I(\omega)$ fulfills the sum rule~\eqref{27}. \\
%
%% --------------------------  FIG. 14 ----------------------------------

%% -----------------------------  FIG. 15 ----------------------------------

\begin{figure}[t]
\includegraphics[width=0.22\textwidth]{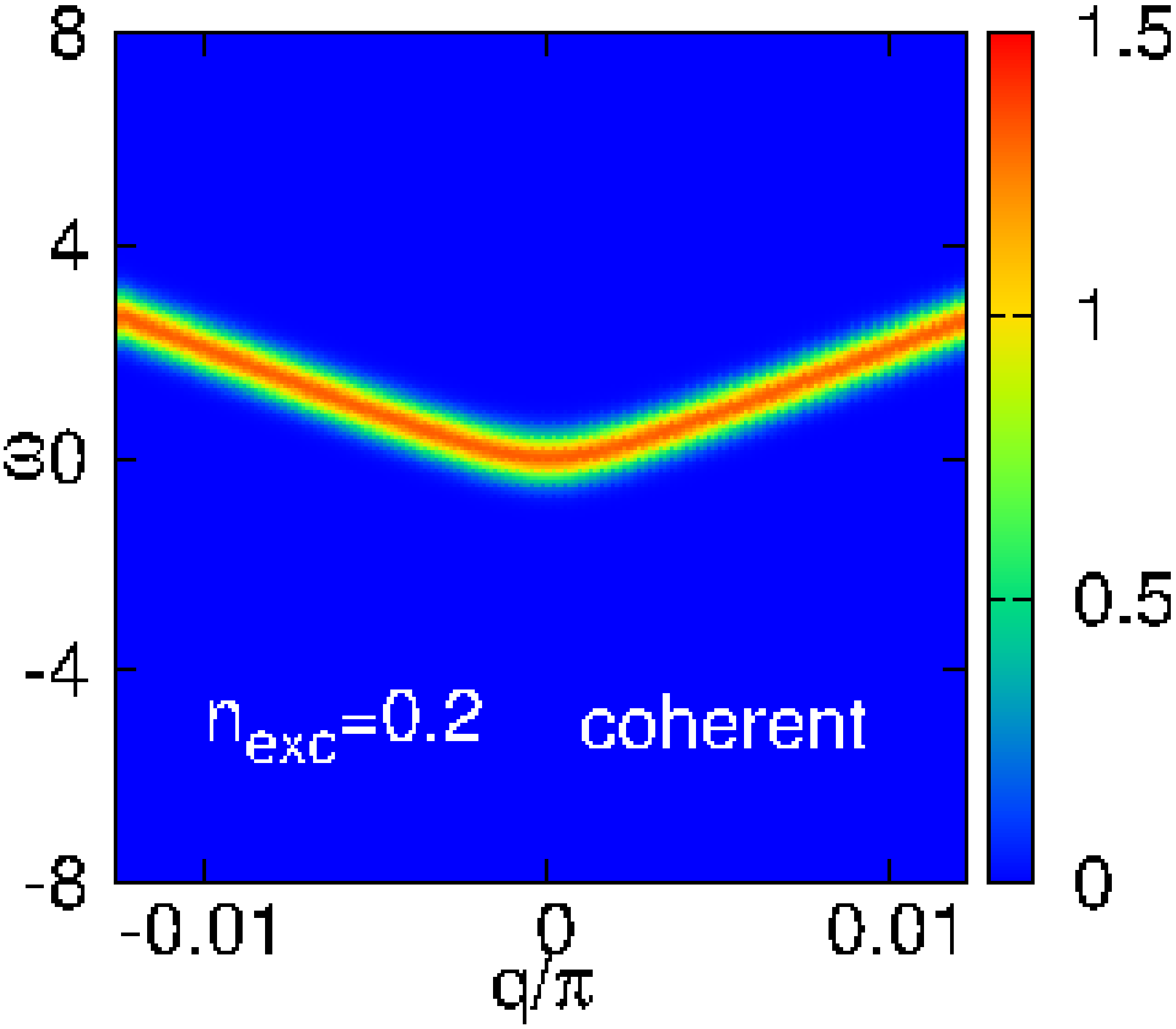}
\includegraphics[width=0.23\textwidth]{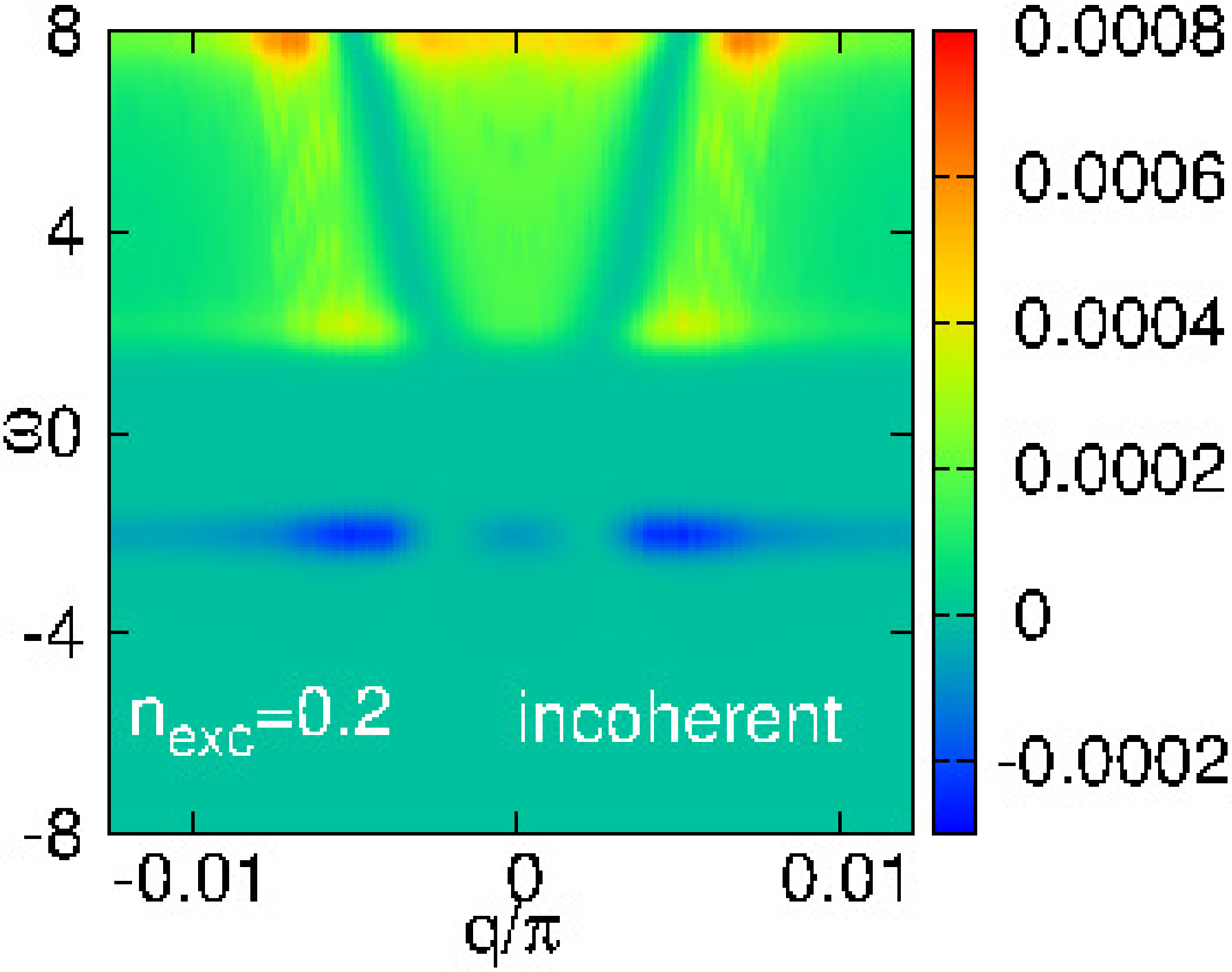}\\
\includegraphics[width=0.22\textwidth]{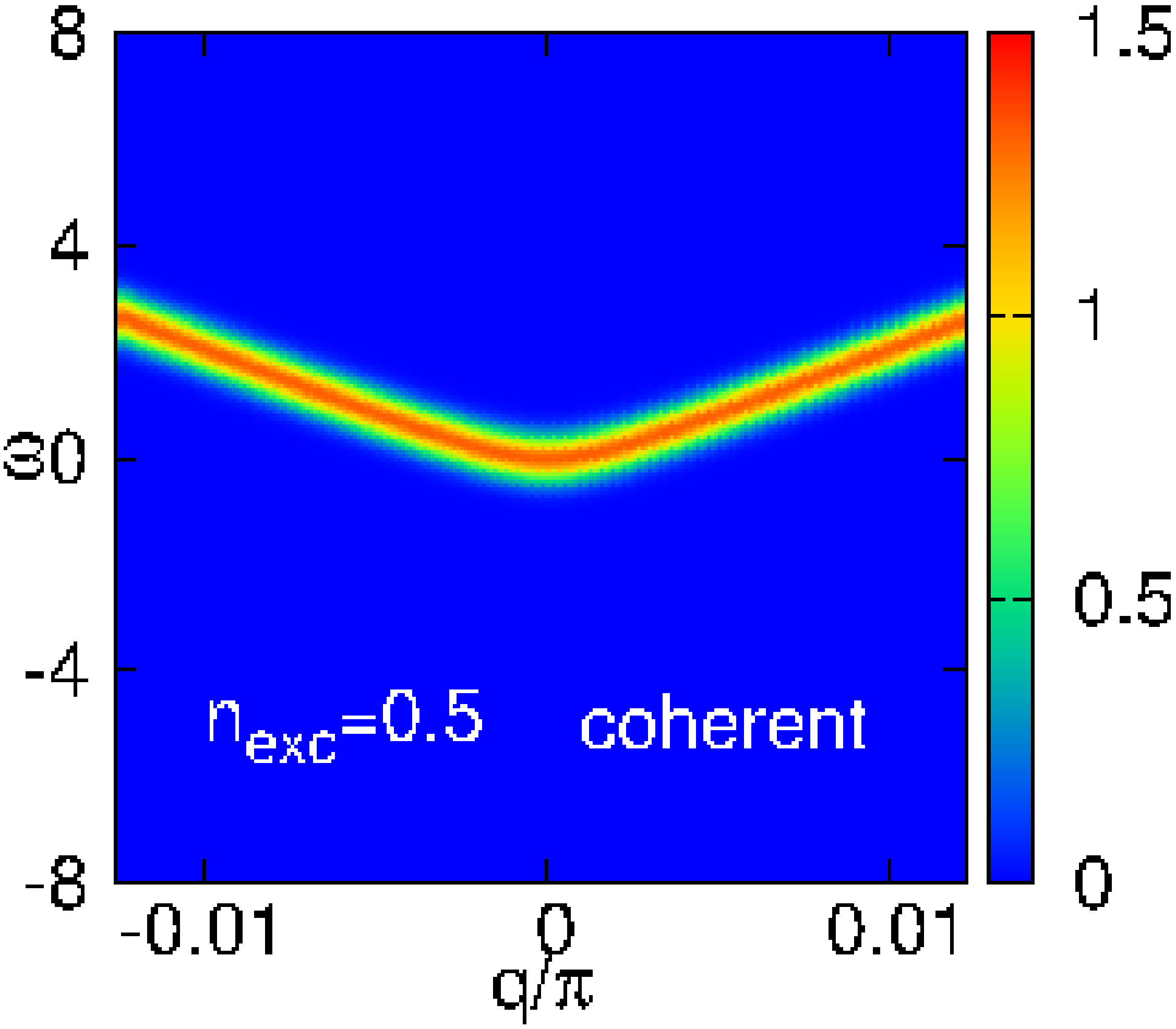}
\includegraphics[width=0.23\textwidth]{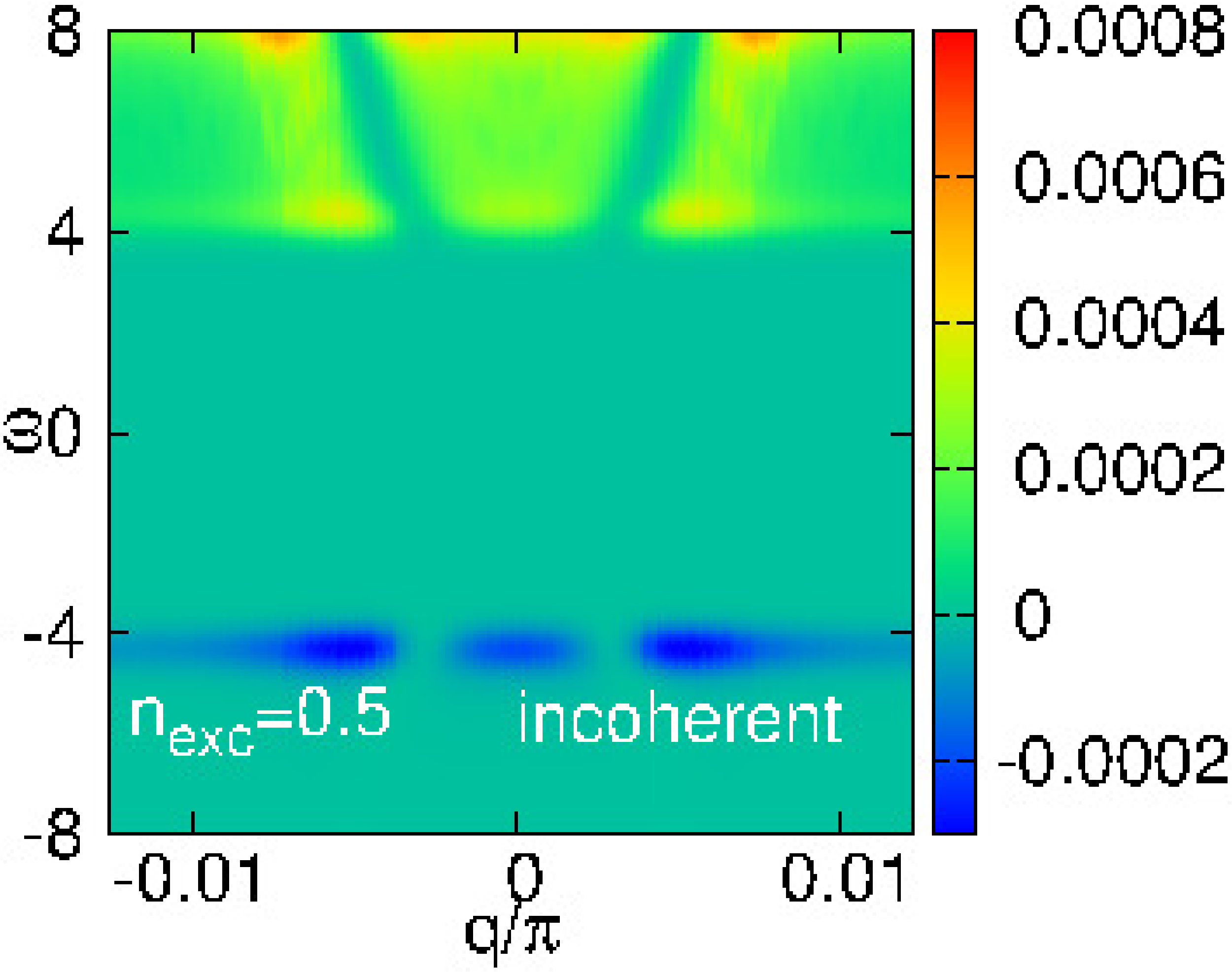}
\caption{(Color online) Intensity plot of the luminescence spectral function $B({\bf q}, \omega)$ for $n_\textrm{exc}=0.2$ (upper panels) and $n_\textrm{exc}=0.5$ (lower panels) at small detuning $d=-0.5$. Other parameters and notations as in Fig.~\ref{fig10}. }
\label{fig12i}
\end{figure}

\begin{figure}[t]
\includegraphics[width=0.21\textwidth]{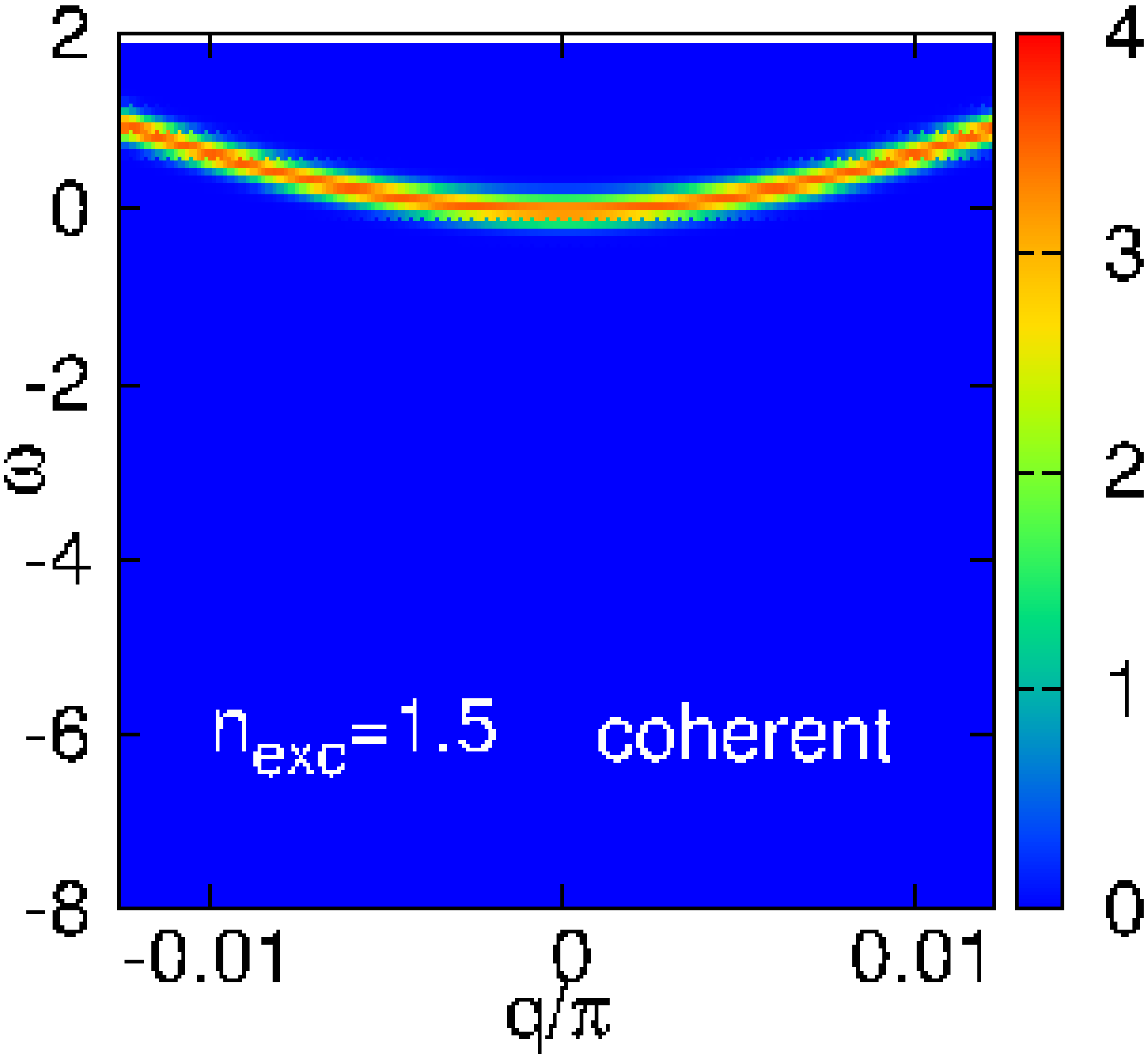}
\includegraphics[width=0.24\textwidth]{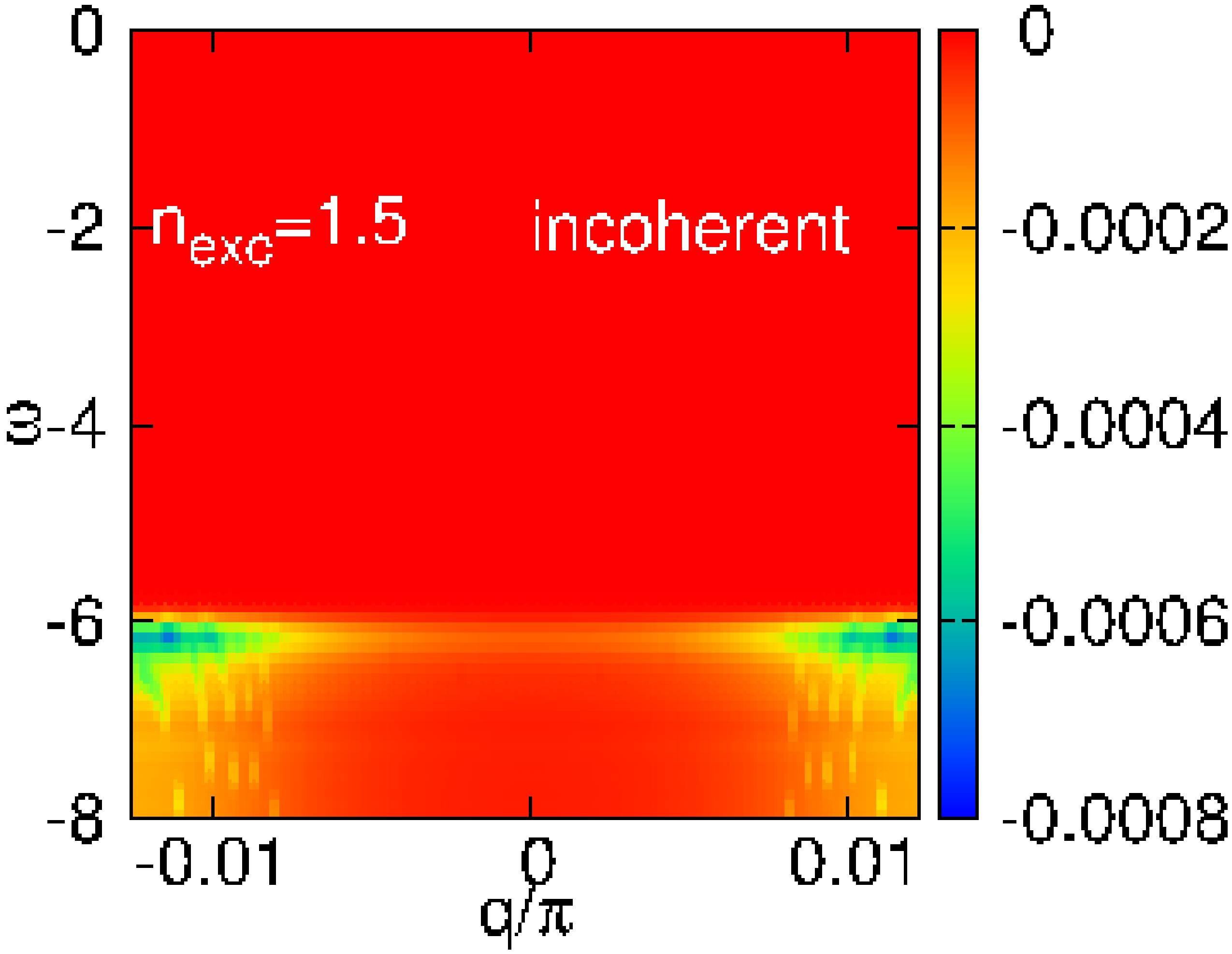}
\caption{(Color online) Intensity plot of the  luminescence function $B({\bf q}, \omega)$ for the same parameters 
as in Fig.\ref{fig12}.}
%$n_\textrm{exc}=1.5$, where $U=2$, and $g=0.2$. 
% Now $\omega_c=4.5$ and $E_g=-3$, resulting in a detuning $d=7.5$.}
\label{fig13i}
\end{figure}

 Finally, we also consider the luminescence spectral function $B(\vec q, \omega)$. The results  for  
small and large detunings are shown in Figs.~\ref{fig12i} and \ref{fig13i}. 
In both cases, the coherent parts  of the spectrum are dominant and follow 
 the renormalized  photon excitation $\omega_{\vec q}$, 
 whereas the incoherent excitations are of minor importance.  
 Note that because of the steep increase with $\vec q$ of the photonic dispersion $\omega_{\vec q}$ [Eq.~\eqref{5}], Figs. 16 and 17 focus
on the small-$\vec q$ interval around $\vec q=0$. 
As anticipated from Appendix \ref{A},  the onsets of the incoherent excitations of $B(\vec q,\omega)$
correspond to those of the coherent parts of  $A(\vec k, \omega)$. However, due to the restricted
$\vec q$ range in Figs.~\ref{fig12i} and \ref{fig13i}, this equivalence  
is hardly seen except in the dark blue horizontal  regions 
of low intensities  in the right panels of Fig.~\ref{fig12i} and the left panels of Fig.~\ref{10}. 
Moreover, the spectral weights of the coherent excitations 
 of $B(\vec q, \omega)$ in Figs.~\ref{fig12i} and \ref{fig13i}  are almost independent of $\vec q$.  However, there seems to be a 
 contradiction to the outcome in Fig.~\ref{fig5}. There, for a small $\omega_c=0.5$,  an intensity plot of the 
 photon density $\langle \psi_{\vec q}^\dag \psi_{\vec q} \rangle$  
in momentum space  is shown,  revealing a strongly peaked intensity around $\vec q=0$ only. 
  This apparent  contradiction can easily be resolved by help of the dissipation-fluctuation theorem:
 \begin{equation}
 \label{32}
   \langle \psi_{\vec q}^\dag \psi_{\vec q}\rangle = \int_{-\infty}^\infty d\omega \, \frac{B(\vec q, \omega)}{\displaystyle e^{\beta \omega} -1}  \,.
  \end{equation} 
   Exploiting the fact that the coherent part of $B(\vec q, \omega)$ is dominant, 
   $B(\vec q, \omega) \approx |\tilde{z}_{\vec q}|^2 \, \delta(\omega - \tilde{\omega}_{\vec q})$, ($|\tilde{z}_{\vec q}|^2 \approx 1$),
   one finds
 \begin{equation}
 \label{33}
   \langle \psi_{\vec q}^\dag \psi_{\vec q}\rangle \approx |\tilde{z}_{\vec q}|^2 \, \frac{1}{\displaystyle e^{\beta \tilde{\omega}_{\vec q}} -1} \, .
   \end{equation} 
 Obviously, for small temperatures (large $\beta$) wave vectors around $\vec q=0$ contribute most,
  since $\tilde{\omega}_{\vec q}$  is smallest there: This is particularly true  for the case of 
 Fig.~\ref{fig5}, where a small zero-point cavity frequency $\omega_c =0.5$ was used. 
 When we calculate the the expectation value
 $\langle \psi^\dag_{\vec q} \psi_{\vec q}\rangle$ for a large photon frequency $\omega_c=4.5$ (and $d=7.5$) the intensity of the photon density
 is smeared out, of course, in momentum space (not shown).
%
%% -------------------- FIG.16 ----------------------------
%

%%------------------------  FIG. 17 ------------------------------------------

%------------------------------------------------------------------------------------------------------------------------
\section{Conclusions}

To summarize, we have adapted the PRM (projective renormalization method) to investigate 
an exciton-polarition microcavity model with regard to the formation of Bose-Einstein condensates. Thereby, correlation and fluctuation effects were included. The PRM allows to derive analytical expressions for the excitonic and photonic (BEC) order parameters, the partial excitation densities of excitons and photons, the fully renormalized quasiparticle band structure, and the luminescence spectrum  in the whole parameter regime of detuning, excitation density, Coulomb interaction, and light-matter coupling. The nature of the condensate changes from an exciton to a polariton and finally to a photon dominated ground state when the density of excitations grows. For large detuning, the exciton condensate shows  a crossover from a BEC- to a BCS-type pairing, mainly because of Fermi-surface and Pauli-blocking effects. In this regime, also a clear onset density is observed for the photonic fraction, when  the total excitation is increased.  At the same time, the carrier density saturates. For small detuning, a strong mixture of electron and photon degrees of freedom takes place, right from starting to increase the excitation density. In this regime, pronounced polariton signatures can be found. The photonic (laser-like) behavior shows  a smooth onset  and dominates the physics at very large excitation densities.  In this way, our more elaborated PRM approach  confirms the exciton-polariton-photon crossover scenario obtained in the framework of a variational (mean-field) treatment~\cite{KO11}. The luminescence   and excitonic polarization spectra presented for the different parameter regimes support this behavior of the microcavity system as well. To analyze the influence of a trap potential~\cite{SSKSKSSNKF12} on the excitonic luminescence would be a worthwhile goal of forthcoming studies. Equally interesting would be to extend the PRM scheme to the study of exciton-polariton systems in non-equilibrium, e.g., with a focus on the description  of lasing.  \\

 Note that this study for the luminescence spectrum differs from those in  literature on  microcavity polaritons, since 
 there the exciton degrees of freedom are often described by local two-level systems (see for instance  Refs.~\onlinecite{KESL05,CC14}). Instead, 
 in the present study a coherent set of conduction electrons and valence holes for the exciton degrees of freedom 
 is considered which has strong influence on the 
 excitonic polarization $A(\vec k,\omega)$, though rather little influence on  the luminescence 
 function $B(\vec q, \omega)$. On the other hand, in this study, the contributions 
 of Goldstone modes to the spectra were neglected. In principle, they should show up 
 since the  continuous gauge symmetry $U(\alpha) \mathcal H U^{-1}(\alpha)= \mathcal H$ with 
 $U(\alpha)= \exp{(- i \alpha \mathcal N_\textrm{exc})}$  is violated in the condensed phase.  Then, in a linearized equation of motion method,
 a coupled set of equations  for the photonic variables $\psi_{\vec q}^\dag, \psi_{-\vec q}$ and for
 the particle-hole excitations $\{e^\dag_{\vec k + \vec q}h^\dag_{-\vec k}\}$, $\{h_{-\vec k} e_{\vec k- \vec q}\}$,  
 $\{e^\dag_{\vec k + \vec q}e_{-\vec k}\}$, and $\{h^\dag_{\vec k + \vec q} h_{-\vec k}\}$ (for all $\vec k$) would have to be solved. 
 Such a study is left for the future. For now, one might speculate that the influence  of Goldstone modes on the spectra 
 is of minor importance since their respective coupling strengths in $A(\vec k, \omega)$ and  $B(\vec q, \omega)$ are of higher order 
 in the interaction parameter $g$.

%--------------------------------------------------------
\acknowledgements
The authors would like to thank D. Semkat, H. Stolz, and B. Zenker for valuable discussions. 
This work was funded by Vietnam National Foundation
for Science and Technology Development (NAFOSTED)
under Grant No.103.01-2014.05 and by Deutsche Forschungsgemeinschaft (Germany) through the Collaborative Research Center 652, Projects B5 and B14.

\begin{appendix}

\section{Mean-field approximation}
\label{MF}
The mean-field approximation is obtained by neglecting the fluctuation part $\mathcal H_1$ in Eq.~\eqref{12}, i.e.~the Hamiltonian
reduces to 
\begin{eqnarray}
\label{MF.1}
\mathcal H_\textrm{MF} &=& \sum_{\vec{k}}\hat{\varepsilon}_{\vec{k}}^{e}e_{\vec{k}}^{\dagger}e_{\vec{k}}^{}+\sum_{\vec{k}}\hat{\varepsilon}_{\vec{k}}^{h}h_{\vec{k}}^{\dagger}h_{\vec{k}}^{}
+\sum_{\vec{q}}\omega_{\vec{q}}\Psi_{\vec{q}}^{\dagger}\Psi_{\vec{q}}^{}\nonumber \\
&+&\Delta\sum_{\vec{k}}(e_{\vec{k}}^{\dagger}h_{-\vec{k}}^{\dagger}+\textrm{H.c.}) \, .
\end{eqnarray}
Here $\hat{\varepsilon}_{\vec{k}}^{e}$, $\hat{\varepsilon}_{\vec{k}}^{h}$, $\Delta$, and $\Psi_{\vec{q}}^{\dagger}$ are 
given by Eqs.~\eqref{15}, \eqref{16}, \eqref{18}, and \eqref{21}. The electronic part of $\mathcal H_\textrm{MF}$ is easily diagonalized. 
By introducing 
\begin{eqnarray}
\label{MF.2}
C_{1\vec{k}}^{\dagger} & = & \xi_{\vec{k}}e_{\vec{k}}^{\dagger}+\eta_{\vec{k}}h_{-\vec{k}}^ {}\,, \\
\label{MF.3}
C_{2\vec{k}}^{\dagger} & = & - \eta_{\vec{k}}e_{\vec{k}}^{\dagger}+\xi_{\vec{k}}
h_{-\vec{k}}^ {} \,,
\end{eqnarray}
with  ($\xi_{\vec k}$, $\eta_{\vec k}$ real) 
\begin{eqnarray}
\label{MF.4} 
 && 
 \xi_{\vec{k}}^{2}=\frac{1}{2}\left[1+\textrm{sgn}(\hat{\varepsilon}_{\vec{k}}^{e}+\hat{\varepsilon}_{\vec{k}}^{h})\frac{\hat{\varepsilon}_{\vec{k}}^{e}+\hat{\varepsilon}_{\vec{k}}^{h}}{W_{\vec{k}}}\right]\,, \\
 &&
 \label{MF.5}
 \eta_{\vec{k}}^{2}=\frac{1}{2}\left[1-\textrm{sgn}(\hat{\varepsilon}_{\vec{k}}^{e}+\hat{\varepsilon}_{\vec{k}}^{h})
 \frac{\hat{\varepsilon}_{\vec{k}}^{e}+\hat{\varepsilon}_{\vec{k}}^{h}}{W_{\vec{k}}}\right]\,,
\\
&& \label{MF.6}
W_{\vec{k}}=\sqrt{(\hat{\varepsilon}_{\vec{k}}^{e}+\hat{\varepsilon}_{\vec{k}}^{h})^{2}+4|  \Delta|^{2}}
\, ,
\end{eqnarray}
ones arrives at  
\begin{align}
\label{MF.7}
{\mathcal H}_\textrm{MF}= & \sum_{\vec{k}} E_{\vec{k}}^{1}C_{1\vec{k}}^{\dagger}C_{1\vec{k}}^ {}+\sum_{\vec{k}}
E_{\vec{k}}^{2}C_{2\vec{k}}^{\dagger}C_{2\vec{k}}^ {}
+\sum_{\vec{q}}{\omega}_{\vec{q}}  {\Psi}_{\vec{q}}^{\dagger}
 {\Psi}_{\vec{q}}^ {},
\end{align}
with
\begin{equation}
\label{MF.8}
 E_{\vec{k}}^{1,2}=\frac{\hat{\varepsilon}_{\vec{k}}^{e}-\hat{\varepsilon}_{\vec{k}}^{h}}{2}\pm\textrm{sgn}(\hat{\varepsilon}_{\vec{k}}^{e}+\hat{\varepsilon}_{\vec{k}}^{h})\frac{W_{\vec{k}}}{2}\,.
\end{equation}
The diagonal form \eqref{MF.7} allows to evaluate all physical quantities in mean-field approximation. For instance,
\begin{eqnarray} 
 \label{MF.9}
&& \langle n_{\vec k}^e \rangle = |\xi_{\vec k}|^2 f(E_\vec k^1) + |\eta_{\vec k}|^2 f(E_\vec k^2), \\
 \label{MF.10}
 && \langle n_{\vec k}^h \rangle = 1- |\eta_{\vec k}|^2 f(E_\vec k^1) - |\xi_{\vec k}|^2 f(E_\vec k^2),\\
  \label{MF.11}
&& d_{\vec k} = \textrm{sgn}({E^1_{\vec k} - E^2_{\vec k}}) \big( f(E_\vec k^1) - f(E_\vec k^2)\big) \frac{\Delta}{W_{\vec k}},   \\
\label{MF.12}
&& \langle \psi_{\vec q=0}\rangle = - \frac{\sqrt N \Gamma}{\omega_{\vec q=0}},  \\
\label{MF.13}
&& \langle \psi^\dag_{\vec q} \psi_{\vec q}\rangle = p(\omega_\vec q) + \frac{N\Gamma^2}{\omega_{\vec q=0}^2} \, ,
\end{eqnarray}
where $p(\omega_{\vec q})$ is the bosonic distribution functions.   Note that the phase factor 
$ \textrm{sgn}({E^1_{\vec k} - E^2_{\vec k}})$ in Eq.~\eqref{MF.11} is found by comparing 
the exact expression for $d_k= \langle e^\dag_{\vec k} h^\dag_{-\vec k}\rangle$ 
with the perturbative result  of $d_k$ to lowest order in the coupling term 
$\Delta \sum_{\vec k} (e^\dag_{\vec k} h^\dag_{-\vec k} + \textrm{H.c.})$ of \eqref{MF.1}.  Equations.~\eqref{MF.9}-\eqref{MF.13} lead to the 
mean-field expressions for the order parameters $\Delta= -({g}/{\sqrt N}) \langle \psi_0\rangle - 
({U}/{N}) \sum_{\vec k} d_{\vec k}$ and $\Gamma= - ({g}/{N}) \sum_{\vec k} d_{\vec k}$, whereas
the total density $n_\textrm{exc}$ is given by
\begin{eqnarray}
\label{MF.14}
&& n_\textrm{exc} = \frac{1}{N} \sum_{\vec q} p(\omega_{\vec q}) + \frac{\Gamma^2}{\omega_{\vec q=0}^2} + \nonumber \\
 &&+
\frac{1}{2N} \sum_{\vec k} \Big[ 1 - \textrm{sgn}(\hat{\varepsilon}_{\vec{k}}^{e}+\hat{\varepsilon}_{\vec{k}}^{h})
\frac{\hat{\varepsilon}_{\vec{k}}^{e}+\hat{\varepsilon}_{\vec{k}}^{h}}{W_{\vec k}} 
\big( f(E^1_{\vec k}) - f(E^2_{\vec k}) \big)
\Big] \nonumber \, . \\
&&
\end{eqnarray}
It describes a mean-field condensate of coupled photons and exciton
polarization, where the term $ {\Gamma^2}/{\omega_{\vec q=0}^2}= {\langle \psi_0\rangle^2}/{N}$
is the density of photons in the condensate. 
The luminescence functions, defined in Eqs.~\eqref{24} and \eqref{27a}, become
\begin{align}
\label{MF.15}
A&(\vec k,\omega)= \frac{1}{N} \sum_{\vec p} \nonumber\\
\times&\Big( |\xi_{\vec k+\vec p}\eta_{\vec p}|^2[f( E^1_{\vec p})-f(E^1_{\vec k+\vec p})]\delta(\omega
- E^1_{\vec k+\vec p}+ E^1_{\vec p})\nonumber\\
&+ |\eta_{\vec k+\vec p}\eta_{\vec p}|^2[f( E^1_{\vec p})-f( E^2_{\vec k+\vec p})]\delta(\omega
- E^2_{\vec k+\vec p}+ E^1_{\vec p})\nonumber\\
&+|\xi_{\vec k+\vec p}\xi_{\vec p}|^2[f(E^2_{\vec p})-f( E^1_{\vec k+\vec p})]\delta(\omega
- E^1_{\vec k+\vec p}+ E^2_{\vec p})\nonumber\\
&+|\eta_{\vec k+\vec p}\xi_{\vec p}|^2 [f( E^2_{\vec p})-f(E^2_{\vec k+\vec p})]\delta(\omega
-E^2_{\vec k+\vec p}+ E^2_{\vec p})\Big) 
\nonumber \\
&
\end{align}
and 
\begin{align}
\label{MF.16}
 & B(\vec q, \omega) = \delta \big( \omega- \omega_{\vec q}\big). \\
 & \nonumber 
\end{align}
 Note the formal similarity of result
\eqref{MF.15}  to the coherent part $A^\textrm{coh}(\vec k, \omega)$ of the PRM expression \eqref{B.6} 
for $A(\vec k, \omega)$ in Appendix \ref{App B}. For the cavity photon spectral function $B(\vec q, \omega)$, the mean-field result 
reduces to a sole $\delta$ function, whereas all contributions from fluctuations in the PRM result \eqref{B.19} are of course missing.

%-----------------------------------------------------------------------------------------
\section{Projector-based Renormalization Method: General concepts}
\label{A}

In this appendix, we show how  the  complete Hamiltonian $\mathcal H$ can be  solved
by means of the PRM. So far, the PRM was successfully applied to a number of different models. Prominent examples are the one-dimensional Holstein model~\cite{SHBWF05}, the Edwards model~\cite{SBF10}, or   the
extended Falicov-Kimball model~\cite{PBF10}. The starting point is always a  decomposition  
of the many-particle Hamiltonian $\mathcal H$ into an ``unperturbed'' part  
$\mathcal{H}_0$ and into a ``perturbation'' $\mathcal{H}_1$, 
where the unperturbed part $\mathcal{H}_0$ is  solvable  [compare Eqs.~\eqref{21} and \eqref{22}].  
The perturbation $\mathcal H_1$  is responsible for transitions between the eigenstates of 
$\mathcal H_0$ with non-vanishing transition energies $|E^n_0 -E^m_0|$.  Here, $E^n_{0}$ and $E^m_{0}$ 
denote the energies of $\mathcal H_{0}$ between which the transitions take place.
The basic idea of the PRM method is to integrate out the interaction $\mathcal{H}_1$ 
by a sequence of discrete unitary transformations~\cite{BHS02}. 
Thereby, the PRM procedure starts from the largest transition energy of the original Hamiltonian $\mathcal H_0$, 
which will be called  $\Lambda$, and proceeds in small steps $\Delta \lambda$ to lower values of the
transition energy $\lambda$.  Every  
step is performed by means of a small unitary transformation, where all excitations between 
$\lambda$ and $\lambda - \Delta \lambda$ are eliminated:
\begin{eqnarray}
 \label{A.1} 
 \mathcal H_{\lambda -\Delta \lambda} &=& e^{X_{\lambda, \Delta \lambda}} \, \mathcal H_\lambda \,
 e^{-X_{\lambda, \Delta \lambda}} \, .
\end{eqnarray}
Here, the operator  $X_{\lambda, \Delta \lambda} =- X_{\lambda, \Delta \lambda}^\dag$ 
is the generator of the unitary transformation. 
Note that for  sufficiently small $\Delta \lambda$, the evaluation of transformation \eqref{A.1}
can be restricted to low orders in $\mathcal H_1$ which usually limits the validity of the 
approach to values of $\mathcal H_1$ of the same magnitude as those of $\mathcal H_0$. 
After each step, the unperturbed part as well as  the perturbation part of the Hamiltonian become 
renormalized and thus depend on the cutoff $\lambda$. One
arrives at a renormalized Hamiltonian $\mathcal H_\lambda = \mathcal H_{0,\lambda} + \mathcal H_{1,\lambda}$, 
where $\mathcal H_{1,\lambda}$ now only accounts for transitions with energies smaller than $\lambda$.  
Proceeding the renormalization stepwise up to zero transition energy $\lambda=0$ all transitions 
with energies different from zero have been integrated out. Thus, one finally arrives 
at a  renormalized Hamiltonian $\mathcal H_{\lambda=0}$, which is diagonal (or at least quasi-diagonal), since all 
transitions from $\mathcal H_1$ with non-zero energies  have been used up.

%-----------------------------------------------------------------------------------------------------------
\subsection{Hamiltonian $\mathcal H_\lambda$}
\label{App A}

Let us assume that all transitions with energies larger than $\lambda$ have already been integrated out. 
An appropriate {\it ansatz} for the transformed Hamiltonian $\mathcal H_\lambda$ reads as 
$\mathcal H_\lambda= \mathcal H_{0,\lambda} + \mathcal H_{1,\lambda}$ with  
 \begin{align}
\label{A.2}
\mathcal{H}_{0,\lambda}&= \sum_{\vec{k}}{\varepsilon}_{\vec{k},\lambda}^{e}
e_{\vec{k}}^{\dagger}e_{\vec{k}}^ {}
+\sum_{\vec{k}}{\varepsilon}_{\vec{k},\lambda}^{h} h_{\vec{k}}^{\dagger}h_{\vec{k}}^ {}  \\
&+\sum_{\vec{k}}\Delta_{\vec k, \lambda}(e_{\vec{k}}^{\dagger}h_{-\vec{k}}^{\dagger}+\textrm{H.c.}) 
+\sum_{\vec{p}}\omega_{\vec{q},\lambda}
\Psi_{\vec{q},\lambda}^{\dagger}\Psi_{\vec{q},\lambda}, \nonumber  \\
\label{A.3}
\mathcal{H}_{1,\lambda}&
=-\frac{g}{\sqrt{N}} \sum_{\vec{k}\vec{q}} \mathbf P_\lambda \big[:e_{\vec{k}+\vec{q}}^{\dagger}h_{-\vec{k}}^{\dagger}\Psi_{\vec{q},\lambda}:+\textrm{H.c.} \big]\nonumber \\
&-\frac{U}{N} \sum_{\vec{k}_{1}\vec{k}_{2}\vec{k}_{3}}
\mathbf P_\lambda\big[:e_{\vec{k}_{1}}^{\dagger}e_{\vec{k}_{2}}h_{\vec{k}_{3}}^{\dagger}h_{\vec{k}_{1}+\vec{k}_{3}-\vec{k}_{2}}:  \big]\, . 
\end{align}
Clearly, all parameters of $\mathcal H_{0,\lambda}$ now depend on the cutoff $\lambda$,
and  $\Delta_{\vec k, \lambda}$ has acquired an additional momentum dependence. 
Moreover, we have introduced a $\lambda$-dependent photon operator  
 \begin{eqnarray}
 \label{A.4}
\Psi^\dag_{\bf q,\lambda} = \psi_{\bf q}^\dag + 
\frac{\sqrt N \Gamma_{ \lambda}}{\omega_{\bf q=0,\lambda}}  \delta_{\bf q, 0} \, ,
 \end{eqnarray}
which is a slight generalization of the former definition \eqref{20}. Finally, the quantity $\mathbf P_\lambda$ 
in Eq.~\eqref{A.3} is a generalized projector, which projects on all transitions 
  with energies smaller than $\lambda$ (with respect to $\mathcal H_{0,\lambda}$).
Note that the coupling strength $g$ of $\mathcal H_{1,\lambda}$ remains
$\lambda$ independent, which is a consequence of the present restriction to
renormalization contributions up to order $g^2$ and $U^2$.  
 
Next, $\mathbf P_\lambda$ has to be applied 
to the operators in $\mathcal H_{1,\lambda}$,
which requires the decomposition of the operators in the squared brackets 
into dynamical eigenmodes of $\mathcal H_{0,\lambda}$. 
As long as one is only interested in renormalization equations up to linear order in the order parameters, 
one finds
 \begin{eqnarray}
\label{A.5}
&&\mathcal{H}_{1,\lambda}
=-\frac{g}{\sqrt{N}} \sum_{\vec{k}\vec{q}} \Theta_{\vec k \vec q, \lambda} \big[:e_{\vec{k}+\vec{q}}^{\dagger}h_{-\vec{k}}^{\dagger}\Psi_{\vec{q},\lambda}:+\textrm{H.c.} \big]\nonumber \\
&& \qquad -\frac{U}{N} \sum_{\vec{k}_{1}\vec{k}_{2}\vec{k}_{3}}
\Theta_{\vec k_1 \vec  k_2 \vec k_3, \lambda}:e_{\vec{k}_{1}}^{\dagger}e_{\vec{k}_{2}}h_{\vec{k}_{3}}^{\dagger}h_{\vec{k}_{1}+\vec{k}_{3}-\vec{k}_{2}}:  , \nonumber \\
&&
 \end{eqnarray}
  where we have introduced two $\Theta$ functions
 \begin{eqnarray}
\label{A.6}
 && \Theta_{\vec{k}\vec{q},\lambda}=\Theta(\lambda-|\varepsilon_{{\vec k+\vec q,\lambda}}^{e}
 +\varepsilon_{-{\bf k,\lambda}}^{h}-\omega_{{\bf q,\lambda}}|), \\
 \label{A.7}
 && \Theta_{\vec{k}_{1}\vec{k}_{2}\vec{k}_{3},\lambda} \nonumber \\
 && \quad =\Theta(\lambda-|\varepsilon_{{\bf {\bf k_{1},\lambda}}}^{e}-\varepsilon_{{\bf k_{2},\lambda}}^{e}+\varepsilon_{{\bf {\bf k_{3},\lambda}}}^{h}-\varepsilon_{\vec{k}_{1}+{\bf k_{3}-k_{2},\lambda}}^{h}|) \, .
 \nonumber  \\
 &&
 \end{eqnarray}
They restrict transitions to excitation energies smaller than $\lambda$.  
Next, one constructs the generator  $X_{\lambda, \Delta \lambda}$ of the unitary
transformation \eqref{A.1}. 
According to Ref.~\onlinecite{BHS02}, the lowest order
for $X_{\lambda, \Delta \lambda}$ is given by
\begin{eqnarray}
\label{A.8}
X_{\lambda, \Delta \lambda} = \frac{1}{ \mathbf L_{0,\lambda}} \mathbf Q_{\lambda- \Delta \lambda} \mathcal H_{1,\lambda}
\, ,
\end{eqnarray}
where $ \mathbf L_{0,\lambda}$ is the Liouville operator of the unperturbed 
Hamiltonian $\mathcal H_{0,\lambda}$. It is 
defined by $ \mathbf L_{0,\lambda} \mathcal A = [\mathcal{H}_{0, \lambda}, \mathcal A]$ 
for any operator quantity 
$\mathcal A$, and $\mathbf Q_{\lambda - \Delta \lambda} =1 - \mathbf P_{\lambda - \Delta \lambda}$ is the complement 
projector to $\mathbf P_{\lambda - \Delta \lambda}$, i.e., $\mathbf Q_{\lambda - \Delta \lambda}$  
projects on all transitions with energies 
larger than $\lambda -\Delta \lambda$. With Eqs.~\eqref{A.5} and \eqref{A.2} one finds
%------
\begin{align}
\label{A.9}
X_{\lambda,\Delta\lambda}&=-\frac{g}{\sqrt{N}}\sum_{\vec{k}\vec{q}}A_{{\bf k{\bf q}}}(\lambda,\Delta\lambda)\big[:e_{\mathbf{k+q}}^{\dagger}h_{-\mathbf{k}}^{\dagger}\Psi_{\mathbf{q},\lambda}:-\textrm{H.c.}\big]\nonumber\\
 &-\frac{U}{N}\sum_{\vec{k}_{1}\vec{k}_{2}\vec{k}_{3}}B_{{\bf k{\bf _{1}k_{2}k_{3}}}}(\lambda,\Delta\lambda):e_{\vec{k}_{1}}^{\dagger}e_{\vec{k}_{2}}h_{\vec{k}_{3}}^{\dagger}h_{\vec{k}_{1}+\vec{k}_{3}-\vec{k}_{2}}:
\end{align}
with the definitions
\begin{align}
\label{A.10}
 & A_{{\bf k{\bf q}}}^ {}(\lambda,\Delta\lambda)=\frac{\Theta_{{\bf k{\bf q,\lambda}}}\big(1-\Theta_{{\bf k{\bf q,\lambda-\Delta\lambda}}}\big)}{\varepsilon_{\vec k+\vec q,\lambda}^{e}+\varepsilon_{-{\bf k,\lambda}}^{h}-\omega_{\vec q,\lambda}} \,,\\
 \label{A.11}
 & B_{{\bf k{\bf _{1}k_{2}k_{3}}}}(\lambda,\Delta\lambda)=\frac{\Theta_{{\bf {\bf k{\bf _{1}k_{2}k_{3}},\lambda}}}\big(1-\Theta_{{\bf k{\bf _{1}k_{2}k_{3}},\lambda-\Delta\lambda}}\big)}{\varepsilon_{{\bf {\bf k_{1},\lambda}}}^{e}-\varepsilon_{{\bf k_{2},\lambda}}^{e}
 +\varepsilon_{{\bf {\bf k_{3},\lambda}}}^{h}-\varepsilon_{\vec{k}_{1}+{\bf k_{3}-k_{2},\lambda}}^h}\,.
\end{align}
Here, the products of the two $\Theta$ functions in $A_{\bf k \bf q}(\lambda, \Delta \lambda)$ 
and $B_{\bf k \bf _1 \bf k_2 \bf k_3}(\lambda,\Delta\lambda)$
assure that only excitations between $\lambda$
and $\lambda -\Delta \lambda$ are eliminated by the unitary transformation \eqref{A.1}.  
In principle, the Liouville operator $\mathbf L_{0,\lambda}$ in 
$X_{\lambda, \Delta \lambda}$ (and the projector 
$\mathbf P_\lambda$ in $\mathcal H_{1,\lambda}$) should have been 
defined with the full unperturbed Hamiltonian 
$\mathcal H_{0,\lambda}$ of Eq.~\eqref{A.2} and not by leaving out the term $\propto\Delta_{\vec k, \lambda}$.
However, its inclusion would only give rise to 
smaller higher-order corrections to $\Delta_{\vec k,\lambda}$ and is not important. 

%------------------------------------------------------------------------------------------------------

\subsection{Renormalization equations}
\label{A:2}

The $\lambda$ dependence of the parameters of $\mathcal H_\lambda$ is found from 
transformation \eqref{A.1}.  For small enough width $\Delta \lambda$ of the transformation steps, an expansion of  \eqref{A.1} in $g$ and $U$ can be limited to ${\cal O}(g^2)$ and ${\cal O}(U^2)$
terms. One obtains
\begin{eqnarray}
  \label{A.12} 
\mathcal H_{\lambda -\Delta \lambda} &=&  \mathcal H_{0,\lambda} + \mathbf P_{\lambda - \Delta \lambda} 
\mathcal H_{1,\lambda}
+ [X_{\lambda, \Delta \lambda} , \mathcal H_{1,\lambda}]  \nonumber \\
&-& \frac{1}{2} [X_{\lambda, \Delta \lambda} , \mathbf Q_{\lambda- \Delta \lambda} \mathcal H_{1, \lambda}] + \cdots \, ,
   \end{eqnarray}
where Eq.~\eqref{A.8} has been used. Renormalization contributions to $\mathcal H_{\lambda - \Delta \lambda}$ 
arise from the last two commutators which have to be evaluated explicitly. The result
must be compared with the generic forms \eqref{A.2} and \eqref{A.5} of $\mathcal H_\lambda$ (with $\lambda$ 
replaced by $\lambda - \Delta \lambda$) when it is written in terms of the original  
$\lambda$-independent variables 
$e^\dag_{\bf k}$, $h^\dag_{\bf k}$, and $\psi^{\dag}_{\bf q}$.  
This leads to the 
 following renormalization equations for the parameters of $\mathcal H_{0,\lambda}$:
\begin{eqnarray}
\label{A.13}
&&\varepsilon_{\vec{k},\lambda-\Delta\lambda}^{e} =\varepsilon_{\vec{k},\lambda}^{e}+\frac{2g^{2}}{N}\sum_{\vec{q}}A_{\vec{q},\vec{k}-\vec{q}}(\lambda,\Delta\lambda)(n_{\vec{q}}^{\Psi}+n_{\vec{q}-\vec{k}}^{h})\nonumber \\
 && \;\; + \frac{U^{2}}{N^{2}}\sum_{\vec{k}_{1}\vec{k}_{2}}B_{\vec k_1\vec k\vec k_2}(\lambda,\Delta\lambda)(1-2n_{\vec{k}_{1}}^{e}) %\nonumber\\
 %&&\qquad \times
 (n_{\vec{k}_{2}}^{h}-n_{\vec k_1+\vec{k}_{2}-\vec{k}}^{h})\nonumber \\
 &&\;\;+ \frac{U^{2}}{N^{2}}\sum_{\vec{k}_{1}\vec{k}_{2}}B_{\vec k,\vec k+\vec k_1-\vec k_2,\vec k_1}(\lambda,\Delta\lambda)\nonumber\\
 &&  \qquad \times[n^h_{\vec{k}_2}(1-n^h_{\vec k_1})+n^h_{\vec{k}_1}(1-n_{\vec{k}_{2}}^{h})]\,,
 \end{eqnarray}
\begin{eqnarray}
\label{A.14}
&& \varepsilon_{\vec{k},\lambda-\Delta\lambda}^{h} =\varepsilon_{\vec{k},\lambda}^{h}+\frac{2g^{2}}{N}\sum_{\vec{q}}A_{\vec{q},-\vec{k}}(\lambda,\Delta\lambda)(n_{\vec{q}}^{\Psi}+n_{\vec{q}-\vec{k}}^{e})\nonumber \\
 &&\;\;+  \frac{U^{2}}{N^{2}}\sum_{\vec{k}_{1}\vec{k}_{2}}B_{\vec k_1,\vec k_1+\vec k_2-\vec k,\vec k_2}(\lambda,\Delta\lambda) \nonumber \\
 && \qquad \times (1-2n^h_{\vec k_2})(n^e_{\vec{k}_1}-n^e_{\vec k_1+\vec{k}_2-\vec{k}})
\nonumber \\
&&\;\;+ \frac{U^{2}}{N^{2}}\sum_{\vec{k}_{1}\vec{k}_{2}}B_{\vec k_1\vec k_2\vec k}(\lambda,\Delta\lambda)
\nonumber \\
 && \qquad \times [n^e_{\vec{k}_2}(1-n^e_{\vec k_1})+n^e_{\vec{k}_1}(1-n_{\vec{k}_{2}}^{e}]  \,,
\end{eqnarray}
and 
\begin{align}
\label{A.15}
\omega_{\vec{k},\lambda-\Delta\lambda}^ {} & =\omega_{\vec{k},\lambda}^{}+\frac{2g^{2}}{N}\sum_{\vec{q}}A_{\vec{k},\vec{q}}(\lambda,\Delta\lambda)(n_{\vec{q}+\vec{k}}^{e}+n_{-\vec{q}}^{h}-1),\nonumber \\
&& \\
\label{A.16}
\Gamma_{\lambda-\Delta\lambda}^ {} & =\Gamma_{\lambda}-\frac{2g^{2}}{N\sqrt{N}}\sum_{\vec{q}}A_{\vec{0},\vec{q}}(\lambda,\Delta\lambda)\langle\psi_{0}\rangle\nonumber\\
&\qquad\qquad\times (n_{\vec{q}}^{e}+n_{-\vec{q}}^{h}-1),
\end{align}
\begin{align}
\label{A.17}
\Delta_{\vec{k},\lambda-\Delta\lambda} &=\Delta_{\vec{k},\lambda}-\frac{U^{2}}{N^{2}}\sum_{\vec{k}_{1}\vec{k}_{2}}\Big[\Gamma^{\vec k_1 \vec k, -\vec k_2}_{\vec k_1 \vec k_2, -\vec k}(\lambda, \Delta \lambda) \, \nonumber\\
&\qquad\qquad+\Gamma^{\vec k_1 \vec k, -\vec k_1}_{\vec k_1 \vec k_2, -\vec k_1}(\lambda, \Delta \lambda)
\Big](2 n^e_{\vec k_1} -1) d_{\vec k_2}\nonumber\\
&-\frac{U^2}{N^2}\sum_{\vec k_1 \vec k_2} \Big[
\Gamma^{\vec k_2, \vec k_1 +\vec k_2+ \vec k, \vec k_1}_{\vec k, \vec k_1 +\vec k_2+ \vec k, \vec k_1}(\lambda, \Delta \lambda)\nonumber\\
&\qquad\qquad+ \Gamma^{\vec k_2,- \vec k_1, -\vec k_2}_{\vec k, -\vec k_1, -\vec k}(\lambda, \Delta \lambda)
\Big](2 n^h_{\vec k_1} -1)d_{\vec k_2}\nonumber\\
&+\frac{2U^2}{N^2} \sum_{\vec{k}_{1}\vec{k}_{2}} 
 \Gamma^{\vec k_{1}k_{2},- \vec k_1}_{\vec{k}_{1}\vec{k}, -\vec{k}_{1}}(\lambda,\Delta\lambda)\nonumber\\
&\qquad\qquad\times (1- n^e_{\vec  k_1} - n^h_{-{\vec k_1}})d_{\vec k_2}\nonumber\\
&-\frac{U}{N}\sum_{\vec{k}_{1}}B_{\vec k\vec k_1,-\vec k}(\lambda,\Delta\lambda)\Delta_{\vec{k}_{1},\lambda}(1-n_{-\vec{k}_{1}}^{h}-n_{\vec{k}_{1}}^{e})\,.
\end{align}
The quantities $n^e_{\bf k}$ and $ n^h_{\bf k}$
are the occupation numbers for electrons and holes from Eq.~\eqref{16}, and $d_{\vec k}$  was defined in 
Eq.~\eqref{19}. Following, we shall also use the 
photonic occupation number $n_{\vec{q},\lambda}^{\Psi}$, 
\begin{eqnarray}
\label{A.18}
n_{\vec{q},\lambda}^{\Psi}&=&\langle\delta\Psi_{\vec{q},\lambda}^{\dag}\delta\Psi_{\vec{q},\lambda}^ {}\rangle=\langle\Psi_{\vec{q},\lambda}^{\dag}\Psi_{\vec{q},\lambda}^ {}\rangle-\langle\Psi_{\vec{q},\lambda}^{\dag}\rangle\langle\Psi_{\vec{q},\lambda}\rangle \nonumber \\
&=&\langle\delta\psi_{\vec{q}}^{\dag}\delta\psi_{\vec{q}}^ {}\rangle= n^\psi_{\bf q} \, ,
\end{eqnarray}
which is independent of  $\lambda$. 
In Eq.~\eqref{A.17}, we have also defined
\begin{align}
\label{A.19}
 \Gamma_{\vec k'_{1}\vec k'_{2}\vec k'_{3}}^{\vec{k}_{1}\vec{k}_{2}\vec{k}_{3}}(\lambda,\Delta\lambda) =&
 \frac{1}{2} \big[
  B_{\vec k'_{1}\vec k'_{2}\vec k'_{3}}(\lambda,\Delta\lambda) \, \Theta_{\vec{k}_{1}\vec{k}_{2}\vec{k}_{3},\lambda} \nonumber\\
&+ B_{\vec k_{1}\vec k_{2}\vec k_{3}}(\lambda,\Delta\lambda) \, \Theta_{\vec{k}'_{1}\vec{k}'_{2}\vec{k}'_{3},\lambda} 
 \big]\,. \\
 && \nonumber
\end{align}

For the numerical solution of the renormalization equations, the initial parameter values are 
 those of the original model $\mathcal H$  ($\lambda = \Lambda$):   
\begin{equation}
\label{A.20}
\varepsilon_{\vec{k},\Lambda}^{e}=\hat{\varepsilon}_{\vec{k}}^{e}, \quad\varepsilon_{\vec{k},\Lambda}^{h}=\hat{\varepsilon}_{\vec{k}}^{h},\quad\omega_{\vec{k},\Lambda}=\omega_{\vec{k}}\,,
\end{equation}
and 
\begin{eqnarray}
\label{A.21}
\Delta_{\vec{k},\Lambda}&=&\Delta = -\frac{g}{\sqrt{N}}\langle\psi_{0}\rangle-\frac{U}{N}\sum_{\vec{k}}d_{\vec k} \,, \\
\Gamma_{\Lambda} &=& \Gamma = -\frac{g}{N}\sum_{\vec{k}}d_{\vec k}   \,,
  \end{eqnarray}
  with $\langle\psi_{0}\rangle= 0^+$,   $d_{\vec k}=\langle e^\dagger_{\vec k}h^\dagger_{-\vec k} \rangle=0^+$. 
  Suppose the expectation values in \eqref{A.13}--\eqref{A.17} would already be known, the renormalization equations 
can be integrated between $\lambda = \Lambda$ and $0$. 
In this way, we obtain the fully renormalized Hamiltonian $\tilde{\mathcal{H}}:= \mathcal H_{\lambda=0} =  
\mathcal H_{0, \lambda=0}$, as was already stated in Eq.~\eqref{23}:
\begin{align}
\label{A.22}
\tilde{\mathcal H}= & \sum_{\vec{k}}\tilde{\varepsilon}_{\vec{k}}^{e}e_{\vec{k}}^{\dagger}e_{\vec{k}}^ {}+\sum_{\vec{k}}\tilde{\varepsilon}_{\vec{k}}^{h}h_{\vec{k}}^{\dagger}h_{\vec{k}} 
+\sum_{\vec{k}}\tilde{\Delta}_{\vec{k}}(e_{\vec{k}}^{\dagger}h_{-\vec{k}}^{\dagger}+\textrm{H.c.})\nonumber\\
&+\sum_{\vec{q}}\tilde{\omega}_{\vec{q}}\tilde{\Psi}_{\vec{q}}^{\dagger}\tilde{\Psi}_{\vec{q}}\,. 
\end{align}
The  tilde symbols denote the fully renormalized quantities at  $\lambda =0$ as before.  All  excitations from
$\mathcal H_{1,\lambda}$ with non-zero energies have been eliminated.
They give rises to the renormalization of $\mathcal H_{0,\lambda}$. 

Finally, the electronic part of $\tilde{\mathcal H}$ will  be diagonalized by a 
Bogoliubov transformation in close analogy to Appendix~\ref{MF}. Defining again new linear combinations 
\begin{eqnarray}
\label{A.23}
C_{1\vec{k}}^{\dagger} & = & \xi_{\vec{k}}e_{\vec{k}}^{\dagger}+\eta_{\vec{k}}h_{-\vec{k}}^ {}\,, \\
C_{2\vec{k}}^{\dagger} & = & -\eta_{\vec{k}}e_{\vec{k}}^{\dagger}+\xi_{\vec{k}}h_{-\vec{k}}^ {}
\end{eqnarray}
(with $\eta_{\vec k}, \xi_{\vec k}$ assumed to be real), where  now the renormalized one-particles energies 
$\tilde{\varepsilon}_{\vec{k}}^{e}$ and  $\tilde{\varepsilon}_{\vec{k}}^{h}$ enter the prefactors $\xi_{\vec k}$ and $\eta_{\vec k}$, 
\begin{eqnarray}
\label{A.24} 
 && 
 \xi_{\vec{k}}^{2}=\frac{1}{2}\left[1+\textrm{sgn}(\tilde{\varepsilon}_{\vec{k}}^{e}+\tilde{\varepsilon}_{\vec{k}}^{h})\frac{\tilde{\varepsilon}_{\vec{k}}^{e}+\tilde{\varepsilon}_{\vec{k}}^{h}}{W_{\vec{k}}}\right]\,, \\
 &&
 \eta_{\vec{k}}^{2}=\frac{1}{2}\left[1-\textrm{sgn}(\tilde{\varepsilon}_{\vec{k}}^{e}+\tilde{\varepsilon}_{\vec{k}}^{h})
 \frac{\tilde{\varepsilon}_{\vec{k}}^{e}+\tilde{\varepsilon}_{\vec{k}}^{h}}{W_{\vec{k}}}\right]\,,
\\
&& \label{A.25}
W_{\vec{k}}=\sqrt{(\tilde{\varepsilon}_{\vec{k}}^{e}+\tilde{\varepsilon}_{\vec{k}}^{h})^{2}+4|\tilde \Delta_{\vec{k}}|^{2}}
\, ,
\end{eqnarray}
 one finds
\begin{align}
\label{A.26}
\tilde{\mathcal{H}}= & \sum_{\vec{k}}\tilde E_{\vec{k}}^{1}C_{1\vec{k}}^{\dagger}C_{1\vec{k}}^ {}+\sum_{\vec{k}}
\tilde E_{\vec{k}}^{2}C_{2\vec{k}}^{\dagger}C_{2\vec{k}}^ {}
+\sum_{\vec{q}}{\tilde \omega}_{\vec{q}} \tilde {\Psi}_{\vec{q}}^{\dagger}
\tilde {\Psi}_{\vec{q}}^ {},
\end{align}
with
\begin{equation}
\label{A.27}
\tilde E_{\vec{k}}^{1,2}=\frac{\tilde{\varepsilon}_{\vec{k}}^{e}-\tilde{\varepsilon}_{\vec{k}}^{h}}{2}\pm\textrm{sgn}(\tilde{\varepsilon}_{\vec{k}}^{e}+\tilde{\varepsilon}_{\vec{k}}^{h})\frac{W_{\vec{k}}}{2}\,.
\end{equation}
Here, the electronic quasiparticle energies $\tilde E^{(1,2)}_{\bf k}$ and 
the quasiparticle modes  $C^{(\dag)}_{1\bf k}$, $C^{(\dag)}_{2\bf k}$
are  renormalized quantities as well. The quadratic form of Eq.~\eqref{A.26} allows to compute any expectation value formed with $\tilde{\mathcal H}$.
Finally, we note that the diagonalization \eqref{A.23}  runs along the same lines as the 
former Bogoliubov transformation of expression \eqref{21} for $\mathcal H_0$, 
except that the renormalized quantities have to be replaced by the unrenormalized ones.

%----------------------------------------------------------------------------------------------------------------
\subsection{Expectation values}
%-------------------------------------------------------------------------------------------------------------------
\label{A:3}

Also, expectation values $\langle  \mathcal A \rangle$, formed with the full $\mathcal H$, can be  
evaluated in the framework of the PRM.
As already stated in Sec.~\ref{III}, they are found by 
exploiting the unitary invariance of operator expressions below a trace, 
$\langle {\mathcal A} \rangle = \langle \mathcal{A}(\lambda)\rangle_{\mathcal H_\lambda}  =
 \langle \tilde{\mathcal A} \rangle_{\tilde{\mathcal H}}$, 
where $\mathcal A(\lambda)= e^{X_\lambda} {\mathcal A}e^{-X_\lambda}$, and  $\tilde{\mathcal A} = \mathcal A(\lambda =0)$.
$X_\lambda$ is the generator for the unitary transformation between cutoff $\Lambda$ and $\lambda$.
To find the expectation values of Eqs.~\eqref{A.13}-\eqref{A.17}, one best starts from an appropriate 
{\it ansatz} for the single-fermion operators 
\begin{align}
\label{A.28}
e_{\mathbf{k}}^{\dagger}(\lambda) & =x_{\mathbf{k},\lambda}e_{\mathbf{k}}^{\dagger}+\frac{1}{\sqrt{N}}\sum_{\mathbf{q}}t_{\mathbf{k-q,q},\lambda}h_{-\mathbf{q}}^ {}:\Psi_{\vec{k}-\mathbf{q},\lambda}^{\dagger}:\nonumber \\
 & +\frac{1}{N}\sum_{\vec{k}_{1}\vec{k}_{2}}\alpha_{\vec{k}_{1}\vec{k}\vec{k}_{2},\lambda}:e_{\vec{k}_{1}}^{\dagger}
 h_{\vec{k}_{2}}^{\dagger}h_{\vec{k}_{1}+\vec{k}_{2}-\vec{k}}:,\\
\label{A.29}
h_{\mathbf{k}}^{\dagger}(\lambda) & =y_{\mathbf{k},\lambda}h_{\mathbf{k}}^{\dagger}+\frac{1}{\sqrt{N}}\sum_{\mathbf{q}}u_{\mathbf{q},-\vec{k},\lambda}e_{\vec{q}-\mathbf{k}}^ {}:\Psi_{\vec{q},\lambda}^{\dagger}:\nonumber \\
 & +\frac{1}{N}\sum_{\vec{k}_{1}\vec{k}_{2}}\beta_{\vec{k}_{1}\vec{k}_{2},\vec{k}-\vec{k}_{1}+\vec{k}_{2},\lambda}:e_{\vec{k}_{1}}^{\dagger}e_{\vec{k}_{2}}h_{\vec{k}-\vec{k}_{1}+\vec{k}_{2}}^{\dagger}:, 
 \end{align}
(where $:\Psi^\dag_{\vec k,\lambda}:= :\psi^\dag_{\bf k}:$),  and for the photon operator 
 \begin{align}
 \label{A.30}
\psi_{\mathbf{q}}^{\dagger}(\lambda) & =z_{\mathbf{q},\lambda}\psi_{\mathbf{q}}^{\dagger}+\frac{1}{\sqrt{N}}\sum_{\mathbf{k}}v_{\mathbf{qk},\lambda}:e_{\vec{k}+\vec{q}}^{\dagger}h_{-\mathbf{k}}^{\dagger}:, 
\end{align}
where again the operator structures of 
\eqref{A.28}--\eqref{A.30} were taken over from a small-$X_\lambda$ 
expansion.
In analogy to the renormalization equations for the parameters of $\mathcal H_\lambda$, 
one derives the following set of renormalization equations for the $\lambda$-dependent coefficients 
 $t_{\mathbf{k-q,q},\lambda}^ {}$,
$u_{\mathbf{q},-\vec{k},\lambda}^ {}$, $v_{\mathbf{k,q},\lambda}$,
$\alpha_{\vec{k}_{1}\vec{k}_{2}\vec{k}_{3},\lambda}$ and $\beta_{\vec{k}_{1}\vec{k}_{2}\vec{k}_{3},\lambda}$:
\begin{align}
\label{A.31a}
&t_{\mathbf{k-q,q},\lambda-\Delta\lambda}^ {}  =  t_{\mathbf{k-q,q},\lambda}^ {}+gx_{\mathbf{k},\lambda}^ {}A_{\mathbf{k-q,q}}^ {}(\lambda,\Delta\lambda)\,,\\
\label{A.31b}
&u_{\mathbf{q,-k},\lambda-\Delta\lambda}^ {}  =  u_{\mathbf{q,-k},\lambda}^ {}-gy_{\mathbf{k},\lambda}^ {}A_{\mathbf{q,-k}}^ {}(\lambda,\Delta\lambda)\,,\\
\label{A.31c}
&v_{\mathbf{kq},\lambda-\Delta\lambda}  =  v_{\mathbf{kq},\lambda}-gz_{\mathbf{k},\lambda}A_{\mathbf{kq}}(\lambda,\Delta\lambda)\,,\\
\label{A.31d}
&\alpha_{\mathbf{k_{1}kk_{2}},\lambda-\Delta\lambda}  =  \alpha_{\mathbf{k_{1}kk_{2}},\lambda}-Ux_{\mathbf{k},\lambda}B_{\mathbf{k_{1}kk}_{2}}(\lambda,\Delta\lambda)\,,\\
\label{A.31e}
&\beta_{\mathbf{k_{1}k_{2},k-k_{1}+k_{2}},\lambda-\Delta\lambda}  =  \beta_{\mathbf{\mathbf{k_{1}k_{2},k-k_{1}+k_{2}}},\lambda}\nonumber\\
&\hspace*{2.5cm}-Uy_{\mathbf{k},\lambda}B_{\mathbf{k_{1}k_{2},k-k_{1}+k_{2}}}(\lambda,\Delta\lambda)\,. 
\end{align}
Using the anticommutation relations for fermion operators and the commutation relations for boson operators  
(as for instance $[e^\dagger_{\mathbf{k}}(\lambda),e_{\mathbf{k}}(\lambda)]_+=1$, valid for any $\lambda$), 
one arrives at 
\begin{align}
\label{A.32}
|x_{\mathbf{k},\lambda}|^{2}= & 1-\frac{1}{N}\sum_{\mathbf{q}}|t_{\mathbf{k-q,q},\lambda}^ {}|^{2}(n_{\mathbf{k-q},\lambda}^{\Psi}+n_{\mathbf{-q}}^{h})\nonumber \\
 & -\frac{1}{N^{2}}\sum_{\vec{k}_{1}\vec{k}_{2}}|\alpha_{\vec{k}_{1}\vec{k}\vec{k}_{2},\lambda}|^{2}\Big[n_{\vec{k}_{1}+\vec{k}_{2}-\vec{k}}^{h}(1-n_{\vec{k}_{2}}^{h})\nonumber\\
 &\qquad\qquad-n_{\vec{k}_{1}}^{e}(n_{\vec{k}_{1}+\vec{k}_{2}-\vec{k}}^{h}-n_{\vec{k}_{2}}^{h})\Big]\,,\\
 \label{A.33}
|y_{\mathbf{k},\lambda}|^{2}= & 1-\frac{1}{N}\sum_{\mathbf{q}}|u_{\mathbf{q,-k},\lambda}^ {}|^{2}(n_{\mathbf{q},\lambda}^{\Psi}+n_{\mathbf{q-k}}^{e})\nonumber \\
 & -\frac{1}{N^{2}}\sum_{\vec{k}_{1}\vec{k}_{2}}|\beta_{\vec{k}_{1}\vec{k}_{2},\vec{k}-\vec{k}_{1}+\vec{k}_{2},\lambda}|^{2}\Big[n_{\vec{k}_{1}}^{e}(1-n_{\vec{k}_{2}}^{e})\nonumber\\
 &\qquad\qquad+(1-n_{\vec{k}-\vec{k}_{1}+\vec{k}_{2}}^{h}(n_{\vec{k}_{2}}^{e}-n_{\vec{k}_{1}}^{e})\Big]\,,\\
 \label{A.34}
|z_{\mathbf{k},\lambda}|^{2}= & 1-\frac{1}{N}\sum_{\mathbf{q}}|v_{\mathbf{qk},\lambda}^ {}|^{2}(1-n_{\mathbf{-k}}^{h}-n_{\mathbf{k+q}}^{e})\,.
\end{align}
Equations~\eqref{A.31a}--\eqref{A.31e} together with the new set  \eqref{A.32}--\eqref{A.34}, 
taken at $\lambda \rightarrow \lambda -\Delta \lambda$, represents
a complete set of renormalization  equations for all $\lambda$-dependent coefficients in Eqs.~\eqref{A.28}--\eqref{A.30}. 
They combine the parameter values at $\lambda$ with those at $\lambda - \Delta \lambda$. Their initial values at $\lambda=\Lambda$ are:
\begin{align}
\label{A.35}
&\{x_{\vec{k},\Lambda},y_{\vec{k},\Lambda},z_{\vec{k},\Lambda}\}=1\,,\\
&\{t_{\vec{k}\vec{q},\Lambda},u_{\vec{k}\vec{q},\Lambda},v_{\vec{k}\vec{q},\Lambda},\alpha_{\vec{k}_{1}\vec{k}\vec{k}_{2},\lambda},\beta_{\vec{k}_{1}\vec{k}\vec{k}_{2},\lambda}\}=0\,.
\end{align}
By integrating the full set of renormalization equations between $\Lambda$ 
and $\lambda =0$, one is led to the fully renormalized one-particle operators:
\begin{align}
\label{A.36}
& \tilde{e}_{\mathbf{k}}^{\dagger} =\tilde{x}_{\mathbf{k}}e_{\mathbf{k}}^{\dagger}+\frac{1}{\sqrt{N}}\sum_{\mathbf{q}}\tilde{t}_{\mathbf{k-q,q}}h_{-\mathbf{q}}^ {}
:{\psi}_{\vec{k}-\mathbf{q}}^{\dagger}:\nonumber\\
&+\frac{1}{N}\sum_{\vec{k}_{1}\vec{k}_{2}}\tilde{\alpha}_{\vec{k}_{1}\vec{k}\vec{k}_{2}}:e_{\vec{k}_{1}}^{\dagger}h_{\vec{k}_{2}}^{\dagger}h_{\vec{k}_{1}+\vec{k}_{2}-\vec{k}}:, \\
\label{A.37}
& \tilde{h}_{\mathbf{k}}^{\dagger} =\tilde{y}_{\mathbf{k}}h_{\mathbf{k}}^{\dagger}+\frac{1}{\sqrt{N}}\sum_{\mathbf{q}}\tilde{u}_{\mathbf{q},-\vec{k}}e_{\vec{q}-\mathbf{k}}^ {}
:{\psi}_{\vec{q}}^{\dagger}:\nonumber\\
&+\frac{1}{N}\sum_{\vec{k}_{1}\vec{k}_{2}}\tilde{\beta}_{\vec{k}_{1}\vec{k}_{2},\vec{k}-\vec{k}_{1}+\vec{k}_{2}}:e_{\vec{k}_{1}}^{\dagger}e_{\vec{k}_{2}}h_{\vec{k}-\vec{k}_{1}+\vec{k}_{2}}^{\dagger}:,\\
\label{A.38}
& \tilde{\psi}_{\mathbf{k}}^{\dagger}  =\tilde{z}_{\mathbf{k}}\psi_{\mathbf{k}}^{\dagger}+\frac{1}{\sqrt{N}}\sum_{\mathbf{q}}\tilde{v}_{\mathbf{kq}}:e_{\vec{q}+\vec{k}}^{\dagger}h_{-\mathbf{q}}^{\dagger}:.
\end{align}
Again, tilde symbols denote the fully renormalized quantities. 
With Eqs.~\eqref{A.36}--\eqref{A.38} the expectation values $n_{\vec k}^e$, $n^h_{\vec k}$, $d_{\vec k}$, and 
$n^{\psi}_{\vec k}$   can be evaluated.  Thus, for the fermionic quantities one obtains
 up to order ${\cal O}(g_{\mathbf{k}}^2)$ 
and ${\cal O}(U_{\mathbf{k}}^2)$:  
\begin{align}
\label{A.39}
&n_{\vec{k}}^{e} =|\tilde{x}_{\vec{k}}|^{2}\tilde{n}_{\vec{k}}^{e}+\frac{1}{N}\sum_{\vec{q}}|\tilde{t}_{\vec{k}-\vec{q},\vec{q}}|^{2}(1-\tilde{n}_{-\vec{q}}^{h})\tilde{n}_{\vec{k}-\vec{q}}^{\psi}\nonumber \\
&\quad +\frac{1}{N^{2}}\sum_{\vec{k}_{1}{\vec{k}_{2}}}|\tilde{\alpha}_{\vec{k}_{1}\vec{k}\vec{k}_{2}}|^{2}\tilde{n}_{\vec{k}_{1}}^{e}\tilde{n}_{\vec{k}_{2}}^{h}(1-\tilde{n}_{\vec{k}_{1}+\vec{k}_{2}-\vec{k}}^{h}),\\
\label{A.40}
&n_{\vec{k}}^{h}=|\tilde{y}_{\vec{k}}|^{2}\tilde{n}_{\vec{k}}^{h}+\frac{1}{N}\sum_{\vec{q}}|\tilde{u}_{\vec{q},-\vec{k}}|^{2}(1-\tilde{n}_{\vec{q}-\vec{k}}^{e}) \tilde{n}_{\vec{q}}^{\psi}\nonumber \\
&\quad+\frac{1}{N^{2}}\sum_{\vec{k}_{1}\vec{k}_{2}}|\tilde{\beta}^{}_{\vec{k}_{1}\vec{k}_{2},\vec{k}-\vec{k}_{1}+\vec{k}_{2}}|^{2}\tilde{n}_{\vec{k}_{1}-\vec{k}_{1}+\vec{k}_{2}}^{h}\tilde{n}_{\vec{k}_{1}}^{c}(1-\tilde{n}_{\vec{k}_{2}}^{c}),\\
\label{A.41}
 & d_{\vec{k}}=x_{\vec{k}}y_{\vec{k}}\langle 
 e_{\vec{k}}^{\dagger}h_{-\vec{k}}^{\dagger}\rangle_{\tilde{\mathcal{H}}} \nonumber \\
& \quad\quad-\frac{1}{N^{2}}\sum_{\vec{k}_{1}\vec{k}_{2}}\tilde{\alpha}_{\vec{k}_{1}\vec{k},\vec{k}-\vec{k}_{1}+\vec{k}_{2}}\tilde{\beta}{}_{\vec{k}_{1}\vec{k}_{2},\vec{k}-\vec{k}_{1}+\vec{k}_{2}}\tilde{n}_{\vec{k}_{1}}^{e}\nonumber\\
&\qquad\qquad\times(1-\tilde{n}_{\vec{k}-\vec{k}_{1}+\vec{k}_{2}}^{h})\langle e_{\vec{k}_{2}}^{\dagger}h_{-\vec{k}_{2}}^{\dagger}\rangle_{\tilde{\mathcal{H}}}\,.
\end{align}
On the right-hand sides the expectation values, formed  with $\tilde{\mathcal H}$, can easily be  evaluated, 
\begin{align}
\label{A.42a}
 & \tilde{n}_{\vec{k}}^{e}=\xi_{\vec{k}}^{2}f(\tilde E_{\vec{k}}^{1})
 +\eta_{\vec{k}}^{2}f(\tilde E_{\vec{k}}^{2})\,,\\\label{A.42b}
 & \tilde{n}_{\vec{k}}^{h}=1-\eta_{\vec{k}}^{2}f(\tilde E_{\vec{k}}^{1})-\xi_{\vec{k}}^{2}f(\tilde E_{\vec{k}}^{2}),\\\label{A.42c}
 & \langle e^\dag_{\vec{k}}h^\dag_{-\vec{k}}\rangle_{\tilde{\mathcal{H}}}=
 \textrm{sgn}( \tilde E^1_{\vec{k}}- \tilde E^2_{\vec{k}})\,
 [f(\tilde E_{\vec{k}}^{1}) -f(\tilde E_{\vec{k}}^{2})] \, \frac{\tilde{\Delta}_{\vec k}}{W_{\vec{k}}} \, ,
\end{align}
where $f(E)$ is the Fermi function. The prefactors $\xi_{\bf k}$ and $\eta_{\bf k}$ are the coefficients from the Bogoliubov transformation \eqref{A.23}.

Finally,  the 
bosonic expectation value $n_{\bf q}^\psi$ is given by 
\begin{align}
\label{A.43}
&n^\psi_{\bf q}  = 
\langle \delta \psi^\dag_{\bf q} \delta \psi_{\bf q}\rangle =
\langle \psi^\dag_{\bf q} \psi_{\bf q}\rangle - 
\langle \psi^\dag_{\bf q} \rangle  \langle \psi_{\bf q} \rangle \delta_{\bf q= 0}\, ,
\end{align}
 where from \eqref{A.38}
 \begin{eqnarray}
\label{A.44}
 \langle\psi_{\vec{q}}^{\dagger}\psi_{\vec{q}}\rangle &=&|\tilde{z}_{\vec{q}}|^{2}\langle\psi_{\vec{q}}^{\dagger}\psi_{\vec{q}}^ {}\rangle_{\tilde{\mathcal{H}}}+\frac{1}{N}\sum_{\vec{k}}|\tilde{v}_{\vec{q}\vec{k}}|^{2}\tilde{n}_{-\vec{k}}^{h}\tilde{n}_{\vec{k}+\vec{q}}^{e}\,,\nonumber\\&& \\
 \langle \psi^\dag_{\bf q} \rangle &\simeq&\tilde z_{\bf q}    \langle \psi^\dag_{\bf q} \rangle_{\tilde{\mathcal H}} \, .
\end{eqnarray}
Note that in $\langle \psi^\dag_{\bf q}\rangle$ a smaller 
contribution from  \eqref{A.38} has been neglected.  
Thus  
\begin{align}
\label{A.44a}
 n^\psi_{\bf q} & = |\tilde{z}_{\vec{q}}|^{2} \big(\langle\psi_{\vec{q}}^{\dagger}\psi_{\vec{q}}^ {}\rangle_{\tilde{\mathcal{H}}}  -  \langle \psi^\dag_{\bf q} \rangle_{\tilde{\mathcal H}} 
 \langle \psi_{\bf q} \rangle_{\tilde{\mathcal H}}
\big)  \nonumber \\
%
%&+& 
& + \frac{1}{N}\sum_{\vec{k}}|\tilde{v}_{\vec{q}\vec{k}}|^{2}\tilde{n}_{-\vec{k}}^{h}\tilde{n}_{\vec{k}+\vec{q}}^{e} \, ,
\end{align}
where the expectation values on the right-hand side are formed with $\tilde{\mathcal H}$.
 With Eq.~\eqref{A.4} they become  
\begin{align}
\label{A.45}
\langle\psi_{\vec{q}}^{\dagger}\psi_{\vec{q}}^ {}\rangle_{\tilde{\mathcal{H}}} 
&=\langle \Psi^\dagger_\mathbf{q} \Psi^{}_\mathbf{q}\rangle_{\tilde{\mathcal{H}}}
-\frac{\sqrt{N}\tilde{\Gamma}}{\tilde{\omega}_\mathbf{q}}
\langle \Psi^\dagger_\mathbf{q} + \Psi^{}_\mathbf{q}\rangle_{\tilde{\mathcal{H}}}\delta_{\mathbf{q,0}}
+\frac{N\tilde{\Gamma}^2}{\tilde{\omega}^2_\mathbf{k}}\delta_{\mathbf{k,0}}\nonumber\\
&=p(\tilde{\omega}_\mathbf{q})+\frac{N\tilde{\Gamma}^2}{\tilde{\omega}^2_\mathbf{q}}\delta_{\mathbf{q,0}}\,,
\end{align}
and 
\begin{equation}
\label{A.46}
 \langle \psi^\dag_{\bf q} \rangle_{\tilde{\mathcal H}} = \Big[\langle \Psi^\dag_{\bf q} \rangle_{\tilde{\mathcal H}} 
 - \frac{\sqrt N  \, \tilde \Gamma}{\tilde \omega_{\bf q}} \Big] \delta_{\bf q, 0} =  
 - \frac{\sqrt N \, \tilde \Gamma}{\tilde \omega_{\bf q}}  \delta_{\bf q, 0} \, ,
\end{equation} 
where we have used $\langle \Psi^\dag_{\bf q} \rangle_{\tilde{\mathcal H}} =0$, and  
$p(\tilde{\omega}_{\mathbf q})$ is the bosonic distribution function. 
Inserting Eqs.~\eqref{A.45} and \eqref{A.46} into \eqref{A.44a}, one finally arrives at
\begin{align}
\label{A.47}
n_{\vec{q}}^{\psi}=|\tilde{z}_{\vec{q}}^ {}|^{2}p(\tilde{\omega}_{\vec{q}}^ {})+\frac{1}{N}\sum_{\vec{k}}|\tilde{v}_{\vec{q}\vec{k}}|^{2}\tilde{n}_{-\vec{k}}^{h}\tilde{n}_{\vec{k}+\vec{q}}^{e}\,,
\end{align} 
and similarly
\begin{align}
\label{A.47a}
&\tilde n^\psi_{\vec q} = \langle \delta \psi^\dag_{\bf q} \delta \psi_{\bf q}\rangle_{\tilde{\mathcal H}}
\approx p(\tilde{\omega}_{\bf q}) \, .
\end{align} 
Obviously, the electronic order parameter $d_{\bf k}$ and the photonic 
order parameter $\tilde\Delta_{\vec k}$ are intimately related. Due to \eqref{A.41} and  \eqref{A.42c}, 
$d_{\bf k}$ is proportional to $\tilde{\Delta}_{\vec k}$, so that both order parameters are 
mutually dependent. 

Note that in Sec.~\ref{IV} the numerical outcome of $n^e_{\vec k}$ 
and $n^h_{\vec k}$ will turn out to be the same. 
The reason for this is the assumed symmetric dispersions for the electron and hole bands 
in Eq.~\eqref{2}, $\varepsilon_{\vec k}^e = \varepsilon_{\vec k}^h$.  
As a consequence, also the original Hamiltonian \eqref{1} shows a certain symmetry: Replacing 
all electron operators $e^{(\dag)}_{\vec k}$ by hole operators $h^{(\dag)}_{\vec k}$ and vice versa,
Hamiltonian \eqref{1} remains the same, except of the sign of the prefactor $g$, i.e.,
\begin{equation}
\label{A.48}
 {\mathcal H}(\{ e^{(\dag)}_{\vec k} \},   \{ h^{(\dag)}_{\vec k} \}, g , U ) =
  {\mathcal H}(\{ h^{(\dag)}_{\vec k} \},   \{ e^{(\dag)}_{\vec k} \}, -g , U ) \,.
 \end{equation}
 A closer inspection shows that the former {\it ansatz} \eqref{A.28} for $e^\dag_{\vec k}(\lambda)$  can be
 transformed to the {\it ansatz} \eqref{A.29} for $h^\dag_{\vec k}(\lambda)$. The same is true for the 
 corresponding renormalization equations of the prefactors in \eqref{A.36} and \eqref{A.37}.  
 Note that
 the property $n^e_{\vec k} = n^h_{\vec k}$ would no longer be valid in case different  dispersions 
 $\varepsilon_{\vec k}^e \neq \varepsilon_{\vec k}^h$ are used. 
 However, also for the latter case the above renormalization equations remain valid.

%---------------------------------------------------------------------------------------------
\section{Luminescence  functions}
\label{App B}

Let us first evaluate the response function for the excitonic polarization  \eqref{24} which reads as after the unitary invariance has been employed
 \begin{eqnarray}
\label{B.1}
A({\bf k}, \omega) &=&  \frac{1}{2\pi} \int_{-\infty}^\infty 
\langle [\tilde{b}_{\bf k}(t), \tilde{b}^\dag_{\bf k}]_-\rangle_{\tilde{\mathcal H}} \,  e^{i\omega t} dt  \, .
\end{eqnarray}
The expectation value is formed with the fully renormalized Hamiltonian $\tilde{\mathcal H}$.
The quantity  $\tilde{b}^\dag_{\bf k}$ is the transformed exciton  creation operator 
\begin{equation}
 \label{B.2}
\tilde{b}_{\vec{k}}^{\dagger}=\frac{1}{\sqrt{N}}\sum_{\vec{p}} \tilde e_{\vec{k}+\vec{p}}^{\dagger}
\tilde h_{-\vec{p}}^{\dagger}
\, ,
\end{equation}
where the unitary transformation has been applied separately to the two one-particle operators $\tilde{e}_{\mathbf{k}}^{\dagger}$ and $\tilde{h}_{\mathbf{k}}^{\dagger}$. Inserting Eqs.~\eqref{A.36} and \eqref{A.37} into expressions
\eqref{B.1} and \eqref{B.2}, one obtains  for $A(\vec k, \omega)$:
\begin{align}
\label{B.5}
A(\vec k,\omega)=A^{coh}(\vec k,\omega) + A^{inc}(\vec k,\omega)\,,
\end{align}
 where the two parts will henceforth be denoted as coherent and incoherent.  The coherent part is given by
\begin{align}
\label{B.6}
A^{coh}&(\vec k,\omega)= \frac{1}{N} \sum_{\vec p}|\tilde{x}_{\vec k+\vec p}\tilde{y}_{-\vec p}|^2\nonumber\\
\times&\{|\xi_{\vec k+\vec p}\eta_{\vec p}|^2[f(\tilde E^1_{\vec p})-f(\tilde E^1_{\vec k+\vec p})]\delta(\omega
-\tilde E^1_{\vec k+\vec p}+\tilde E^1_{\vec p})\nonumber\\
&+ |\eta_{\vec k+\vec p}\eta_{\vec p}|^2[f(\tilde E^1_{\vec p})-f(\tilde E^2_{\vec k+\vec p})]\delta(\omega
-\tilde E^2_{\vec k+\vec p}+\tilde E^1_{\vec p})\nonumber\\
&+|\xi_{\vec k+\vec p}\xi_{\vec p}|^2[f(\tilde E^2_{\vec p})-f(\tilde E^1_{\vec k+\vec p})]\delta(\omega
-\tilde E^1_{\vec k+\vec p}+\tilde E^2_{\vec p})\nonumber\\
&+|\eta_{\vec k+\vec p}\xi_{\vec p}|^2 [f(\tilde E^2_{\vec p})-f(\tilde E^2_{\vec k+\vec p})]\delta(\omega
-\tilde E^2_{\vec k+\vec p}+\tilde E^2_{\vec p})\}. 
\nonumber \\
&
\end{align}
It follows from the dominant contributions $ \propto \tilde x_{\vec k}e^{\dag}_{\vec k}$ 
and $\propto \tilde y_{\vec k} h^\dag_{\vec k }$  in  Eqs.~\eqref{A.36} and \eqref{A.37}. 
In addition, the one-particle operators $e^{(\dag)}_{\vec k}$ 
and $h^{(\dag)}_{\vec k}$ have to be expressed by the dynamical eigenvectors $C^{1,2}_{\vec k}$, 
which leads to the appearance of the Bogoliubov coefficients $\xi_{\vec k}$ and $\eta_{\vec k}$ in \eqref{B.6}. 

The incoherent part $A^{inc}(\vec k,\omega)$ of the response function \eqref{B.1} reads to order $\mathcal O(g^2)$ and $\mathcal O(U^2)$:
\begin{align}
\label{B.7}
A^{inc}(\vec{k},\omega)= &\Pi^0_{\vec k}\delta[\omega-\tilde{\omega}(\vec k)]
-\frac{1}{N}\sum_{\vec p}\Pi^1_{\vec p\vec k}\delta[\omega-E^1_{\vec p}(\vec k)]\nonumber\\
&+\frac{1}{N^2}\sum_{i,\vec p\vec k_1}\Pi^{2,i}_{\vec p\vec k_1,\vec k}\delta[\omega-E^{2,i}_{\vec p\vec k_1}(\vec k)]\nonumber\\
&+\frac{1}{N^3}\sum_{i,\vec p\vec k_1\vec k_2}\Pi^{3,i}_{\vec p\vec k_1\vec k_2,\vec k}
\delta[\omega-E^{3,i}_{\vec p\vec k_1\vec k_2}(\vec k)],
\end{align}
with
\begin{align}
\label{B.8}
&E^1_{\vec p}(\vec k)=\tilde{\varepsilon}^e_{\vec k+\vec p}+\tilde{\varepsilon}^h_{-\vec p},\\
&E^{2,1}_{\vec p\vec k_1}(\vec k)=\tilde{\varepsilon}^h_{-\vec p}-\tilde{\varepsilon}^h_{-\vec k_1}
+\tilde{\omega}_{\vec k+\vec p-\vec k_1}, \\
&E^{2,2}_{\vec p\vec k_1}(\vec k)=\tilde{\varepsilon}^e_{\vec k+\vec p}-\tilde{\varepsilon}^e_{\vec p+\vec k_1}+\tilde{\omega}_{\vec p}, \\
&E^{3,1}_{\vec p\vec k_1\vec k_2}(\vec k)=\tilde{\varepsilon}^h_{-\vec p}-\tilde{\varepsilon}^h_{\vec k_1-\vec k_2
-\vec k-\vec p}+\tilde{\varepsilon}^h_{\vec k_2}+\tilde{\varepsilon}^e_{\vec k_1}, \\
&E^{3,2}_{\vec p\vec k_1\vec k_2}(\vec k)=\tilde{\varepsilon}^h_{-\vec p-\vec k_1+\vec k_2}-\tilde{\varepsilon}^e_{\vec k_2}+\tilde{\varepsilon}^e_{\vec k_1}+\tilde{\varepsilon}^e_{\vec k+\vec p}, 
\end{align}
and
\begin{widetext}
\begin{align}
\label{B.9}
&\Pi^0_{\vec k}=\frac{2}{N^2}\sum_{\vec p\vec k_1}\tilde{x}_{\vec k+\vec p}\tilde{y}_{-\vec k_1}\tilde{u}_{\vec k\vec p}\tilde{t}_{\vec k\vec k_1} \tilde n^e_{\vec p+\vec k}(1-\tilde n^h_{-\vec k_1})\,,\\
&\Pi^1_{\vec p}(\vec k)=2\tilde{x}_{\vec k+\vec p}\tilde{y}_{-\vec k}\frac{1}{N}
\sum_{\vec k_1}\left[\tilde{x}_{\vec k_1+\vec k}\tilde{\beta}_{\vec k+\vec p,\vec k_1+\vec k,-\vec p}
\tilde n^e_{\vec k_1+\vec k}-\tilde{y}_{-\vec k_1}\alpha_{\vec k+\vec p,\vec k_1+\vec k,-\vec p}(1
-\tilde n^h_{-\vec k_1})\right](1-\tilde n^e_{\vec k_1}-\tilde n^h_{-\vec p})\nonumber\\
&\qquad\qquad+\frac{1}{N^2}\sum_{\vec k_1\vec k_2}\Big\{2\tilde{x}_{\vec k+\vec p}\tilde{y}_{-\vec p}\alpha_{\vec k_2,\vec k_1+\vec k,-\vec p}\beta_{\vec k+\vec p,\vec k_2,-\vec k-\vec p+\vec k_2-\vec k_1}
\tilde n^e_{\vec k_2}(1-\tilde n^h_{-\vec k-\vec p+\vec k_2-\vec k_1})\nonumber\\
&\qquad\qquad\qquad\qquad+2\tilde{x}_{\vec k_1
+\vec k}\tilde{y}_{-\vec k_1}\tilde{\alpha}_{\vec k+\vec p,\vec k_1
+\vec k,-\vec p}\tilde{\beta}_{\vec k+\vec p,\vec k_2+\vec k,-\vec p}(1-\tilde n^h_{-\vec p})
\tilde n^e_{\vec k_2+\vec k}\nonumber\\
&\qquad\qquad\qquad\qquad-\tilde{y}_{-\vec k_1}\tilde{y}_{-\vec k_2}\tilde{\alpha}_{\vec k+\vec p,\vec k_1+\vec p,-\vec p}\tilde{\alpha}_{\vec k+\vec p,\vec k_2+\vec k,-\vec p}(1-\tilde n^h_{-\vec p})
(1-\tilde n^h_{-\vec k_2})\nonumber\\
&\qquad\qquad\qquad\qquad-\tilde{x}_{\vec k+\vec p}\tilde{x}_{\vec k_2
+\vec k}\tilde{\beta}_{\vec k_1,\vec k+\vec p,-\vec k_1
+\vec k}\tilde{\beta}_{\vec k_1,\vec k_2+\vec k,-\vec k_2+\vec k}n^e_{\vec k_2
+\vec k}\tilde n^e_{\vec k+\vec p}\Big\}(1-\tilde n^e_{\vec k+\vec p}-\tilde n^h_{-\vec p})\,,\\
&\Pi^{2,1}_{\vec p \vec k_1,\vec k}=|\tilde{y}_{-\vec p}\tilde{t}_{\vec k+\vec p-\vec k_1,\vec k_1}|^2
[\tilde n^h_{-\vec k_1}(1-\tilde n^h_{-\vec p})-\tilde n^\psi_{\vec k+\vec p-\vec k_1}
(\tilde n^h_{-\vec k_1}-\tilde n^h_{-\vec p})],\\\
&\Pi^{2,2}_{\vec p\vec k_1,\vec k}=|\tilde{x}_{\vec k+\vec p}\tilde{u}_{\vec k_1\vec p}|^2
[\tilde n^e_{\vec k+\vec p}(1-\tilde n^e_{\vec p+\vec k_1})-\tilde n^\psi_{\vec k_1}(\tilde n^e_{\vec p+\vec k_1}
-\tilde n^e_{\vec k+\vec p})],\\
&\Pi^{3,1}_{\vec p\vec k_1\vec k_2,\vec k}=\left(|\tilde{y}_{-\vec p}\tilde{\alpha}_{\vec k_1,\vec k+\vec p,\vec k_2}|^2-\tilde{y}_{-\vec p}\tilde{y}_{\vec k_2}\alpha_{\vec k_1,\vec k+\vec p,\vec k_2}\tilde{\alpha}_{\vec k_1,-\vec k_2+\vec k,-\vec p}\right)\nonumber\\
&\qquad\qquad\quad\times\left[(1-\tilde n^e_{\vec k_1})(1-\tilde n^h_{-\vec p})(\tilde n^h_{\vec k_1+\vec k_2-\vec k-\vec p}-\tilde n^h_{-\vec k_1})+\tilde n^h_{\vec k_2}(1-\tilde n^h_{\vec k_1+\vec k_2-\vec k-\vec p})(1-\tilde n^e_{\vec k_1}-\tilde n^h_{-\vec p})\right],\\
&\Pi^{3,2}_{\vec p\vec k_1\vec k_2,\vec k}=\left(|\tilde{x}_{\vec k+\vec p}\tilde{\beta}_{\vec k_1\vec k_2,-\vec p-\vec k_1+\vec k_2}|^2-\tilde{x}_{\vec k+\vec p}\tilde{x}_{\vec k_1}\beta_{\vec k_1\vec k_2,-\vec p-\vec k_1+\vec k_2}\tilde{\beta}_{\vec k+\vec p,\vec k_2,-\vec p-\vec k_1+\vec k_2}\right)\nonumber\\
&\qquad\qquad\quad\times\left[\tilde n^e_{\vec k+\vec p}\tilde n^h_{-\vec p-\vec k_1+\vec k_2}
(\tilde n^e_{\vec k+\vec p}-\tilde n^e_{\vec k_2})+\tilde n^e_{\vec k_2}(1-\tilde n^e_{\vec k_1})(1-\tilde n^e_{\vec k+\vec p}-\tilde n^h_{-\vec p-\vec k_1+\vec k_2})\right] \, .
\end{align}
\end{widetext}
Again all  expectation values on the right-hand sides are formed with the 
renormalized Hamiltonian $\tilde{\mathcal H}$.  Note that for simplicity $A^{inc}(\vec{p},\omega)$ 
 was calculated without use of the Bogoliubov transformation \eqref{A.23}.  The reason for this approximation 
results from the fact that $A^{inc}(\vec{p},\omega)$ turns out to be quite small compared to the coherent part
of $A(\vec k, \omega)$. Moreover, 
the additional sums in \eqref{B.9} tend to cover the influence of $\tilde \Delta_{\vec k}$
in $W_{\vec k}$ [compare Eq.~\eqref{A.25}]. \\

 Finally, we consider the response function for the cavity photon mode
\begin{eqnarray} 
\label{B.17}
B(\vec q, \omega)  &=&  \frac{1}{2\pi} \int_{-\infty}^\infty 
\langle [\tilde{\psi}_{\bf q}(t), \tilde{\psi}^\dag_{\bf q}]_-\rangle_{\tilde{\mathcal H}} \,  e^{i\omega t} dt  \, ,
\end{eqnarray}
where $\tilde{\psi}_{\vec q}^{\dag}$ is again the fully renormalized quantity.  According to \eqref{A.38}
we have 
\begin{eqnarray}
\label{B.18}
& \tilde{\psi}_{\mathbf{q}}^{\dagger}  =\tilde{z}_{\mathbf{q}}\psi_{\mathbf{q}}^{\dagger}+\frac{1}{\sqrt{N}}\sum_{\mathbf{k}}
\tilde{v}_{\mathbf{qk}}:e_{\vec{q}+\vec{k}}^{\dagger}h_{-\mathbf{k}}^{\dagger}:.
%\end{align} 
\end{eqnarray}
Using Eqs.~\eqref{23} and \eqref{A.26}, one easily finds
\begin{eqnarray}
\label{B.19}
&& B({\bf q}, \omega) = |\tilde z_{\bf q}|^2 \delta(\omega - \tilde{\omega}_{\bf q}) + \frac{1}{N} \sum_{\bf k} |\tilde v_{\vec q \vec k}|^2
\nonumber \\
&& \quad \times \big\{|\xi_{\vec k+\vec q}\eta_{\vec k}|^2[f(\tilde E^1_{\vec k})-f(\tilde E^1_{\vec k+\vec q})]
\delta(\omega -\tilde E^1_{\vec k+\vec q}+\tilde E^1_{\vec k})\nonumber\\
&&\quad + |\eta_{\vec k+\vec q}\eta_{\vec k}|^2[f(\tilde E^1_{\vec k})-f(\tilde E^2_{\vec k+\vec q})]
\delta(\omega -\tilde E^2_{\vec k+\vec q}+\tilde E^1_{\vec k})\nonumber\\
&& \quad +|\xi_{\vec k+\vec q}\xi_{\vec k}|^2[f(\tilde E^2_{\vec k})-f(\tilde E^1_{\vec k+\vec q})]\delta(\omega
-\tilde E^1_{\vec k+\vec q}+\tilde E^2_{\vec k})\nonumber\\
&&\quad +|\eta_{\vec k+\vec q}\xi_{\vec k}|^2 [f(\tilde E^2_{\vec k})-f(\tilde E^2_{\vec k+\vec q})]\delta(\omega
-\tilde E^2_{\vec k+\vec q}+\tilde E^2_{\vec k})\big \}. \nonumber \\
&& \\
&& \nonumber 
\end{eqnarray}
Note that, apart from the first $\delta$ function and the prefactor under the sum, the
result for $B(\mathbf q, \omega)$ resembles  that of the coherent contribution $A^{coh}(\mathbf k, \omega)$ of the excitonic polarization.

\end{appendix}
\bibliography{ref}
\bibliographystyle{apsrev}

\end{document}